\documentclass[%
notitlepage,
twocolumn,
 amsmath,amssymb,
]{revtex4-1}

\usepackage{graphicx}% Include figure files
\usepackage{dcolumn}% Align table columns on decimal point
\usepackage{bm}% bold math
\usepackage{units}
\usepackage{ulem}		% to use \sout command

\usepackage{xcolor}
\usepackage{mathtools}
\usepackage{bbold}

\usepackage{empheq}

\usepackage[most]{tcolorbox}

\tcbset{colback=yellow!10!white, colframe=black!50!black, 
        highlight math style= {enhanced, %<-- needed for the ’remember’ options
            colframe=black,colback=red!10!white,boxsep=0pt}
        }

\newcommand{\bv}[1]{\mathbf{#1}}
\newcommand{\kk}{\mathbf{k}}

\newcommand{\QQ}{\mathbf{Q}}
\newcommand{\qq}{\mathbf{q}}
\newcommand{\rr}{\mathbf{r}}

\newcommand{\EE}{\mathbf{E}}
\newcommand{\PP}{\mathbf{P}}

\newcommand{\p}[1]{\frac{\partial}{\partial #1}}
\newcommand{\pp}[1]{\frac{\partial^2}{\partial {#1} ^2}}
\newcommand{\pll}{_\parallel}

\definecolor{amber}{rgb}{0.82, 0.1, 0.26}

\newcommand{\changes}[1]{{\color[rgb]{0,0,0}{#1}}}

\begin{document}
%\Large{Key results: Förster-type interaction between molecule and TMDC}
\normalsize
%\tableofcontents
\preprint{APS/123-QED}

\title{Impact of dark excitons on F\"orster type resonant energy transfer between dye molecules and atomically thin semiconductors}

\author{Manuel Katzer$^{1}$}
\email{manuel.katzer@physik.tu-berlin.de}
\author{Sviatoslav Kovalchuk$^2$}
\author{Kyrylo Greben$^2$}
\author{Kirill I. Bolotin$^2$}
\author{Malte Selig$^1$}
\author{Andreas Knorr$^1$}
\affiliation{$^1$Technische Universit\"at Berlin, Institut f\"ur Theoretische Physik, Nichtlineare Optik und Quantenelektronik, Hardenbergstra{\ss}e 36, 10623 Berlin, Germany}
\affiliation{$^2$Freie Universit\"at Berlin, Department of Physics, Quantum Nanoelectronics of 2D Materials, Arnimallee 14, 14195 Berlin, Germany}

\date{\today}

\begin{abstract}
Interfaces of dye molecules and two-dimensional transition metal dichalcogenides (TMDCs) combine strong molecular dipole excitations with high carrier mobilities in semiconductors. F\"orster type energy transfer is one key mechanism for the coupling between both constituents. We report microscopic calculations of a spectrally resolved F\"orster induced transition rate from dye molecules to a TMDC layer. Our approach is based on microscopic Bloch equations which are solved self-consistently together with Maxwells equations. This approach allows to incorporate the dielectric environment of a TMDC semiconductor, sandwiched between donor molecules and a substrate. Our analysis reveals transfer rates in the meV range for typical dye molecules in closely stacked structures, with a non-trivial dependence of the F\"orster rate on the molecular transition energy resulting from unique signatures of dark, momentum forbidden TMDC excitons.
\end{abstract}

\maketitle

\section{Introduction}
Hybrid inorganic and organic systems (HIOS) are a promising platform for future optoelectronic applications since they combine appealing properties of two different material classes. Organic molecules show highly tunable transition energies and provide large optical dipole moments as the respective excitons are of the Frenkel type~\cite{blumstengel2006OM,nurmikkogroup2007OM,friendgruppe2012OM,koppensgruppe2013OM,kochgruppe2015OM,koppensgruppe2015OM,smithgroup2017OM,judith2018,Zhao2019,plochocka2021HIOS,kochgruppe2021,feierabend2021dark}.
Transition metal dichalcogenites (TMDCs) are inorganic, atomically thin semiconductors that have stimulated research in the last years: Their atomic thickness leads to a reduced screening of the Coulomb interaction and consequently to the formation of stable, bound electron hole pairs, namely Wannier type excitons, which dominate the optical properties in the vicinity of the band edge~\cite{chernikov2014rydbergseries,gunnar2014wannier,Steinhoff2017fission,chernikov2018RevModPhys,selig2020suppression,reiter2020phonon,Trovatello2020excitonproperties,katsch2020exciton,shoudong2021,selig2022impact,holler2022interlayer}. The large oscillator strength of optically bright excitons makes TMDC monolayers ideal resonant substrates for energetically tunable molecules, but also for quantum dots~\cite{Bolotingruppe2015QD,raja2016energy,zhu2019QD,akshaygruppe2020QD}, NV~centers \cite{Tisler2011NVcenter} and plasmonic structures \cite{baumberggruppe2017plasmo,stengergruppegoldaufws2_2019,zhi-heng2021plasmo,carlson2021strong}. On the other hand, it is known that optically dark, momentum forbidden TMDC excitons contribute strongly to the excitonic dynamics \cite{zhang2015experimental,arora2015excitonic,selig2016excitonic,christiansen2017phononen,selig2018dark,selig2019PRR,christiansen2021strong} and may become visible if the translational invariance is broken by a zero-dimensional molecular emitter.
\begin{figure}[t!]
 \begin{center}
 \includegraphics[width=\linewidth]{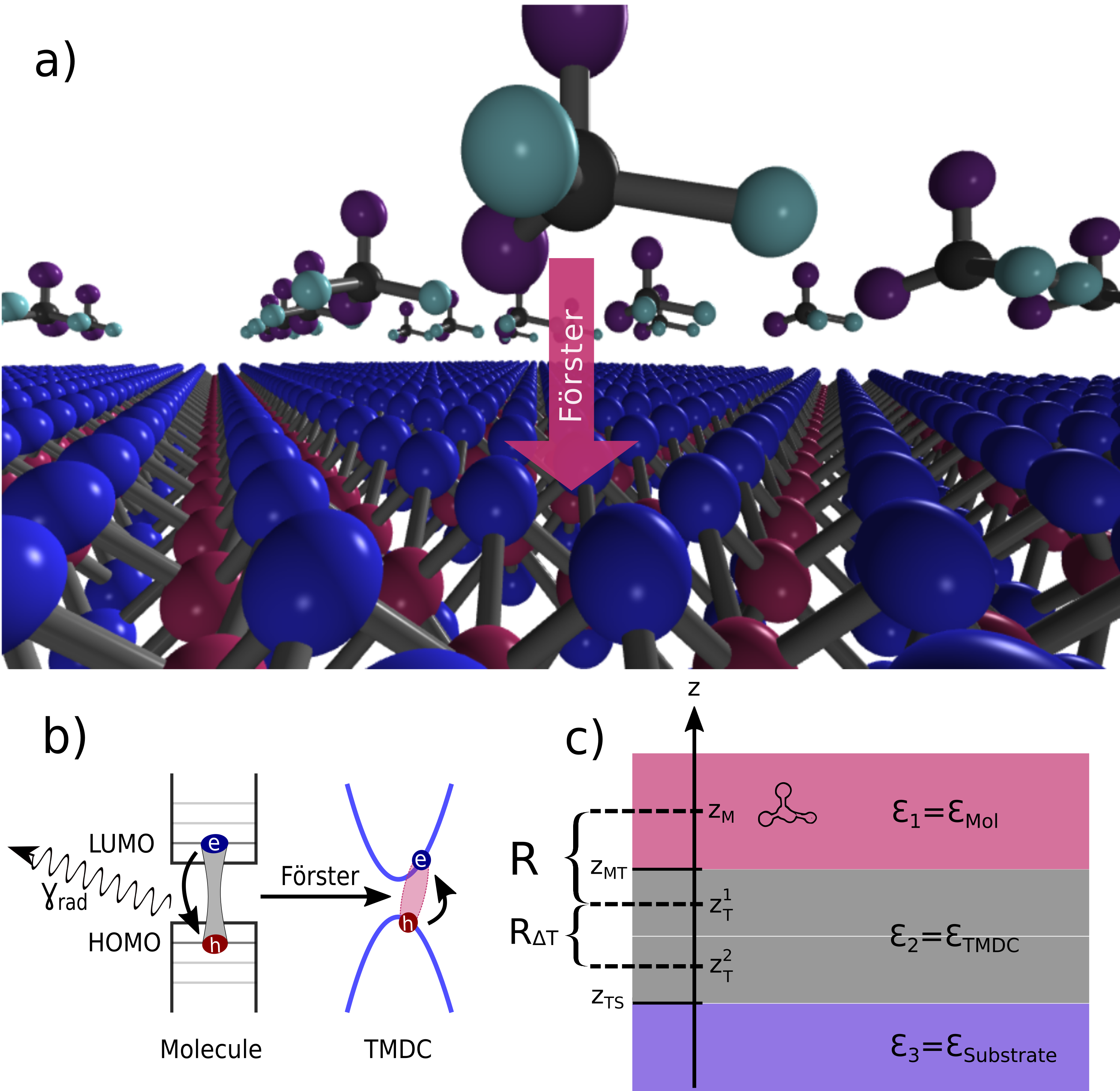}\\
 \end{center}
\caption{a) Schematic illustration: HIOS of a sheet of dye molecules above one or two layers of TMDC, b) scheme of the F\"orster energy transfer from DCM molecules to the MoS$_2$ TMDC in momentum space, c) sketch of the dielectric environment in real space.}
 \label{fig:sketch}
\end{figure}
The HIOS considered here is built from organic donor molecules, e.g. dichloromethane (DCM) as adsorbants on a TMDC mono- or bilayer, e.g. MoS$_2$, which serves as an acceptor for the energy transfer.
 While the TMDC Wannier excitons start to interact at relatively low densities, molecular Frenkel excitons interact more weakly but can produce comparably stronger optical signatures. For more general reviews on mixed-dimensional heterojunctions, the key interaction processes and application perspectives, see also \cite{hersamgroup2017natmatreview,weissgruppe2020review0d2d}. 
The process of F\"orster resonant energy transfer (FRET) \cite{foersteroriginal1948}, is one among several excitation exchange processes known to be important between thin film materials \cite{froehlicher2018}. While other processes like Dexter- and tunneling transfer require an electronic wavefunction-overlap, Förster transfer is conveyed by spatially non-overlapping dipole-dipole contributions of the Coulomb interaction and is more flexible with respect to geometry, and observable at much larger distances compared to processes with direct charge transfer, which makes it interesting for a plethora of technical applications~\cite{DEMIR2011FRETQDLED,junkergruppe2018fret}.
Already in 2009, Swathi and Sebastian \cite{Swathi2009} argued that resonant energy transfer from a localized donor no longer shows the typical $R^{-6}$-distance dependence in the case of a delocalized acceptor. They find an exponential dependence for short distances and a $R^{-4}$-power law dependence for longer distances. Their work was later expanded for a molecule-graphene structure \cite{Malic2014,berciaudgruppe2015distance} and more recently also for a graphene-TMDC heterostack \cite{PhysRevB.99.035420}. The $R^{-4}$-dependence is also reflected in experiments, see e.g.~\cite{koppensgruppe2013OM,mazzamuto2014single,Bolotingruppe2015QD}. 
In this manuscript, we present a microscopic study on the F\"orster coupling at interfaces of organic molecules and TMDC monolayers (compare Fig.~\ref{fig:sketch}), and discuss significant deviations from the mentioned $R^{-4}$ limit for small distances in the range of few nanometers between the constituents. \changes{In particular, our spectrally resolved approach opens the possibility to directly measure the energy distribution of momentum-dark excitonic states in the TMDC, which are only indirectly accessible in optical experiments \cite{selig2018dark,Malic2018Dark,lindlau2018role,brem2020phonon,shoudong2021}, due to the neglible in-plane momentum of far field plane wave excitation. We show that the symmetry breaking of the translational invariance due to a finite momentum distribution introduced by the spatially localized molecular scatterers allows the excitation of those momentum-dark states by scattered light, initializing the Förster process. This provides a more direct access to states outside of the lightcone of optical experiments, and thus spectrally resolved FRET rates can make those states visible in linear absorption and photoluminescence experiments.}
The paper is structured as follows: In section~\ref{Sec:Theo}, the theoretical model is sketched, necessary mathematical details are provided in the appendix. In section~\ref{Sec:Foe}, we discuss the energy transfer rate from the organic molecule to the TMDC layer, and discuss its dependence on the molecular transition energy and the spatial separation of both layers. We find F\"orster rates in the range of meV for the exemplary interface of DCM on MoS$_2$. The rate shows a maximum well above the optically active transition energies, which originates from formerly momentum dark TMDC excitons, activated by breaking the translational invariance in HIOS. In order to connect our results to experiments, in section~\ref{Sec:Lin}, we discuss the impact of F\"orster coupling on the linear optical properties such as reflection and transmission. We find a significant broadening of the spectra for molecular transition energies above the excitonic resonance due to the coupling to momentum dark TMDC excitons. In section~\ref{Sec:Photo} we calculate the photoluminescence emission of the molecules, and find pronounced dips in the molecular PL due to the interplay between the F\"orster induced recombination and the radiative decay of optically excited molecular transitions. In section~\ref{Sec:Concl} we conclude with a summary.
\section{Theoretical Model}\label{Sec:Theo}
\subsection{Hamiltonian and Bloch equations}
As an interface we propose a layer of molecules distributed over a mono- or bilayer TMDC, on top of a substrate, compare Fig.~\ref{fig:sketch}, (a) real space, (b) momentum space, (c) dielectric environment. Throughout this work we assume a sparse and random distribution of the molecules parallel to the TMDC layer, brought in position by evaporation~\cite{jarvinen2013molecular,Tsai2015dipolealignment,zhao2019strong}. A structured spatial distribution is rather the exception than the rule as long as the inter-molecular spacing is large compared to the distance of each molecule to the TMDC~\cite{Tsai2015dipolealignment}. This allows to calculate the coupling for a single molecule on top of the TMDC layer, as inter-molecular interactions are negligible, and assume an inhomogenous broadening due to the varying transition energies of the molecules~\cite{bondarev2004fluorescence}.
The Hamiltonian of the electronic excitations in the heterostructure is given by
\begin{align}
    H  
    &=\sum_{\mu\ell\QQ\pll} \mathcal{E}_\mathbf{Q_\parallel}^{\mu\ell} \hat p_\mathbf{Q_\parallel}^{\dagger\mu\ell } \hat p_\mathbf{Q_\parallel}^{\mu\ell} 
    - \sum_{\mu\ell\QQ\pll} \mathbf{E}_{\QQ\pll} (z_T^\ell) \cdot \Big( \bv{d}\varphi_{r=0}^{\mu\ell}
     \hat p_\mathbf{Q_\parallel}^{\dagger \mu\ell} 
    + h.c. \Big)\nonumber \\
    &\qquad
    + 
    \mathcal{E}_M^{21} \hat\sigma_{21}
    \hat\sigma_{12}
    - 
    \mathbf{E} (\mathbf{r}_M)\cdot
    \Big(\mathbf{d}^{12} \hat\sigma_{12} + h.c.\Big),\label{eq:hamiltonian}
\end{align}
where the first line accounts for the TMDC sheet and the second for the molecules. The coupling of the compounds is mediated via the electric field $\mathbf{E}$, determined by Maxwell's equations. The first term accounts for the dispersion $\mathcal{E}_\mathbf{Q_\parallel}^{\mu\ell}$ of excitons in the bosonic zero density limit, with excitonic annihilation (creation) operators $\hat p_{\QQ\pll}^{(\dagger)\mu\ell }$~\cite{katsch2018theory} with the Fourier component of the in-plane center of mass motion $\mathbf{Q}_\parallel$. The layer index $\ell$ stands for the different TMDC layers (e.g. monolayer: $\ell=1$, or bilayer: $\ell=1,2$). (Note that only monolayer TMDCs are direct semiconductors, additional layers change the screening and can thus make the semiconductor indirect, and consequently, the exciton ground state becomes dark~\cite{mak2010atomically,chernikov2018RevModPhys}. We therefore examine both mono- and bilayers in this work. For PL measurements, we suggest a bilayer TMDC, where the excitation in the TMDC will quickly decay to the energetically lower lying momentum-indirect intervalley exciton states and thus will not contribute relevantly to the luminescence \cite{splendiani2010emerging,wurstbauer2017light,gerber2019interlayer}. This makes it possible to measure the PL of the molecules without significant contributions from the TMDC, see Sec.~\ref{Sec:Photo}.)
The super index $\mu = (\xi,\lambda,s)$ includes the valley $\xi=K,K'$ of the electron and hole which the exciton $\hat p_\mathbf{Q_\parallel}^{\dagger\mu\ell} $ is built of, the index for the different bound and unbound energetic states $\lambda=(s1,s2..)$ and the spin of those carriers $s=\uparrow,\downarrow$, yielding Rydberg like energy series A and B, respectively~\cite{selig2019PRR}. Note that we focus on excitons in the $K,K$ (and $K',K'$) valley and omit intervalley excitons with electron and hole in different valleys (K,K'), since the F\"orster interaction does not provide sufficiently high momenta to activate these states. The appearing excitonic dispersion $\mathcal{E}_\mathbf{Q_\parallel}^{\mu\ell} = \mathcal{E}^{\mu\ell}+\frac{\hbar^2\QQ\pll^2}{2M}$ is determined by $\mathcal{E}^{\mu\ell}$, the sum of free gap energy and the binding energy for each excitonic state $\mu = (\xi,\lambda,s)$ obtained by evaluating the Wannier equation~\cite{kira2006many}, and additionally by the kinetic energy of the excitons with effective mass $M$. The required parameters for the underlying electronic dispersion are taken from \cite{kormanyos2015k,deilmann2018interlayer} and are listed in the appendix. The second term in Eq.~(\ref{eq:hamiltonian}) represents the field matter interaction at the position $z_T^\ell$ of the TMDC, which in the case of a bilayer gives two positions in the respective layers, compare Fig.~\ref{fig:sketch}~(c). We introduce the Fourier component of the electric field with respect to the in-plane direction $\mathbf{E}(\mathbf{r}_\parallel,z) = \sum_\mathbf{Q_\parallel} e^{i \mathbf{Q}_\parallel \cdot \mathbf{r}_\parallel} \mathbf{E}_{\mathbf{Q}_\parallel}(z)$, and $\bv{d}\varphi_{r=0}^{\mu\ell}$ as the excitonic dipole moment with the wavefunction in real space at position $r=0$~\cite{selig2020suppression} which accounts for the circular dichroism. The strength of the underlying electronic dipole matrix element is taken from ab initio calculations~\cite{xiao2012coupled}. 
The second line in Eq.~(\ref{eq:hamiltonian}) accounts for the molecular Hamiltonian, with a dominant molecular transition $\hat\sigma_{12} = \hat a_1^\dagger \hat a_2$ defined via annihilation (creation) $a^{(\dagger)}_i$ operators for molecular orbitals, where subscript 1 stands for the \changes{ highest occupied molecular orbital (HOMO)} and subscript 2 for the \changes{lowest unoccupied molecular orbital (LUMO)} state, respectively. The first term in the second line thus represents the energy $\mathcal{E}_M^{21}$ of the free molecular electronic transitions~\cite{bondarev2004fluorescence}, described as fermions, a reasonable approach in linear optics~\cite{may2008charge}. The second term describes the interaction of the molecules with the electric field $\mathbf{E} (\mathbf{r}_M)$ at the position $\rr_M$ of the molecule. $\mathbf{d}^{12}$ accounts for the molecular dipole moment, which can be taken from related ab initio calculations, see e.g.~\cite{verdenhalven2014dipolemomentmolek}. % for quarterphenyl (L4P) molecules. 
For a detailed derivation of the field matter interaction Hamiltonian, we refer to App.~\ref{app:fieldmatterappendix}. Note that the Hamiltonian so far does not explicitly contain the coupling between the TMDC excitons and the molecular orbitals, however, both couple to the jointly experienced electromagnetic field $\mathbf{E}$, which mediates the mutual Förster coupling in the near field.
To obtain a macroscopic variable, we focus on the macroscopic polarization $\PP = \langle \hat \PP \rangle$ of the heterostructure, defined by $H_{int} = -\int d^3 r\, \hat\PP\cdot\hat\EE$, which reads
\begin{align}
    \PP(\rr)
    &=
    \big(
    \mathbf{d}^{12}
    \sigma_{12}
    +
    h.c.
    \big)
    \delta(\rr-\rr_{M})\nonumber\\
    &\qquad
    +
    \sum_{\mu\ell\QQ\pll}
    (
    e^{i\QQ\pll\cdot \rr\pll}
    \bv{d}\varphi_{r=0}^{\mu\ell}
    p_{\QQ\pll}^{\mu\ell}
    +
    h.c.)
    \delta(z-z_T^\ell)
\end{align}
Here, the first term accounts for the expectation value of the polarization induced by the molecular transitions $\sigma_{12} = \langle \hat\sigma_{12} \rangle$. The second term accounts for the response of each TMDC exciton governed by the excitonic polarization $p_{\QQ\pll}^{\mu\ell}=\langle \hat p_{\QQ\pll}^{\mu\ell}\rangle$~\cite{katsch2018theory}. The Fourier transform of the macroscopic polarization with respect to the in-plane component of the TMDC layer reads (note that exploiting the cylindrical symmetry of the single molecule problem, we can set $\rr_M^\parallel=0$):
\begin{align}
    \PP_{\QQ\pll}(z)
    &=
    \mathbf{d}^{12}
    \sigma_{12}
    \delta(z-z_{M})\nonumber\\
    &\qquad
    +    
    \sum_{\mu\ell}
    \bv{d}\varphi_{r=0}^{\mu\ell}
    p_{\QQ\pll}^{\mu\ell}
    \delta(z-z_T^\ell)\label{eq:Polarization}
    +
    h.c.
\end{align}
The Bloch equations for the polarization $\sigma_{12}$ of the molecule  and the excitonic polarization $p_{\QQ\pll}^{\mu\ell}$ of the TMDC are derived by exploiting Heisenbergs equation of motion:
\begin{align}\label{eq:sigmaBloch}
    i\hbar \partial_t
    \sigma_{12}
    &=
    \mathcal{E}_M^{21}
    \sigma_{12}
    -
    \bv{d}^{21}
    \cdot
    \EE(\rr_M)\\
    i\hbar \partial_t p_{\QQ\pll}^{\mu\ell}
    &=
    \mathcal{E}_{\QQ\pll}^{\mu\ell}
    p_{\QQ\pll}^{\mu\ell} 
    - (\bv{d}\varphi_{r=0}^{\mu\ell})^*\cdot\EE_{\QQ\pll}(z_T^\ell)\label{eq:pBloch}
\end{align}
The first terms on the right hand side of Eqs.~(\ref{eq:sigmaBloch},\ref{eq:pBloch}) account for the oscillation with the orbital gap $\mathcal{E}^{21}_{M}$ between the HOMO and the LUMO state in the molecule, and the excitonic energy $\mathcal{E}_{\QQ\pll}^{\mu\ell}$ in the TMDC layer, respectively. The second terms describe the coupling of both constituents to the joint electric field $\mathbf{E}$.

\subsection{Near-Field Solution of the Wave Equation}

The Förster transfer between the two mentioned constituents is mediated via the electric field $\EE_{\QQ\pll}(z)$ in Eqs.~(\ref{eq:sigmaBloch},\ref{eq:pBloch}). Assuming a small spacing between the molecular layer and the TMDC layer we ignore retardation effects of the electric field in the near field and start with the quasistatic wave equation.
We take advantage of the translational invariance of the dielectric environment in in-plane direction $\epsilon(\rr)=\epsilon(z)$, assuming that the molecules are sparsely distributed such that their off-resonant electronic transitions do not significantly contribute to the dielectric background, which can therefore be described by the function $\epsilon(z)\in\{\epsilon_1,\epsilon_2,\epsilon_3\}$, compare~Fig~\ref{fig:sketch}~(c). 
We can write (for each layer i seperately):
\begin{align}\label{eq:helmholtzeq}
    \epsilon_i\epsilon_0
    \nabla^2
    \EE_i(\rr)
    =
    -
    \nabla
    \big(\nabla\cdot\PP_i(\rr)\big).
\end{align}
The corresponding Helmholtz equation is solved using a Greens dyade $\mathcal{G}_{\QQ_\parallel}(z,z')$, where the solutions in the different layers are connected via boundary conditions following a Rytova-Keldysh-type approach~\cite{rytova1967,rytova2020}, compare~App.~\ref{app:rytovaappendix}.
The externally applied field is added as the homogeneous solution of Eq.~(\ref{eq:helmholtzeq}), compare \cite{christiansen2022mesoscale}
\begin{align}\label{eq:Greensfunctionapproach}
    \EE_{\QQ_\parallel}(z)
    =
    \int dz'
    \mathcal{G}_{\QQ_\parallel}(z,z')
    \PP_{\QQ_\parallel}(z') 
    + \EE_{\QQ\pll}^0(z),
\end{align}

Eq.~(\ref{eq:Greensfunctionapproach}), together with Eqs.~(\ref{eq:Polarization}), (\ref{eq:sigmaBloch}) and (\ref{eq:pBloch}), gives a coupled set of equations for the polarizations of both the molecular transition $\sigma_{12}$ and the TMDC excitons $p_{\QQ\pll}^{\mu\ell}$
\begin{align}\label{eq:foerster_bloch_eq_sigma}
    i\hbar \partial_t
    \sigma_{12}
    &=
    \Big[\mathcal{E}_{M}^{12}-i\gamma_{M}\Big]
    \sigma_{12}\nonumber\\
    &\qquad
    +
    \sum_{\mu\ell\QQ\pll}
    V_{\QQ\pll \mu\ell 12}^{T\rightarrow M}(z_T^\ell,z_M)
    p_{\QQ\pll}^{\mu\ell}
    -
    \bv{d}^{21}\cdot
    \EE_0(z_M,t)
    \\
    i\hbar \partial_t p_{\QQ\pll}^{\mu\ell}
    &=
    \Big[
    \mathcal{E}_{\QQ\pll}^{\mu\ell}
     -
    i\gamma^{\mu\ell}
    \Big]
    p_{\QQ\pll}^{\mu\ell}\nonumber\\
    &\qquad
    +
    V_{12\QQ\pll \mu\ell}^{M\rightarrow T}(z_M,z_T^\ell)
    \sigma_{12}
    -
    (\bv{d}\varphi_{r=0}^{\mu\ell})^*
    \cdot
    \EE_{\QQ\pll}^0(z_T^\ell,t)
    \label{eq:foerster_bloch_eq_p}
\end{align}
The constants~$\gamma_M$ and $\gamma^{\mu\ell}$ were added phenomenologically to account for radiative and non-radiative dephasing in the molecule and the TMDC, respectively. In the latter, above cryogenic temperatures, dephasing due to exciton-phonon, and exciton-exciton interaction are dominant compared to radiative losses, and can be calculated accordingly~\cite{christiansen2017phononen,selig2019PRR,reiter2020phonon}. 
In Eq~(\ref{eq:foerster_bloch_eq_sigma}) we again make use of $\rr_M^\parallel = 0$ and write for the externally applied electric field $\EE_0(z_M,t)=\sum_{\QQ\pll}\EE_{\QQ\pll}^0(z_M,t)$. The Förster-Hamiltonian that directly gives Eqs.~(\ref{eq:foerster_bloch_eq_sigma},\ref{eq:foerster_bloch_eq_p}) is included in~App.~\ref{app:Hamiltonianappendix}.
The coupling (as shown in App.~\ref{app:rytovaappendix}) is given by
\begin{align}\label{eq:coupling_M}
V_{\QQ\pll \mu\ell 12}^{T\rightarrow M}(z_T^\ell,z_M)
 &=
 \frac{1}{A\epsilon\epsilon_0}
 \bv{d}^{21}
 \cdot
 \mathcal{G}_{\QQ\pll}(z_T^\ell,z_M)
 \cdot
 \bv{d}\varphi_{r=0}^{\mu\ell}\\
V_{12\QQ\pll \mu\ell}^{M\rightarrow T}(z_M,z_T^\ell)
&=
 \frac{1}{\epsilon\epsilon_0}
 (\bv{d}\varphi_{r=0}^{\mu\ell})^*
 \cdot
 \mathcal{G}_{\QQ\pll}(z_M,z_T^\ell)
 \cdot
 \bv{d}^{12}\label{eq:coupling_V}
\end{align}
with the Greens dyade
\begin{align}\label{eq:Greensdyade}
    \mathcal{G}_{\QQ\pll}(z,z')
    =
    \left(
    \begin{array}{cc}
     \QQ^T\pll G_{\QQ\pll}(z,z')\QQ\pll &  
     i \p{z'}
   G_{\QQ_\parallel}(z,z')\QQ\pll \\
  i\QQ^T\pll \p{z'}
   G_{\QQ_\parallel}(z,z')
   & 
   -\pp{z'} G_{\QQ_\parallel}(z,z')
    \end{array}
    \right)
\end{align}
Here, the Greens function $G_{\QQ\pll}$ is the solution of a scalar Poisson equation fulfilling the respective boundary conditions. \changes{Note that Eqs.~(\ref{eq:coupling_M}-\ref{eq:Greensdyade}) are given in a rather general formulation, which allows an application also to related systems, i.e. to for instance also include nonzero z-components of the dipols of donor and acceptor.}

\subsection{Rytova-Keldysh type Greens dyade}\label{subsec:rytovasection}
\changes{The Greens function in~Eq.~(12) takes the dielectric environment of the heterostructure into account, compare~Fig.~1~(c) (layer of molecules in vacuum $\epsilon_1=1.0$, TMDC layers $\epsilon_2=13.36$,  substrate layer $\epsilon_3=3.9$, ab initio values taken from \cite{berkelbach2013theory}). Note that this background screening is present in addition to the resonant response of molecule and TMDC computed from our two band model, and has to be taken into account to give realistic values for the transfer rates, as it accounts for the screening which is caused by off-resonant transitions in the TMDC and the substrate.} The Greens function can be derived in the spirit of Rytova-type solutions, Eqs.~(\ref{eq:coupling_M},\ref{eq:coupling_V}), with boundary conditions~\cite{rytova1967,rytova2020}. The details of the derivation can be found in App.~\ref{app:rytovaappendix}. In Eq.~(\ref{eq:VT-Mabk}) the dielectric constants are abbreviated as $\delta_{21}=\frac{\epsilon_2-\epsilon_1}{\epsilon_2+\epsilon_1}$ and $\delta_{23}=\frac{\epsilon_2-\epsilon_3}{\epsilon_2+\epsilon_3}$, respectively. $z_M$ and $z_T^\ell$ refer to the positions of the molecule (donor) and the TMDC (acceptor) of the transfer process, while $z_{MT}$ and $z_{TS}$ refer to the boundaries between the involved dielectric environments, compare~Fig.~\ref{fig:sketch}~(c) and App.~\ref{app:rytovaappendix}. In agreement with recent observations \cite{sigl2022optical}, we assume that the z-components of the dipoles~$d_z\varphi_{r=0}^{\mu\ell}\approx 0$ in the TMDCs, and consider the molecular dipoles to be aligned with respect to the in-plane coordinate, i.e. $d_z^{12}\approx 0$. \changes{Finite angles between the dipole elements change the strength of the Förster coupling, but not the momentum selection rules which determine the spectral dependence of the rate, see App.~\ref{app:anglesappendix}.} However, the validity of the assumption of aligned dipoles is reflected in experiments~\cite{Tsai2015dipolealignment}. These assumptions lead to the coupling

\changes{
\begin{align}\label{eq:VT-M}
V_{\QQ\pll \mu\ell 12}^{T\rightarrow M}(z_T^\ell,z_M)
&=
\frac{1}{A}
(V_{12\QQ\pll \mu\ell }^{M\rightarrow T}(z_M,z_T^\ell))^*
\nonumber\\
&=
\frac{1}{A}
\mathcal{V}^{\mu\ell}_{ \QQ\pll}
e^{-Q\pll(z_M-z_T^\ell)}
\end{align}}
with 
\begin{align}\label{eq:VT-Mabk}
    &\mathcal{V}^{\mu\ell}_{ \QQ\pll}
    = 
    \frac{\big(
   \bv{d}\pll^{21}
   \cdot    
   \QQ\pll
   \big)
   \big(
   \QQ_\parallel
   \cdot
   \bv{d}\pll\varphi_{r=0}^{\mu\ell}
   \big)
   \big(
    \delta_{23}e^{-2Q\pll (z_M-z_{TS})}
    +
    1
    \big)
   }{Q\pll\epsilon_0(\epsilon_2+\epsilon_1)(1-\delta_{21}\delta_{23}e^{2Q\pll(z_{TS}-z_{MT})})}
\end{align}
 For the following discussion, the thickness of one TMDC layer is referred to as $R_{\Delta T}=z_{MT}-z_{TS}$, for MoS$_2$ we assume a thickness of $R_{\Delta T}=0.6$~nm~\cite{rasmussen_thygesen_2015}. All parameters are listed in~App.~\ref{app:parameterappendix}. The distance $R = z_M - z_T^1$ between molecule and uppermost semiconductor is an important parameter for the Förster rate, compare~Fig.~\ref{fig:sketch}~(c).

\subsection{Analytical approach to the Förster rate}
For optical spectroscopy in frequency space $\omega$, the energy transfer can be quantitatively described by solving Eqs.~(\ref{eq:foerster_bloch_eq_sigma},\ref{eq:foerster_bloch_eq_p}), which gives for the molecular polarization
\begin{align}\label{eq:sigmawitheigen}
    \hbar\omega
    \sigma_{12}
    &=
    \Big[
    \mathcal{E}_{M}^{12}
    -
    i\gamma_{M}
    -
    \Sigma_{F}(\omega)
    \Big]
    \sigma_{12}
    -
    \mathbf{d}^{21} \cdot \EE_0(z_M,\omega)
\end{align}

with the complex self-energy of the Förster transfer from the molecule to the TMDC exciton
\begin{align}\label{eq:foerstereigenenergy}
    \Sigma_{F}(\omega)
    &=
    \mathcal{E}_{renorm}
    +
    i
    \gamma_{F}
    %^{\mathcal{E}_{M}^{12}}
    \nonumber\\
    &
    =
    \sum_{\mu\ell\QQ\pll}
    \frac{V_{\QQ\pll \mu\ell 12}^{T\rightarrow M}(R,R_{\Delta T})
    V_{12\QQ\pll \mu\ell }^{M\rightarrow T}(R,R_{\Delta T})}
    {\hbar\omega-\mathcal{E}_{\QQ\pll}^{\mu\ell}
    +i\gamma^{\mu\ell}}
\end{align}
The real part of this self-energy leads to an energy  renormalization, and thus to a shift of the absorption line, compare Sec.~\ref{Sec:Lin}. The imaginary part, however, provides the rate $\gamma_{F}(\omega) = Im\big(\Sigma_{F}(\omega)\big)$ of the Förster energy transfer from the molecule to the TMDC exciton continuum $\QQ\pll$.
In the limit of negligible additional dephasing of the TMDCs, i.e.  $\gamma^{\mu\ell} \approx 0$, which is a reasonable approximation at low temperatures, Eq.~(\ref{eq:foerstereigenenergy}) can be integrated analytically. Due to the strict resonance condition, the rate then directly depends on the molecular transition energy $\mathcal{E}_{M}^{12}$:
\begin{align}\label{eq:foersterrate}
\gamma_{F}^{\mathcal{E}_{M}^{12}}
    &=\sum_{\mu\ell}
    \frac{(d\pll\varphi_{r=0}^{\mu\ell})^2(d^{12})^2M}{8\epsilon_{0}^2(\epsilon_1+\epsilon_2)^2\hbar^2}
    % \big(
    % 3\cos^2\vartheta
    % -
    % 2
    % \big)
    \Theta(Q_0^\parallel)
    e^{-2Q_0^\parallel R}
    \nonumber\\
    &\quad
    \times
    % \nonumber\\
    % &\qquad\qquad\qquad
    % \times
    \frac{(Q_0^\parallel)^2(1+\delta_{23}
    e^{-Q_0^\parallel(R + \frac{3}{2}R_{\Delta T})})^2
    e^{-2Q_0^\parallel (\ell-1)R_{\Delta T}}}
    {(1-\delta_{21}\delta_{23}e^{4Q_0^\parallel R_{\Delta T}})^2}
\end{align}

Here, the cut-off in-plane momentum $Q_0^\parallel$, resulting from the integration via the energy conserving delta function is given by the detuning of molecular and excitonic resonance:
\begin{align}\label{eq:inplanemomentum-energyconservation}
Q_0^\parallel=\sqrt{
    \frac{2M}{\hbar^2}
    \big(
    \mathcal{E}_{M}^{12}
    -\mathcal{E}^{\mu\ell}
    \big)},
\end{align}
reflecting that the molecular transition energy in the energy conserving Markov approximation must exceed at least the lowest excitonic resonance in order for the Förster energy transfer to be energetically allowed, as only positive detuning $\Delta = \mathcal{E}_{M}^{12}
-\mathcal{E}^{\mu\ell} > 0$ leads to non-zero transition rates. (Note that there is no measurable Förster process taking place in the opposite direction from TMDC to molecule, due to the symmetry breaking between 0d donor and 2d acceptor, see also Sec.~\ref{Sec:Lin}.)
This analytical approximation is in good agreement with the numerical results from Eq.~(\ref{eq:foerstereigenenergy}), as can be seen in~App.~\ref{app:comparenumericalanalytical}. As typical for Förster transfer, the rate depends on the squares of the optical dipole matrix elements $d\pll\varphi_{r=0}^{\mu\ell},d^{12}$ of both constituents. Screening due to the dielectric environment occurs due to corrections $\delta_{12}$, $\delta_{23}$, respectively. This includes the case of homogeneous dielectric environment $\epsilon_1=\epsilon_2=\epsilon_3$, where $\delta_{12}=0=\delta_{23}$ results in the usual $\frac{1}{\epsilon}$-dependence of the Förster rate. Moreover, for small enough distances, the rate shows a combination of exponential and power law decay as a function of the distance $R$ of molecule and TMDC sheet. For larger distances, the dependence is a power law $(R^{-4})$, in agreement with earlier work, compare~App.~\ref{app:distanceappendix}.
In the subsequent sections we use the numerical results for the evaluation of Eq.~(\ref{eq:foerstereigenenergy}) at a finite temperature, which is reflected by a finite non-radiative dephasing $\gamma^{\mu\ell} > 0$ in the TMDC~\cite{selig2016excitonic}, and in Sec.~\ref{Sec:Lin} we also discuss the mentioned energy shift resulting from the real part of the self-energy $Re(\Sigma_F)$. However, the most important physical aspects can also be seen in the analytical result, Eq.~(\ref{eq:foersterrate}), compare~App.~\ref{app:comparenumericalanalytical}. 
\section{Förster rate}\label{Sec:Foe}
\begin{figure}[t!]
 \begin{center}
\includegraphics[width=\linewidth]{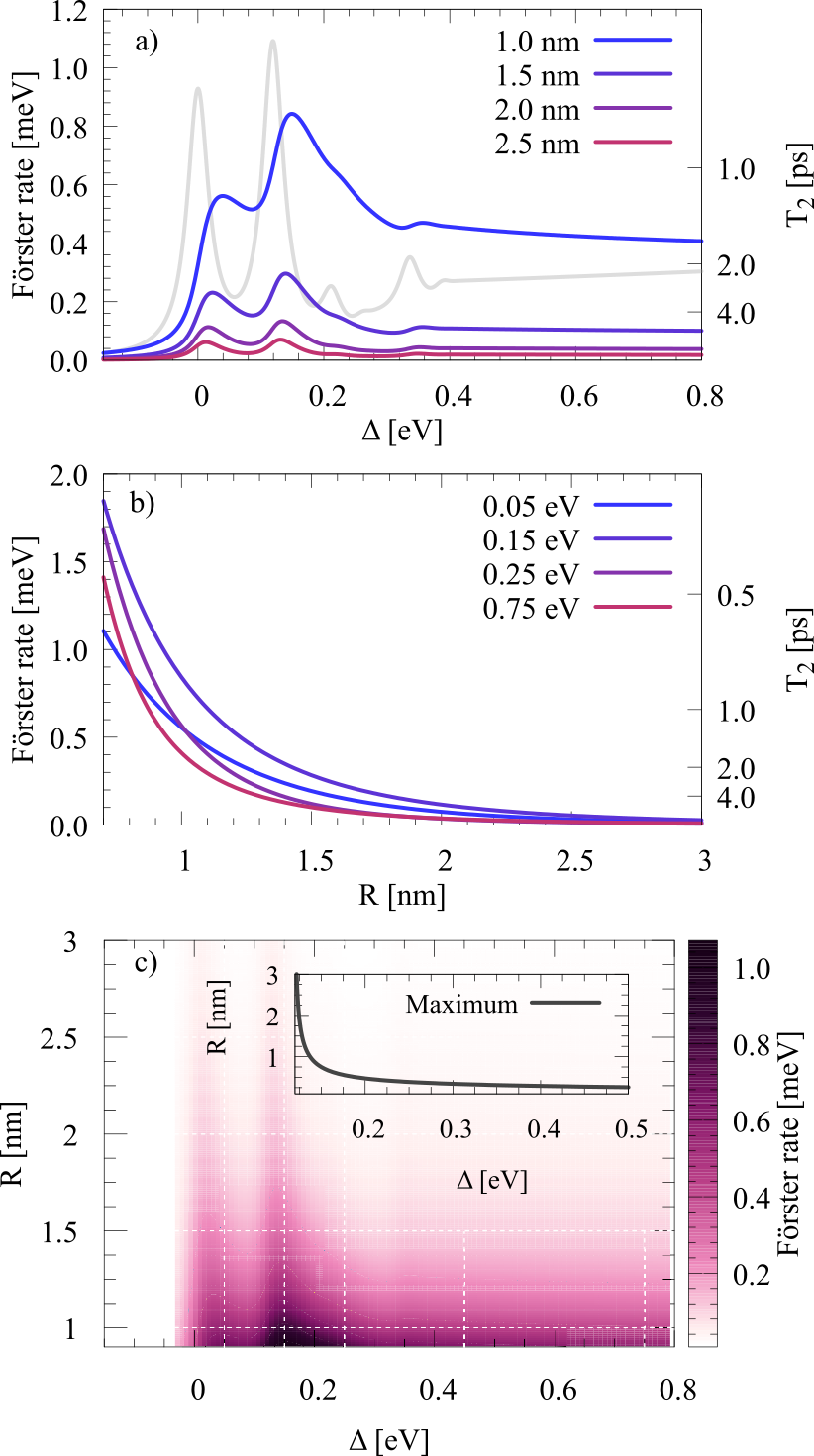}
 \end{center}
\caption{
 F\"orster induced transition rate for the energy transfer from optically pumped molecular excitons into the TMDC monolayer, plotted a) over the detuning~$\Delta$ between the transition energy of the donor and the 1s-resonance energy in the acceptor for different molecule-TMDC distances $R$.
 %The grey line gives the linear absorption of the excitonic Rydberg series (A and B exciton combined) of the acceptor (MoS$_2$ excitons). 
 Clearly, the excitonic states are visible in the spectrally resolved FRET. The plateau between A and B resonance result from the momentum dark excitons above the bright 1s exciton, thus the Förster rate clearly does not resemble the absorption (grey line), which only displays the bright exciton resonances. b) Rate plotted over the distance $R$ between donors (DCM molecules) and acceptor (MoS$_2$-monolayer), for selected values of detuning $\Delta$. 
 %Only physically accessible distances are shown, since the semiconductor has a finite thickness $>0.5$ nm. 
 c) F\"orster rate as a two parameter plot of $\Delta$ and $R$. Dashed lines depict the respective positions of the 1d plots (a,b). The inset shows the Maximum of the rate with respect to distance $R$ and detuning~$\Delta$. For parameters cf.~App.~\ref{app:parameterappendix}.
}
 \label{fig:foerster_mono}
\end{figure} 
For the numerical evaluation of the rate ($\gamma_F^{\mathcal{E}_M^{12}} = Im(\Sigma_F)$ in Eq.~(\ref{eq:foerstereigenenergy})) we choose dichloromethane (DCM) molecules on MoS$_2$, all used material parameters can be found in App.~\ref{app:parameterappendix}. In particular, we include a broadening due to exciton-phonon interaction, i.e. apply a non-radiative dephasing in the TMDC of $\gamma^{\mu\ell}=\unit[20]{meV}$, which is a realistic value at room temperature~\cite{selig2016excitonic,christiansen2017phononen}. For lower temperatures, the lines become sharper but qualitatively show the same characteristics, compare~App.~\ref{app:comparenumericalanalytical}. In the following we present the numerically calculated Förster rates on monolayer MoS$_2$, and then also discuss the differences to a bilayer substrate in the subsequent section. %

\subsection{Monolayer}
Fig.~\ref{fig:foerster_mono} depicts the F\"orster rate for molecules on a monolayer TMDC, as a function of two parameters, namely the detuning~$\Delta =  \mathcal{E}_{M}^{21}-\mathcal{E}_\mathbf{0}^{A,1s}$ of the molecular transition energy with respect to the TMDC exciton 1s~A~resonance, and the spatial distance $R$ of molecule and TMDC layer. All plots show the imaginary part of the self-energy, Eq.~(\ref{eq:foerstereigenenergy}). In Fig.~\ref{fig:foerster_mono}~(a) the F\"orster rate is plotted over the detuning~$\Delta$ for selected distances. We find that the F\"orster rate increases as the detuning becomes positive ($\Delta>\unit[0]{eV}$), reflecting the energy conservation of the transfer of molecular excitation to the momentum dark TMDC excitons above the optically active exciton. At all distances, the rate exhibits pronounced peaks which can be traced back to MoS$_2$ excitonic resonances: the first peak originates from dark ($Q_0^\parallel>0$) 1s A excitons, the second peak from relaxation into 1s B excitons, respectively. Above the 1s B peak also smaller peaks can be recognized which are assigned to higher excitonic bound states (2s, etc.), until the F\"orster rate reaches a constant value when the detuning reaches the excitonic continuum. Interestingly, the observed peaks are blueshifted with respect to the excitonic energies measured in pristine linear optical response, which is shown as a grey curve in Fig.~\ref{fig:foerster_mono}~(a). The reason for this shift is that due to the vanishing excitonic center of mass momentum, for $Q_0^\parallel \rightarrow 0$, the F\"orster matrix element disappears directly at the resonance, compare Eq.~(\ref{eq:inplanemomentum-energyconservation}). A finite center of mass momentum $Q_0^\parallel$ and corresponding kinetic energy have to be provided via excess energy, thus shifting all resonances in the F\"orster rate to slightly higher energies compared to the corresponding lines in bare TMDC linear response. Nevertheless, the spectrally resolved FRET rate reflects the excitonic structure of the substrate. This information can be directly related to results from optical spectroscopy, as analyzed in Secs.~\ref{Sec:Lin} and \ref{Sec:Photo}.
As shown in Fig.~2~(b), the F\"orster rate decreases as a function of the distance $R$ for all detunings, which originates from the exponential $R$ dependence in the F\"orster coupling elements Eq.~(\ref{eq:VT-M}). For short distances, we observe a crossover behavior for detunings large enough to transfer molecular excitation into the TMDC continuum, while for small detunings, only the bound exciton states can be addressed, leading to a rate that prevails for slightly longer distances. For even longer distances of several tens of nm, Eq.~(\ref{eq:foerstereigenenergy}) shows a power law dependence of $R^{-4}$, independent of the detuning, which is in agreement with previous calculations~\cite{Swathi2009,Malic2014}, compare the discussion in App.~\ref{app:distanceappendix}.
Fig.~\ref{fig:foerster_mono}~(c) is a heatmap showing the whole two dimensional parameter space of detuning $\Delta$ and distance $R$ dependence of the Förster rate. Dashed lines show where the cuts for plots in (a) and (b) are located. The inset shows how the maximum of the Förster rate is shifted towards lower detunings as the distance increases, reflecting again the $R$ dependence of the coupling elements.

\subsection{Bilayer}

\begin{figure}[t!]
 \begin{center}
\includegraphics[width=\linewidth]{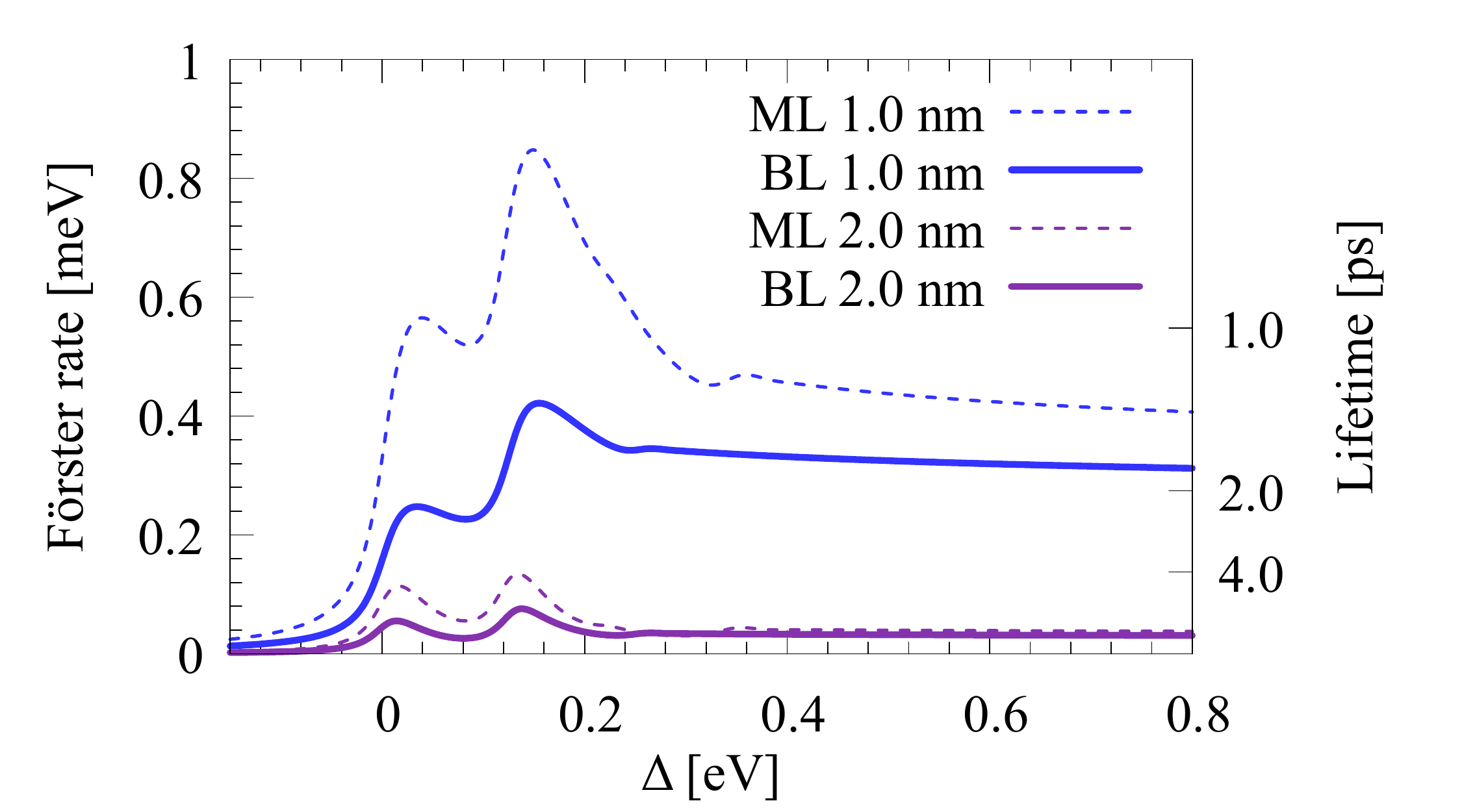}
 \end{center}
\caption{
 Comparison of F\"orster transition rates between TMDC monolayer (ML, dashed lines, from Fig.~\ref{fig:foerster_mono}) and bilayer (BL, full lines) substrates, for two different distances $R=\unit[1]{nm},\unit[2]{nm}$. Although in the BL the number of acceptor states is doubled, the rate is significantly smaller than for the ML, due to strongly increased screening. Besides, the continuum is energetically closer to the s1 resonance, as the screening yields smaller exciton binding energies in both layers.}
 \label{fig:comp_bilayer}
\end{figure} 

As already mentioned, the MoS$_2$ bilayer is an indirect semiconductor, with a ground state energetically below the s1 resonance of the monolayer, which is however not accessible by the Förster transfer due to its  small optical dipole~\cite{mak2010atomically,chernikov2018RevModPhys}. When using a bilayer TMDC as an acceptor, we thus see resonances at the same energies as in the monolayer. Moreover, one might naively expect stronger Förster rates compared to the monolayer, as the number of states possibly excited by the energy transfer are doubled. However, this is overcompensated by the significantly increased screening due to the addition of the second semiconductor layer, compare~Fig.~\ref{fig:comp_bilayer}. The reason is, that in the monolayer, only the substrate screens the electric field. This provides strong binding energies for the excitons of over $\unit[300]{meV}$, and strong optical dipole moments $d\pll\varphi_{r=0}^{\mu\ell}$, which give strong Förster rates, compare~Eq.~(\ref{eq:foersterrate}). In the bilayer, the binding energies are smaller due to the strong screening of the respective other layer, resulting in a continuum which is energetically closer to the s1A-resonance, and an overall Förster rate of only around half the rates expected for the monolayer, compare~Fig.~\ref{fig:comp_bilayer}. 
In order to clarify whether and how the F\"orster induced energy transfer from the dye molecules to the TMDC can be accessed by optical experiments, in the following sections we calculate the linear coherent optical response (Sec.~\ref{Sec:Lin}) and the luminescence (Sec.~\ref{Sec:Photo}) of the molecules both with and without the TMDC substrate.

\section{Linear spectroscopy}\label{Sec:Lin}
\subsection{Single molecule}
\begin{figure}[t!]
 \begin{center}
    \includegraphics[width=\linewidth]{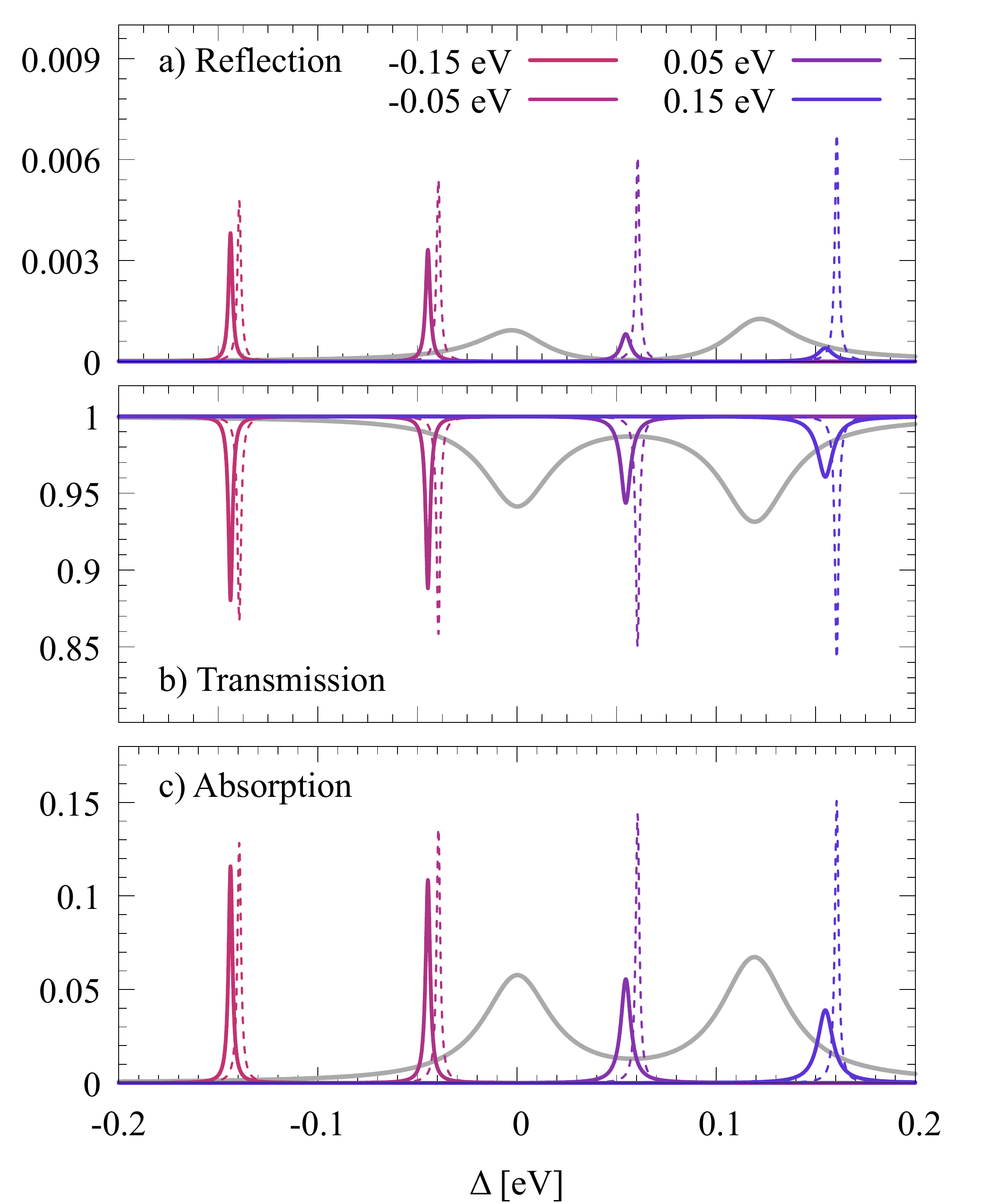} \end{center}
\caption{Calculated (a) reflection, (b) transmission and (c)~absorption of the DCM-MoS$_2$ stack, as a function of the detuning $\Delta$ relative to the energetically lowest 1s A exciton transition energy for a sweep of four different molecular transition energies below and above the pristine semiconductor 1s A-resonances (additionally depicted in gray). The detuning $\Delta$ ranges from $\unit[-0.15]{eV}$ to $\unit[0.15]{eV}$. The linear optical response of the underlying TMDC layer was subtracted for better visibility of the effect, and is depicted in gray. This is justified due to the small $\chi$, cf.~App.~\ref{app:full_spectrum}. For comparison, we illustrate with dashed lines the pristine molecular spectra without TMDC substrate.}
 \label{fig:abs_spec}
\end{figure}
%
%\manu{Andreas hier wolltest du nächstes mal mit dem Lesen beginnen.}
In this section we discuss the accessibility of the spectrally resolved Förster rate in linear, coherent optical response experiments, such as reflection and transmission. We exploit Eqs.~(\ref{eq:Polarization}), (\ref{eq:foerster_bloch_eq_sigma}) and (\ref{eq:foerster_bloch_eq_p}) to derive an expression for the linear optical susceptibility of the combined structure~\cite{kira2006many}. By self-consistent solution of Maxwell's equations, we calculate the transmission $T(\omega)$ and reflection $R(\omega)$ of the heterostack explicitly~\cite{knorr1996theory,stroucken1996coherent}. In the following we find an unidirectional Förster coupling, i.e. even if both constituents are optically excited, the excitation flow follows only the direction from molecule to TMDC, (or at least, only this direction is visible in the linear spectrum). As already mentioned, this effect is connected to the breaking of the translational invariance between 0d donor and 2d acceptor, as we will show in the following, compare~Eqs.~(\ref{eq:chitmdc}-\ref{eq:chimol}). This reflects the character of the TMDC excitons as a reservoir for the molecular excitation.
As the distance $R$ between the layers is small we can assume $e^{i\sqrt{\epsilon}\frac{\omega}{c}R}
	\approx 
	1$ and thus write for the true absorption
\begin{align}\label{eq:absorption}
	\alpha(\omega)
	=
	1-T(\omega)-R(\omega)
	=
	\frac{\frac{\omega}{\sqrt{\epsilon}c}Im\big(\chi_{_{TMDC}}+\chi_{_{Mol}}\big)}
	{|1-
		\frac{i\omega}{2\sqrt{\epsilon}c}\big(\chi_{_{TMDC}}
		+
		\chi_{_{Mol}}\big)|^2}
\end{align}
Strictly speaking, there is a nonlinear dependence on the susceptibilities $\chi=\chi_{_{Mol}}+\chi_{_{TMDC}}$ in~Eq.~(\ref{eq:absorption}). However, the susceptibilities are small, and we can still treat the response as linear in good approximation, see App.~\ref{app:full_spectrum} for the full spectra.
The susceptibility for the TMDC mono- and bilayer is already well documented~\cite{gunnar2014wannier,chernikov2014rydbergseries,chernikov2018RevModPhys}, and reads, when including the self-energy from the Förster transition,
\begin{align}\label{eq:chitmdc}
    \chi_{_{TMDC}}(\omega)
    =
    \frac{1}{\epsilon_0}
    \sum_{\mu\ell}
    \frac{(d\pll\varphi_{r=0}^{\mu\ell})^2}    
    {
    E^{\mu\ell}
    -
    \hbar\omega
    -
    i\gamma_{rad}^{\mu\ell}
    +
    \Sigma_{TMDC}^{\mu\ell,Q\pll=0}(\omega)}.
\end{align}
The self-energy of the TMDC, however, does not contribute to the coherent linear response which measures only bright excitons ($Q\pll=0$). The coupling $V_{\QQ\pll \mu\ell 12}^{T\rightarrow M}(z_T^\ell,z_M)$ vanishes at $Q\pll \rightarrow 0$, compare~Eq.~(\ref{eq:coupling_M}), and thus:
\begin{align}\label{eq:TMDCeigenE}
    &\Sigma_{TMDC}^{\mu\ell,Q\pll=0}(\omega)
    =
    \frac{|V_{\QQ\pll=0, \mu\ell 12}^{T\rightarrow M}(z_T^\ell,z_M)|^2}{\hbar\omega - \mathcal{E}_{M}^{12}
    +i\gamma_M}
    =
    0
\end{align}
Note that unlike in the case of the self energy of the molecule (Eq.~\ref{eq:foerstereigenenergy}), the self energy of the TMDC does not contain a sum over the in-plane momenta $Q\pll$, which reflects the symmetry breaking between 0d donor and 2d acceptor. The Förster transfer, which, as shown before, needs nonzero momentum $Q\pll > 0$, is thus only visible in the direction from molecule to TMDC, and not vice versa. 
The molecular susceptibility for the dominant HOMO-LUMO transition reads
\begin{align}\label{eq:chimol}
    \chi_{_{Mol}}(\omega)
    =
    \frac{1}{\epsilon_0}
    \frac{(d^{12})^2}{
    \mathcal{E}_{M}^{12}
    -
    \hbar\omega
    -
    i\gamma_{M}
    -
    \Sigma_{F}(\omega)},
\end{align}
here the Förster process does enter in the Lorentzian denominator via the already discussed complex self-energy, compare~Eq.~(\ref{eq:foerstereigenenergy}), as the sum over all in-plane momenta also takes non-zero contributions into account.

Fig.~\ref{fig:abs_spec} (a) illustrates the calculated reflection spectra (full lines) for different single molecular transition energy detunings in the range of \unit[-0.15]{eV} to \unit[0.15]{eV} with respect to the transition energy of the 1s A resonance in MoS$_2$. Dashed lines give the pristine response of a molecule with a certain transition energy (color code). The respective full lines show the response with underlying TMDC substrate.
Note that for all of Fig.~\ref{fig:abs_spec}, we show the optical response of the molecules, with the TMDC response subtracted. For the full response, compare~App.~\ref{app:full_spectrum}.
We plot only the monolayer case, as we showed in the previous section that the effect for a bilayer is less pronounced but similar. For all transition energies, the spectra show a redshift and a resonance broadening compared to the pristine case without Förster coupling (dashed lines). The shift is due to the real part of the Förster self-energy, compare~Eq.~(\ref{eq:foerstereigenenergy}). Above the excitonic transition energies (gray line), the Förster rate acts as an additional decay channel for the molecular excitations, which leads to a significant broadening of the reflection peak. This broadening of the molecular line results from the coupling of the molecular transition to the continuum of momentum dark excitons, which are activated by the finite momentum that can be transferred between the spatially localized molecule and the translationally invariant TMDC plane. The first Förster-type decay channel opens as soon as the detuning $\Delta$ becomes positive, i.e. the molecular transition energy exceeds the resonance energy of the lowest semiconductor exciton state (1s of the A exciton). (Fig.~\ref{fig:abs_spec} shows the scenario at room temperature, where non-radiative dephasing smears out the energy conservation, and thus this is already slightly affecting the lines directly beneath the resonance.) When the transition energy exceeds also the lowest resonance of the B exciton, this opens a second Förster type channel into the resonance of the 1s B-exciton, resulting in an even further broadening of the linear response.
Fig.~\ref{fig:abs_spec}~(b) illustrates the transmission spectra of the dye molecules. As for the reflection, we find that for all molecular transition energies the spectra are redshifted and experience a significant broadening when the molecular transition energy exceeds the 1s A transition.
Finally, Fig.~\ref{fig:abs_spec}~(c) illustrates the calculated absorption. Again, we find a redshift of all lines as well as a broadening of the lines for molecular transition energies above the 1s A transition.
\begin{figure}[t!]
\begin{center}
\includegraphics[width=\linewidth]{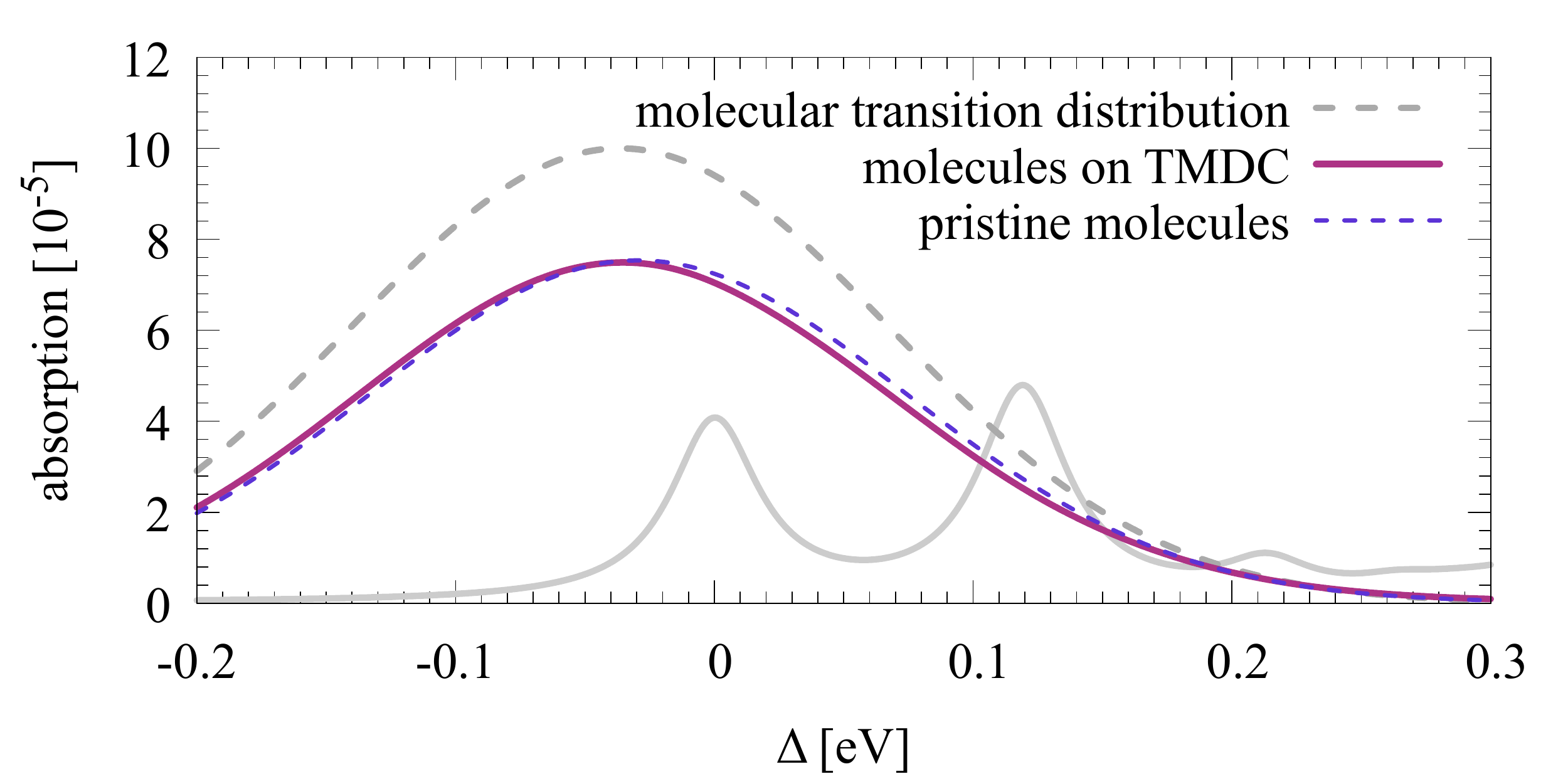} \end{center}
\caption{Absorption for a Gaussian distribution of transition energies (pink) in the molecules, compared to the case for pristine molecules (dashed purple). As for the single molecular case, a slight overall redshift can be recognized, but clearly the effect of the Förster process above the excitonic resonances (gray) can hardly be seen. The Gaussian distribution of tranistion energies is depicted in dashed gray lines. It is evident that linear spectroscopy of molecular ensembles does not provide access to the spectrally resolved Förster rate.}
 \label{fig:abs_dos}
\end{figure}

\subsection{Molecular ensemble}\label{subsec:molecularabsorption}
In experiments, the molecular layer will however show significant inhomogeneous broadening, i.e. a continuous range of molecular transition energies \cite{bondarev2004fluorescence}, and thus the linewidths of single molecules typically cannot be observed directly. Other types of 0d-donors show more controllable transition energies, and thus might provide an opportunity to observe single emitter effects (e.g. DBT molecules \cite{mazzamuto2014single,pazzagli2018self}, or NV centers \cite{Tisler2011NVcenter}). 
In order to simulate a sheet of loosely distributed dye molecules with a Gaussian distribution of inhomogeneously broadenend transition energies, we include a summation j over all molecules in the layer, and introduce a density $n$ of molecules per area, which we estimate to be $n=\unit[0.05]{nm^{-2}}$, a reasonable value for sparsely distributed molecules~\cite{Tsai2015dipolealignment}. This implies susceptibilities small enough to ignore nonlinear effects in Eq.~(\ref{eq:absorption}). Significantly higher densities would make inter-molecular transitions more likely and exceed the scope of this study. In order to account for an energetically dense Gaussian distribution of energies, we furthermore define a density of states (DOS) of transition energies $     DOS(\omega)
\equiv
\sum_j \delta(\omega-\frac{1}{\hbar}\mathcal{E}_{M}^{12,j})
\equiv
ae^{-\frac{(\omega-\omega^j)^2}{b}}$, where the values of a and b are taken from typical experiments \cite{bondarev2004fluorescence}. (Note that the maximum of this DOS can slightly differ for absorption and emission energy, but as this does not affect the results in our theory, it will not be taken into account here.)
With this we can write
\begin{align}\label{eq:dos_chi}
    \chi_{_{Sheet}}
    &=
    \sum_j \chi_{_{Mol}}^j\nonumber\\
    &=
    \frac{n}{\epsilon_0}
    \int d\omega'
    DOS(\omega')
    \frac{(d^{12})^2}{
    \hbar\omega'
    -
    \hbar\omega
    -
    i\gamma_{M}
    -
    \Sigma_{F}(\omega)}
\end{align}

Fig.~\ref{fig:abs_dos} shows a plot of the pristine inhomogeneously broadened absorption, using Eq.~(\ref{eq:dos_chi}) as the susceptibility, and a respective plot with Förster transfer to a monolayer TMDC. Evidently, the Förster transfer cannot be observed directly, which is due to the fact that although the linewidth of the absorption in every molecule is significantly broadend by the Förster rate (Fig.~\ref{fig:abs_spec}), this effect vanishes when integrating with a Gaussian distribution with a FWHM in the range of hundreds of meV of molecular transition energies as it is the case in ensemble experiments \cite{bondarev2004fluorescence}. For such ensembles of molecular transition energies, we thus propose to instead carry out luminescence experiments, as introduced in the following section, which more directly give access to a spectrally resolved measurement of the FRET rate.

\section{Photoluminescence}\label{Sec:Photo}
In this section we discuss how the F\"orster rate may be accessed via a quench in the luminescence spectrum of the dye molecules attached to the TMDC layer. For PL experiments, the broad distribution of the HOMO-LUMO transition energies is a clear advantage over other zero-dimensional donors: As we will demonstrate, it makes the deposited molecules ideal candidates for probing a wide spectral range of momentum dark excitonic states in the TMDC. 
Although we have discussed in Sec.~\ref{Sec:Foe} that the Förster rates are most prominent for a monolayer TMDC acceptor, we suggest a bilayer TMDC for PL measurements, which, as already mentioned in contrast to the monolayer is an indirect semiconductor, where the excitation in the TMDC will quickly decay to the energetically lower lying momentum-indirect intervalley exciton states, which are dark and will not contribute relevantly to the luminescence \cite{mak2010atomically,splendiani2010emerging,wurstbauer2017light,gerber2019interlayer}. This makes it possible to measure (and thus, calculate) the PL of the molecules without significant contributions from the TMDC. We thus assume that the relevant part of the signal stems from the molecular HOMO-LUMO transitions, 
\begin{align}
    I(\Omega_\qq)
    =
    I_{Mol}(\Omega_\qq) 
    +
    I_{TMDC}(\Omega_\qq) 
    \approx 
    I_{Mol}(\Omega_\qq).
\end{align}
The steady state photoluminescence spectrum of the molecular HOMO-LUMO transition reads \cite{Kira1999}
\begin{align}\label{eq:plfromkira}
    I(\Omega_\qq)
    =
    |M_\qq|^2
    \text{Im}\Big(
    \frac{ \sigma_{22}}{ \mathcal{E}_{M}^{12}-\hbar\Omega_\qq-i\gamma}
    \Big)
\end{align}
%Einheit PL: entweder eV/fs, hier aber nur 1/fs.
with $|M_\qq|^2 =  (d^{12})^2\frac{\Omega_\qq}{\epsilon_0V}$, i.e. the molecular dipole moment $d^{12}$ as already defined earlier, the photon frequency $\Omega_\mathbf{q} = c |\mathbf{q}|$ in free space, a broadening $\gamma \approx 0$ which we neglect in the following, and the occupation of the excited state $\sigma_{22}$, which needs to be computed.
Stationary luminescence of dye molecules is typically measured during cw excitation with a pump energy which is large compared to the emitting molecular transition energy $\mathcal{E}_{M}^{12}$. 
As we will show in the following, the Förster process can be made visible in the form of frequency dependent quenches in the PL signal, even without detailed knowledge of the non-radiative energy relaxation pathways inside the molecule from the states excited by the laser to the HOMO-LUMO transition.
We provide two different analytical treatments for the stationary occupation of the excited molecular occupation $\sigma_{22}$ in Eq.~(\ref{eq:plfromkira}), i.e. we calculate the related luminescence of two limiting cases of close-to-resonance (Sec.~\ref{subsec:nearresonant}) and far-from-resonance driving (Sec.~\ref{subsec:offresonant}). Our results show that the quench in the molecular PL due to the Förster rate is robust with respect to different driving scenarios. 
\subsection{Driving near the molecular resonance}\label{subsec:nearresonant}
\begin{figure}[t!]
 \begin{center}
    \includegraphics[width=0.6\linewidth]{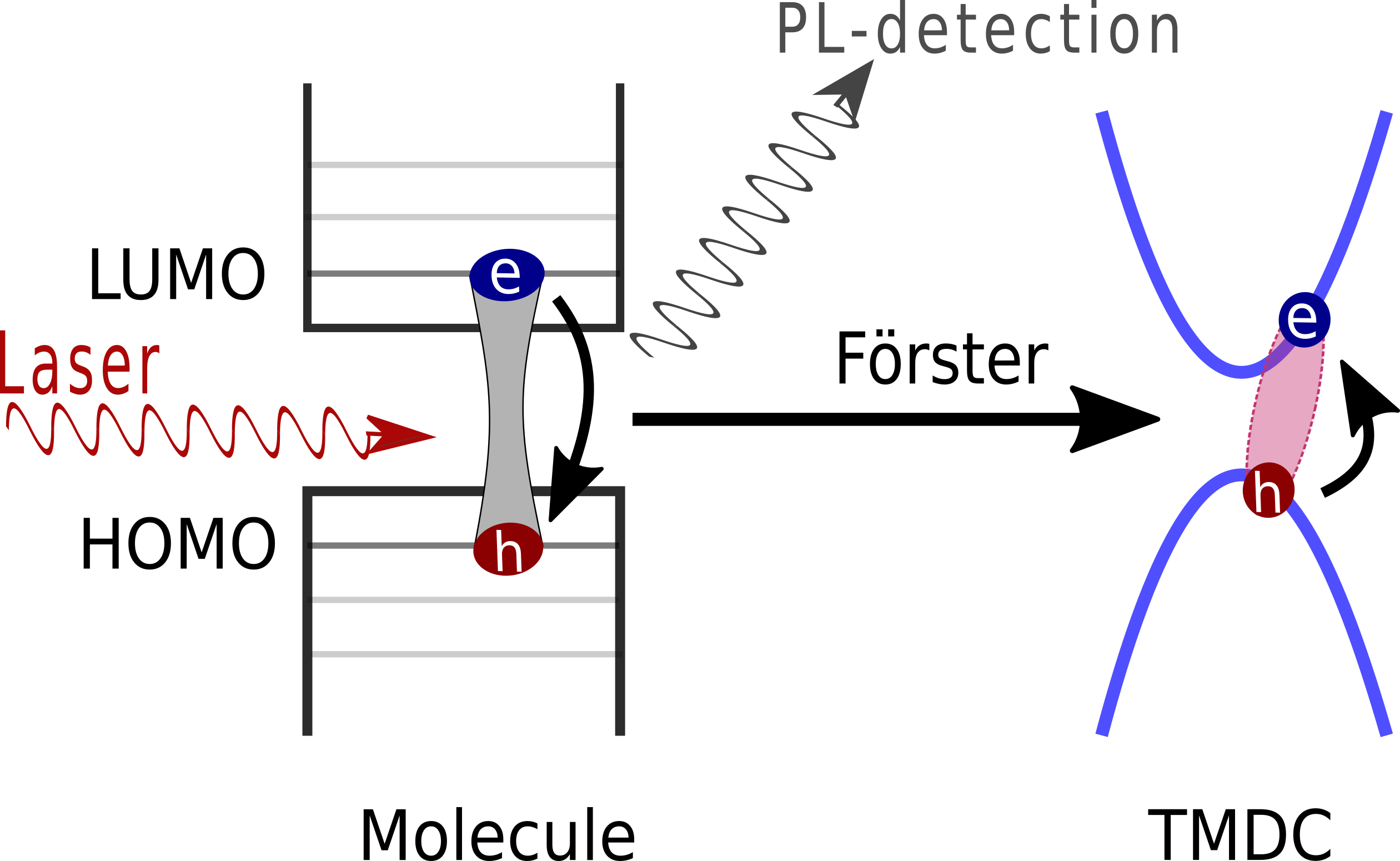} \end{center}
\caption{Near resonance optical excitation. Only those excitons of the molecules with transition energies similar to the driving laser frequency are excited, which leads to luminescence only in a small spectral range, compare Fig.~\ref{fig:linear_pl_ple_near}~(a). After recombination, two scenarios are possible: a creation of a photon which is detectable as PL signal, or a Förster transfer to the TMDC bilayer, where the excitation will vanish to the momentum-indirect intervalley exciton state and thus not be visible in the PL detection.}
\label{fig:sketch_plnear}
\end{figure}
First, we assume that the LUMO is directly excited by resonant laser driving, as depicted in Fig.~\ref{fig:sketch_plnear}. Similar to Eq.~(\ref{eq:sigmawitheigen}), we can find an equation for the occupation density~$\sigma_{22}$ of the LUMO: The factor of 2 accounts for the population lifetime $T_1 = \frac{1}{2}T_2$ (coherence lifetime of the polarization).
\begin{align}\label{eq:sigma22pl}
    i\hbar \partial_t
    \sigma_{22}
    &=
    -2i(\gamma_M
    + \gamma_{F}^{\mathcal{E}_M^{12}}
    )
    \sigma_{22}
    +2iIm
    \EE_0(z_M,t)\cdot(\bv{d}^{12}\sigma_{12})
\\
    i\hbar \partial_t
    \sigma_{12}
    &=
    (
    \mathcal{E}^{12}_M
    -i\gamma_M
    -i \gamma_{F}^{\mathcal{E}_M^{12}}
    )
    \sigma_{12}
    -
    \EE_0(z_M,t)\cdot\mathbf{d}^{21}\label{eq:sigma12pl}
\end{align}
Eq.~(\ref{eq:sigma12pl}) is similar to Eq.~(\ref{eq:sigmawitheigen}), but in the time domain, with only the imaginary part of the self-energy taken into account. Note that we write $\gamma_{F}^{\mathcal{E}_M^{12}}$, and use the analytical solution, Eq.~(\ref{eq:foersterrate}) in Sec.~\ref{Sec:Theo}, which directly depends on the molecular energy $\mathcal{E}_M^{12}$. The corresponding results are very close to the full numerical solution, i.e. the imaginary part of Eq.~(\ref{eq:foerstereigenenergy}).
The coupled Eqs.~(\ref{eq:sigma22pl},\ref{eq:sigma12pl}) can be solved in the rotating frame of the driving laser frequency $\omega_L$ in the steady state, omitting fast oscillating contributions (rotating wave approximation). We find the dependency of the molecular occupation on the laser frequency as
\begin{align}\label{eq:sigmafertig}
    \sigma_{22}(\omega_L,\mathcal{E}_M^{12})
    &=
    \frac{|\hat E_0|^2(d^{12})^2
    }{(\mathcal{E}_M^{12}
    -\hbar\omega_L)^2+(\gamma_M+ \gamma_{F}^{\mathcal{E}_M^{12}})^2},
\end{align}
where $|\hat E_0|^2$ denotes the intensity of the driving light field. Eq.~(\ref{eq:sigmafertig}) an be inserted into Eq.~(\ref{eq:plfromkira}), which for the assumption of completely resonant excitation ($\gamma = 0$) yields for the photoluminescence of a single molecule
\begin{align}
    I(\Omega_\qq)
    =
   %\frac{2}{\hbar}
    \pi
    |M_\qq|^2
    \delta
    (\mathcal{E}_M^{12}-\hbar\Omega_q)
    \sigma_{22}(\omega_L,\mathcal{E}_M^{12})
\end{align}
For the luminescence of a molecular ensemble with a Gaussian distribution of transition energies, we insert a convolution over the frequencies $f(\mathcal{E}_M^{12,j}) = \int d\omega \delta(\omega-\frac{\mathcal{E}_M^{12,j}}{\hbar})f(\hbar\omega)$ and, as before, assume a density of states for the molecular transition energies $\sum_j \delta(\omega-\omega^j)=DOS(\omega)$, which is made to resemble respective experiments~\cite{bondarev2004fluorescence}, compare also Sec.~\ref{subsec:molecularabsorption}. We find for the luminescence of the molecular ensemble
\begin{align}\label{eq:PLensemble}
    &I_{Ensemble}(\Omega_\qq)\nonumber\\
    &=
    DOS(\Omega_\qq)
    \frac{\pi
    |M_\qq|^2|\hat E_0|^2(d^{12})^2
    }{(\hbar\Omega_\qq
    -\hbar\omega_L)^2+(\gamma_{M}+ \gamma_{F}(\Omega_\qq))^2}.
\end{align}
The decay of the occupation by radiative recombination and by the Förster process both contribute to the denominator, which later leads to the quench compared to the pristine signal for frequencies where Förster rates dominantly occur. In the following we discuss the resulting photoluminescence (PL) signal, plotted over the emission frequency $\Omega_\qq$, and then also give the result for photoluminescence excitation (PLE), where the signal is integrated over the whole emission spectrum (i.e. integrated over $\Omega_\qq$) and plotted over the excitation frequency $\omega_L$ of the laser. 

\subsubsection{Photoluminescence (PL)}

\begin{figure}[t!]
 \begin{center}
    \includegraphics[width=\linewidth]{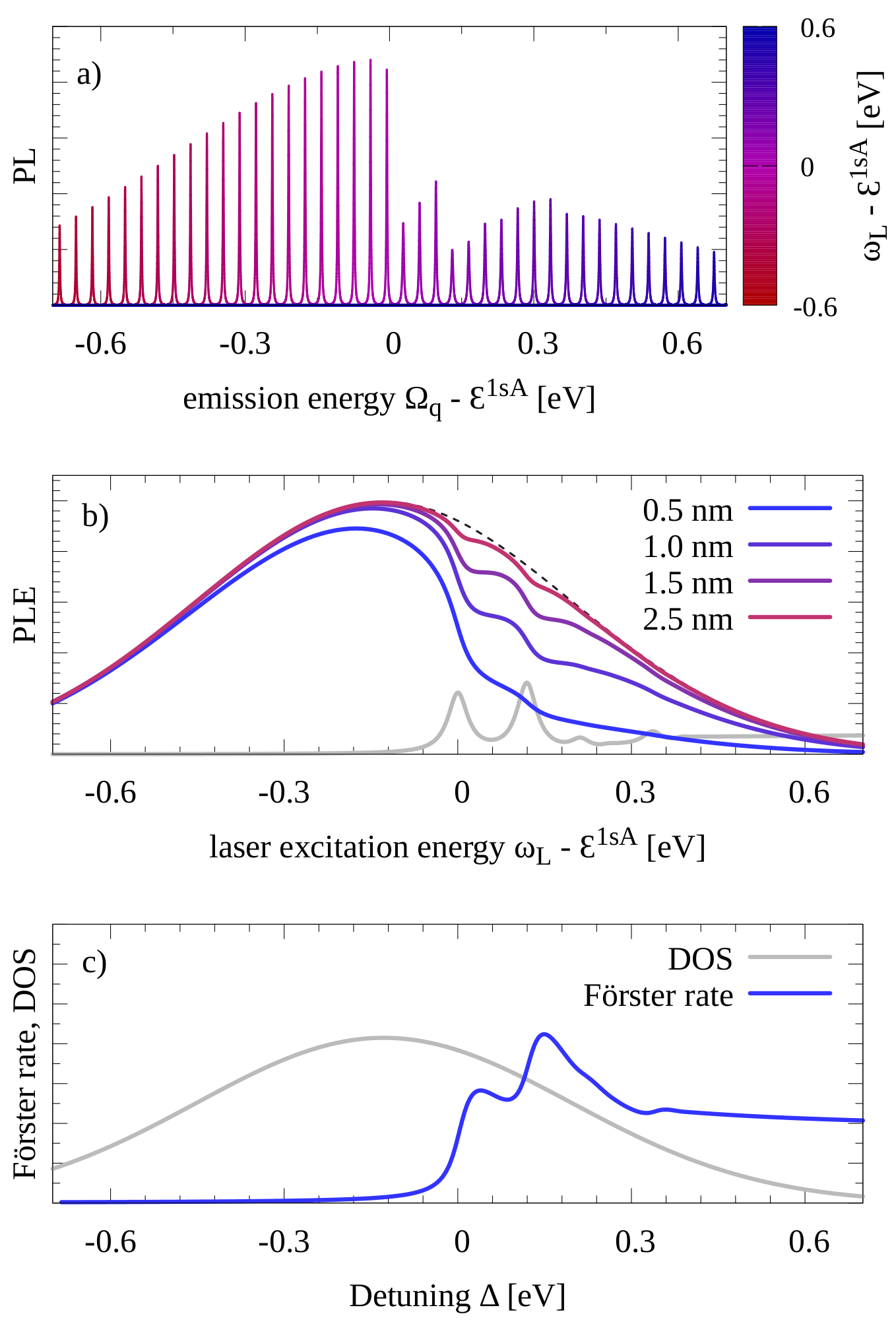} \end{center}
\caption{a) Photoluminescence (PL) of a sheet of molecules with a Gaussian distribution of transition energies $\Omega_\qq$ around the lowest 1s-A-resonance of the semiconductor. The laser energy $\omega_L$ is sweeped over the molecular resonances, with each Lorentzian peak referring to a different laser excitation energy, plotted in one picture but in different colors from red to blue, accordingly, see key on the right side of the plot. This plot is for small non-radiative dephasing $\gamma^{\mu\ell}=\unit[1]{meV}$, for better visibility of the different plots. The effect is however also clearly visible at room temperature. b)~Photoluminescence excitation (PLE) spectroscopy; for each laser excitation energy $\omega_L$ we show the PL signal integrated over the whole emission frequency range $\Omega_\qq$. The dashed line shows the PLE of the pristine molecules without the TMDC substrate. The linear response of the TMDC excitons is depicted in gray to show the connection of the effect to the energetic resonances in the semiconductor. Both plots make visible, that above the excitonic resonances, the additional decay channels cause dips in the luminescence, as less excitons can recombine radiatively. c) Förster rate (with distance $R=\unit[1]{nm}$) and density of states (DOS) of molecular transition energies for comparison.}
 \label{fig:linear_pl_ple_near}
\end{figure}
Fig.~\ref{fig:linear_pl_ple_near}~(a) shows the calculated PL as a function of the emission frequency $\Omega_\qq$ for many different laser excitation energies $\omega_L$, all plotted into one picture, a single plot is just one of the visible Lorentzian peaks, as only the excitons in molecules with transition energies near the driving energy are excited by the laser. As we assume a continuum of different molecular transition energies, for every laser energy, there will always be molecules addressed resonantly as the laser is tuned through the molecular distribution. Above the lowest TMDC excitonic resonances, the luminescence is quenched by the additional decay through the Förster coupling, thus the dip due to the Förster process above the lowest excitonic resonances is clearly visible. 

\subsubsection{Photoluminescence excitation (PLE)}

To connect the theoretical description (Eqs.~(\ref{eq:sigma22pl}-\ref{eq:PLensemble})) to a photoluminescence excitation (PLE) spectrum, the integrated signal, Eq.~(\ref{eq:PLensemble}) as a function of the excitation energy $\omega_L$ is calculated.
\begin{align}\label{eq:PLE}
    &I_{PLE}(\omega_L)\nonumber\\
    &=
    \int
    d\Omega_\qq
    PL_{Ensemble}(\Omega_\qq)\nonumber\\
    &=
    \int
    d\Omega_\qq
    DOS(\Omega_\qq)
    \frac{\pi
    |M_\qq|^2|\hat E_0|^2(d^{12})^2
    }{(\hbar\Omega_\qq
    -\hbar\omega_L)^2+(\gamma_{M}+\gamma_{F}(\Omega_\qq))^2}
\end{align}

Fig.~\ref{fig:linear_pl_ple_near} (b) shows the calculated PLE, Eq.~(\ref{eq:PLE}), for selected molecule - TMDC distances as a function of the laser energy~$\omega_L$ with respect to the 1s A transition energy~$\mathcal{E}^{1sA}$. For laser excitations below the 1s A exciton resonance the PLE almost follows a Gaussian, which originates from the already discussed density of states of the molecule~\cite{bondarev2004fluorescence}. However, above zero detuning, substantial dips in the PLE can be observed. They stem from F\"orster induced de-excitation of the PL emitting molecules, resulting from a FRET induced relaxation pathway to the momentum dark TMDC excitons, compare Eq.~(\ref{eq:foerstereigenenergy}). As a result, the PLE decreases, giving a relatively direct way to measure also the momentum dark excitonic structure in the TMDC in a spectrally resolved experiment. We further find that the dips vanish for larger distances, originating from the decrease of the F\"orster coupling for increasing distance, compare~Figs.~\ref{fig:foerster_mono} and \ref{fig:comp_bilayer}.

\subsection{Driving far from resonance}\label{subsec:offresonant}

\begin{figure}[t!]
    \centering
    \includegraphics[width=0.7\linewidth]{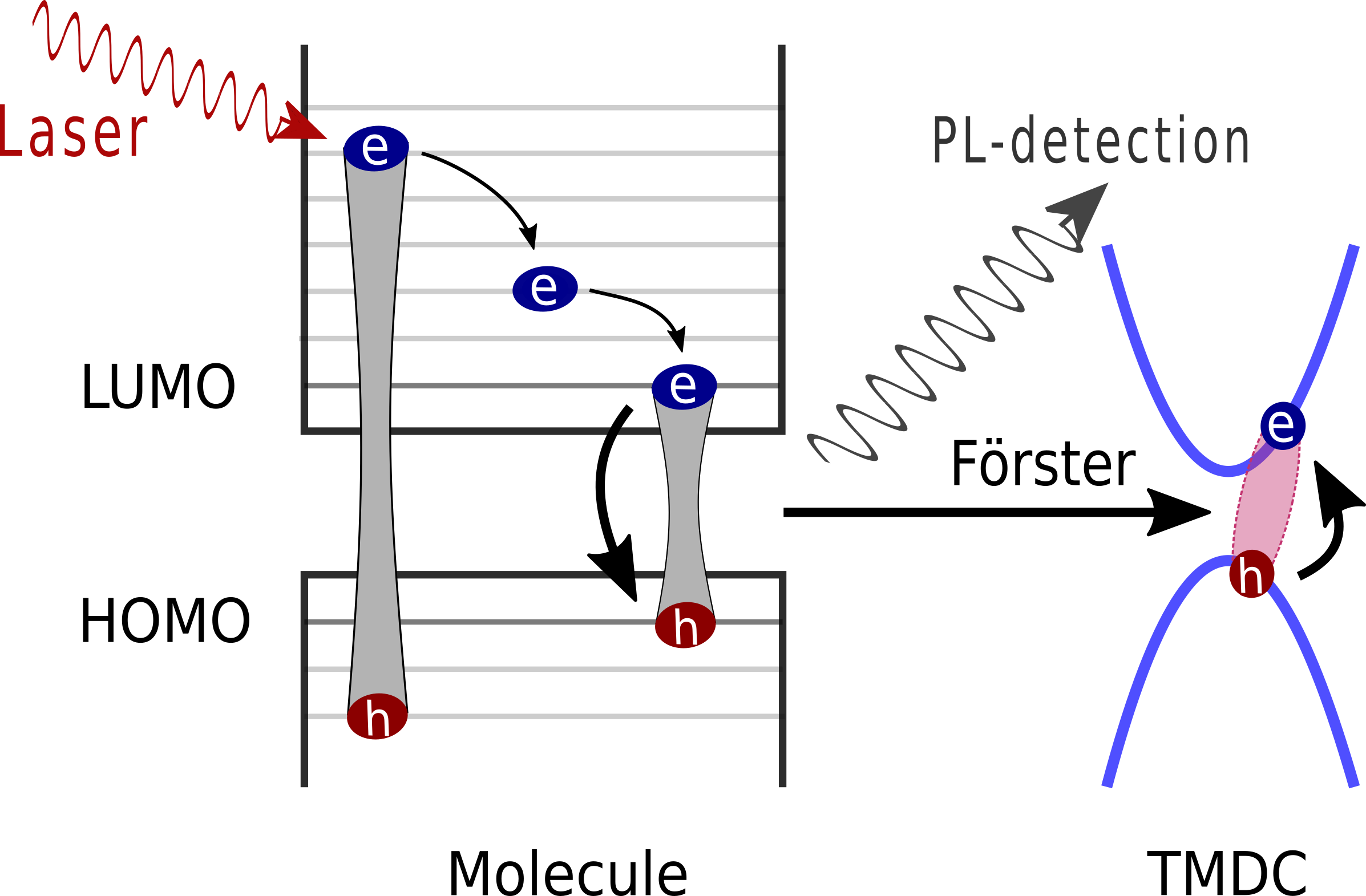}
    \caption{Sketch for the setup with far off-resonant driving. The laser is creating excitons in excited molecular levels. Non-radiative processes lead to a population of the LUMO. There, the recombination takes place, producing either a photon, which can be detected in the PL measurement, or exciting a semiconductur exciton in the TMDC via Förster coupling. \label{fig:sketch_faroff}}
\end{figure}
As depicted in Fig.~\ref{fig:sketch_faroff}, we now assume a laser driving higher levels in the molecule, activating non-radiative relaxation channels which cause incoherent population of the LUMO. The time needed for those non-radiative transitions is longer, the bigger the energy gap to the LUMO level, due to more vibron assisted scattering events (however, this is of no special interest here since we study stationary PL). We assume for the respective transition rate:
\begin{align}
    \Gamma
    \propto
    \frac{1}{\hbar\omega_L-\mathcal{E}_M^{12}}
    \equiv
    \frac{\Gamma_{nonrad}}{\hbar\omega_L-\mathcal{E}_M^{12}}
\end{align}
Then we assume the following set of equations, motivated by the previous section and an additional non-radiative process between the driven level 3 and the LUMO (level 2)
\begin{align}
    i\hbar \partial_t
    \sigma_{33}
    &=
    -
    2i\Gamma
    \sigma_{33}
    +
    \EE_0(z_M,t)\cdot(\bv{d}^{13}\sigma_{13}
    -
    \bv{d}^{31}\sigma_{31})
    \\
    i\hbar \partial_t
    \sigma_{22}
    &=
    (-2i\gamma_M
    -2i\gamma_{F}^{\mathcal{E}_M^{12}}
    )
    \sigma_{22}
    +
    2i\Gamma
    \sigma_{33}
\end{align}
We again assume a quasi-equilibrium for the occupation of the LUMO level, $\partial_t\sigma_{22}=0$
and thus
\begin{align}
    \sigma_{22}
    =
    \frac{\Gamma(\hbar\omega_L-\mathcal{E}_M^{12})}{\gamma_M
    +\gamma_{F}^{\mathcal{E}_M^{12}}}
    \sigma_{33}
\end{align}
Then we can write for the luminescence of one off-resonantly driven molecule
\begin{align}
    I(\Omega_\qq)
    &=
    %\frac{2}{\hbar}
    \pi|M_\qq|^2\delta( \mathcal{E}_{M}^{12}-\hbar\Omega_\qq)
    \frac{\Gamma(\hbar\omega_L-\mathcal{E}_M^{12})}{\gamma_M
    +\gamma_{F}^{\mathcal{E}_M^{12}}}
    \sigma_{33}
\end{align}
For the molecular ensemble, analogous to the previous section, we again apply the convolution $f(\mathcal{E}_M^{12,j})=\int d\omega \delta(\omega-\frac{\mathcal{E}_M^{12,j}}{\hbar})f(\hbar\omega)$ in order to identify the density of states for the transition energies $\sum_j \delta(\omega-\omega^j)=DOS(\omega)$, which yields
\begin{align}\label{eq:pl_far}
    &I_{fardriven}(\Omega_\qq)\nonumber\\
    &=
    DOS(\Omega_\qq)
    %\frac{2}{\hbar}
    \pi|M_\qq|^2
    \frac{\Gamma_{nonrad}}{(\hbar\omega_L-\hbar\Omega_\qq)(\gamma_{M}
    +\gamma_{F}(\hbar\Omega_\qq))}
    \sigma_{33}
\end{align}
\begin{figure}[t!]
    \centering
    \includegraphics[width=\linewidth]{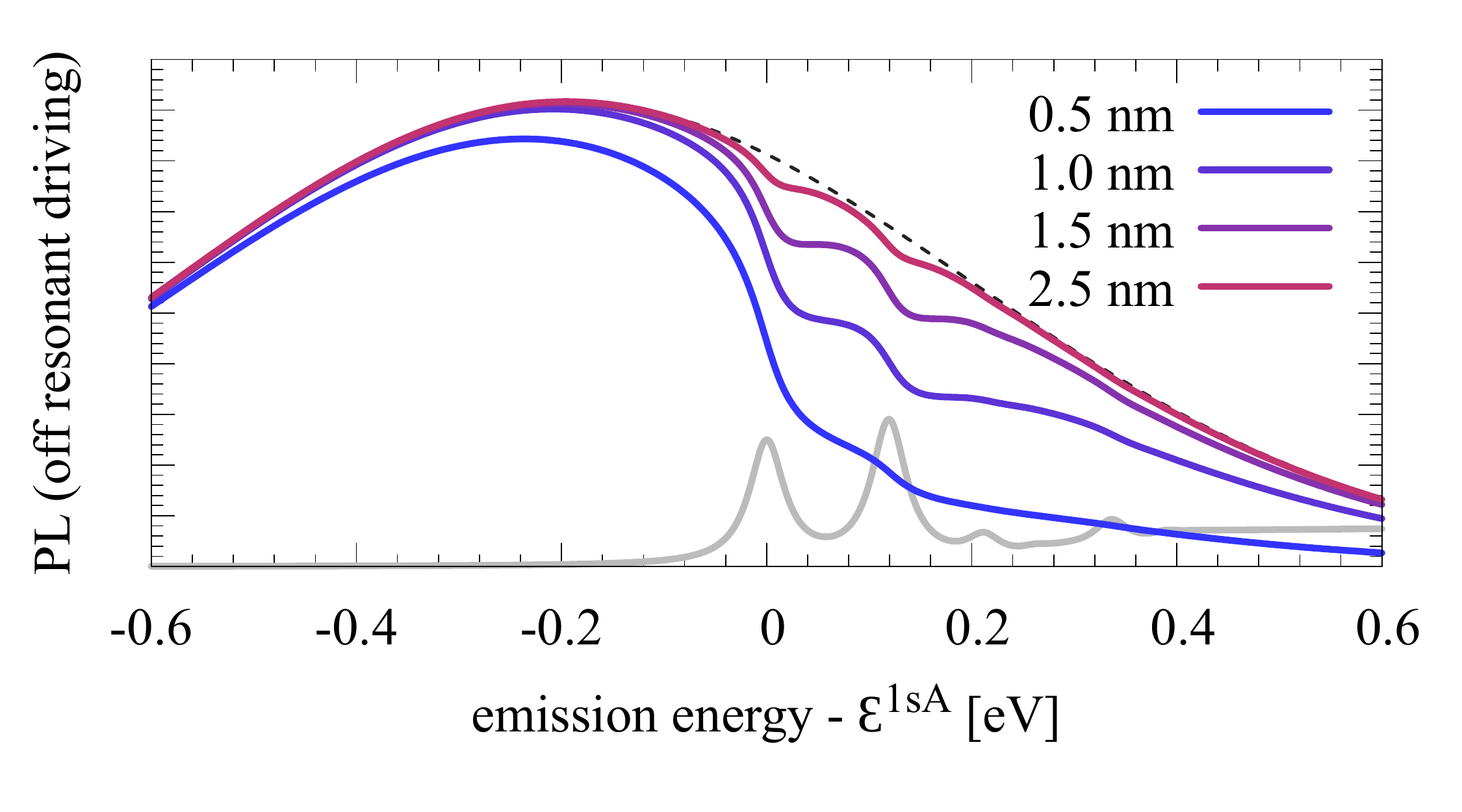}
    \caption{PL spectrum for off resonant driving for different values for the distance $R$ between donor (dye molecules) and acceptor (TMDC). Again the dashed line gives the PL of the pristine molecules without a TMDC substrate, and the excitonic resonances of the TMDC are again depicted in gray for better orientation. The result is very similar to the near-resonant case (Fig.~\ref{fig:linear_pl_ple_near}). This suggests a good experimental accessibility of the Förster process via luminescence experiments, as the signatures proof robust towards different driving scenarios. 
    \label{fig:pl-far}}
\end{figure}
Fig.~\ref{fig:pl-far} illustrates the photoluminescence spectrum after far off-resonant excitation. Similar to the PLE of the near-resonant case (Fig.~\ref{fig:linear_pl_ple_near}~b), we find that the spectrum for emission energies below the 1s A exciton follows the mentioned Gaussian distribution determined via the DOS of the molecules. However, the luminescence exhibits pronounced dips above the 1s A resonance due to F\"orster mediated relaxation of molecular excitons to the TMDC layer, giving again experimental access to the excitonic states in the semiconductor. Similar as for the PLE, the quenching of the luminescence is again most prominent for closely stacked structures, but decreases as the distance is increased. 
As both the near and the far resonant case show very similar results, we suggest to extract the F\"orster induced relaxation rate and its momentum dependence from optical luminescence measurements. Usually, in experiments related to molecular FRET, the efficiency or quantum yield of the Förster rate is given as  $e=\frac{\gamma_F}{\gamma_F+\gamma_M}$, see e.g. \cite{Tisler2011NVcenter,lakowicz2006bookluminescence,wilhelmsson2012FRETmeasure,valeur2013measurefret}. Hence, as a consequence of Eq.~(\ref{eq:pl_far}), we suggest to compute
\begin{align}
    1-\frac{I_{DA}}{I_D}
    =
    \frac{\gamma_F}{\gamma_F+\gamma_M}
\end{align}
to obtain the mentioned efficiency of the F\"orster rate $\gamma_F$ with respect to the pristine relaxation rate $\gamma_M$ in the molecule, with the PL intensity of the interface $I_{DA}$ compared to the intensity $I_D$ of the pristine donor without any FRET acceptor.
\section{Conclusion}\label{Sec:Concl}
In this work we discussed F\"orster type energy transfer at a planar interface between dye molecules and an atomically thin semiconductor. Due to the specific geometry and the breakdown of spatial invariance at the interface, momentum dark excitons are preferably excited. The spectrally resolved FRET rate and the corresponding luminescence signatures occur most prominently at positions above the bright excitonic resonance, opening an interesting opportunity to measure momentum dark exciton states acting as acceptors. In the long distance limit, we reproduced the $R^{-4}$ power law dependence which was already found in related earlier work. For shorter distances, we additionally found an exponential dependence on distance. We also discussed the effects of the energy transfer on linear spectroscopy and derived a scheme for an experimental extraction of the spectrally resolved F\"orster rate in photoluminescence experiments.
\section*{Acknowledgements}
We thank Dominik Christiansen, Robert Salzwedel and Lara Greten (TU Berlin) for fruitful discussions and guidance for the creation of Fig.~\ref{fig:sketch}. 
We gratefully acknowledge support of the Deutsche Forschungsgemeinschaft (DFG) through projects B12 and B15 of the SFB 951, project number 182087777.
%\newpage
%
\appendix

\section{Field-Matter Hamiltonian}\label{app:fieldmatterappendix}

Starting point for the calculation of the F\"orster coupling is the semi-classical field-matter coupling Hamiltonian in dipole approximation \cite{cohen1997photons}. For the molecule it reads
\begin{align}
    H
    =
    -e
    \sum_{i,j\neq i}
    \int
    d^3r
    \phi^{*i}(\rr)
    \rr
    \cdot
    \EE(\rr,t)
    \phi^{j}(\rr)
    a^\dagger_i 
    a_j,
\end{align}
with $e$ the elementary charge, $\EE(\rr,t)$ the electric field, $\phi^i (\mathbf{r})$ molecular orbital wavefunctions and annihilation (creation) operators $a^{(\dagger)}_i$ in the state $i$. For small enough spatial dimensions we can assume a point dipole and thus $\EE(\rr,t) \approx\EE(\rr_0,t)$, and simplify
\begin{align}
    H
    =
    -
    \EE(\rr_0,t)
    \cdot
    \sum_{i,j\neq i}
    \bv{d}^{ij}
    a^\dagger_i
    a_j
\end{align}
with the molecular optical dipole moment
$    \bv{d}^{ij}
    =
   \int
   d^3r
   \phi^{*i}(\rr)
   (e\rr)
   \phi^{j}(\rr)$.
Analogously, we write the field matter coupling for the semiconductor layer
\begin{align}
    H
    &=
    -e
    \sum_{\mathbf{k_\parallel},\mathbf{Q_\parallel},\lambda,\lambda'}
    \int
    d^3r
    \phi^{*\lambda}_\mathbf{k_\parallel+Q_\parallel}(\rr)
    \rr
    \cdot
    \EE(\rr,t)
    \phi^{\lambda'}_\mathbf{k_\parallel}(\rr)\nonumber\\
    &\qquad\qquad\qquad\qquad\qquad
    \times
    a^{\dagger \lambda}_\mathbf{k_\parallel+Q_\parallel}
    a_\mathbf{k_\parallel}^{\lambda'},
\end{align}
with electronic Bloch wave $\phi^{\lambda}_\mathbf{k_\parallel}$ and annihilation (creation) operators $a^{(\dagger) \lambda}_\mathbf{k_\parallel}$ with band index $\lambda$ and in-plane momentum $\mathbf{k}_\parallel$.
To exploit the translational invariance in in-plane direction, we introduce the Fourier transform  of the electric field $\mathbf{E} (\mathbf{r},t) = \sum_\mathbf{Q_\parallel} e^{i\mathbf{Q_\parallel}\cdot \mathbf{r_\parallel}}\mathbf{E}_\mathbf{Q_\parallel} (z)$. We define a dipole element 
$\bv{d}_{\kk\pll+\QQ\pll,\kk\pll}^{\lambda\bar\lambda}=\int d^2\rr\pll \phi_{\kk\pll+\QQ\pll}^{*\lambda}(\rr)e\rr \phi_{\kk\pll}^{\lambda'}(\rr)$
and can thus write
\begin{align}
    H^{(2)}
    =
    -\sum_{\kk\pll\QQ\pll\lambda\lambda'}
    \bv{d}_{\kk\pll+\QQ\pll,\kk\pll}^{\lambda\lambda'}
    \cdot
    \EE_{\QQ\pll}(z_T)
    a_{\kk\pll+\QQ\pll}^{\dagger\lambda}
    a_{\kk\pll}^{\lambda'}
\end{align}
To arrive at the Hamiltonian in the main text (Eq.~(\ref{eq:hamiltonian})), we go to an excitonic basis, with a projection on the solutions of the Wannier equation \cite{gunnar2014wannier,katsch2018theory}, and assume a thin film approximation.
\section{Rytova-type Greens function}\label{app:rytovaappendix}
The derivation of the Rytova-type Greens function is carried out in analogy to \cite{rytova2020}, however, here we consider Greens functions instead of electrostatic potentials. The dielectric environment is depicted in Fig.~\ref{fig:sketch}~(c) of the main text. We thus start with a general set of equations for the three z ranges, and Fourier transform with respect to the in plane coordinates. For the case of a source which is positioned in the intermediate layer ($z_{TS} < z' < z_{MT}$), this reads
\begin{align}\label{ftpoisson}
    &\partial_z^2G_1(z,z')
    -
    Q\pll^2G_1(z,z')
    =0\nonumber\\
    &\partial_z^2G_2(z,z')
    -
    Q\pll^2G_2(z,z')
    =
    -
    \delta(z-z')
    \nonumber\\
    &\partial_z^2G_3(z,z')
    -
    Q\pll^2G_3(z,z')
    =
    0
\end{align}
In order to find the equations of the main text, one also has to solve equations for the case $z'>z_{MT}$. The ansatz for the Greens function is thus different in the three sections and has to obey the following boundary conditions
\begin{align}\label{boundarycond}
   G_1(z_{MT})
    &=
    G_2(z_{MT})\nonumber\\
    \epsilon_1
    \partial_z G_1(z)\big|_{z=z_{MT}}
    &=
    \epsilon_2
    \partial_z G_2(z)\big|_{z=z_{MT}}\nonumber\\\nonumber\\
   G_2(z_{TS})
    &=
    G_3(z_{TS})\nonumber\\
    \epsilon_2
    \partial_z G_2(z)\big|_{z=z_{TS}}
    &=
    \epsilon_3
    \partial_z G_3(z)\big|_{z=z_{TS}}
\end{align}
Besides, we only take into account solutions that obey
\begin{align}
    \lim_{|z|\rightarrow\infty}G_{1,3}(z) \neq \pm\infty
\end{align}
Starting from Eq.~(\ref{eq:Greensfunctionapproach}) in the main text, we find that Eq.~(\ref{eq:helmholtzeq}) in the main text is solved by 

\changes{

\begin{align}\label{eq:M-rytova}
&V_{\QQ\pll \ell\mu 12}^{T\rightarrow M}(z_T^\ell,z_M)
   \nonumber\\
   &=
   \frac{\big(
   \QQ_\parallel
   \cdot
   \bv{d}\pll^{\mu}\varphi_{r=0}^{\mu\ell}
   \big)
   \Big(1+\delta_{23}
    e^{-2Q\pll(z_M-z_{TS})}
    \Big)
    e^{-Q\pll(z_T^\ell-z_M)}}
    {A\epsilon_0(\epsilon_1+\epsilon_2)(1-\delta_{21}\delta_{23}e^{2Q\pll(z_{TS}-z_{MT})})}
    \nonumber\\
    &\qquad
    \times
    \bigg(
    \frac{1}{Q\pll}
   \big(
   \bv{d}\pll^{21}
   \cdot    
   \QQ\pll
   \big)
   +
   i
   d_z^{21}
   \bigg)
\end{align}
}
for the coupling from the TMDC to the molecule, and, analogously, 

\changes{
\begin{align}\label{eq:V-rytova}
   &V_{\QQ\pll \ell\mu 12}^{M\rightarrow T}(z_M,z_T^\ell)
   \nonumber\\
   &=
    \frac{\big(
   \bv{d}\pll^{*\mu}\varphi_{r=0}^{*\mu\ell}
   \cdot    
   \QQ\pll
   \big)
    \Big(
    1
    +
    \delta_{23}e^{-2Q\pll (z_M-z_{TS})}
    \Big)
   e^{Q\pll(z_M-z_T^\ell)}}
    {\epsilon_0(\epsilon_2+\epsilon_1)(1-\delta_{21}\delta_{23}e^{2Q\pll(z_{TS}-z_{MT})})}
    \nonumber\\
    &\qquad
    \times
    \bigg(
    \frac{1}{Q\pll}
   \big(
   \QQ_\parallel
   \cdot
   \bv{d}\pll^{12}
   \big)
   -
   i
   d_z^{12}
   \bigg)
\end{align}
}

for the coupling from the molecule to the TMDC. In order to compute the rates, Eqs.~(\ref{eq:foerstereigenenergy},\ref{eq:foersterrate}) in the main text, one has to parametrize the involved vectors, (with $\xi=-1,1$, reflecting the optical dichroism of the K,K' valley \cite{xiao2012coupled})
\begin{align}\label{eq:parametrization}
&\bv{d}\pll^{\mu}
=
\frac{d\pll^{\mu}}
{\sqrt{2}}
\left[
    \begin{array}{cc}
         1\\
         i\xi 
    \end{array}
    \right]
&\bv{d}^{12}
=
d^{12}
\left[
    \begin{array}{cc}
         \cos{\vartheta}\\
         0\\
         \sin{\vartheta}
    \end{array}
    \right]\\
&\QQ\pll
=
Q\pll 
\left[
    \begin{array}{cc}
         \cos{\varphi}\\
         \sin{\varphi} 
    \end{array}
    \right]
\end{align}
In the main text we assume $\vartheta \approx 0$ as discussed in Sec.~\ref{subsec:rytovasection}.
\changes{
\section{Förster rate for different dipolar angles}\label{app:anglesappendix}
For the discussion in the main text we assume that the molecular dipoles are parallel to the TMDC plane, as it is typical for such interfaces produced by evaporation that the molecular dipoles are randomly distributed on the TMDC, and thus have a random in-plane angle, but tend to align parallel to the sheet they are evaporated on~\cite{Tsai2015dipolealignment}. In principle it is possible to also account for a nonzero angle of $\mathbf{d}^{12}$ with respect to the TMDC plane, i.e. to assume that $\vartheta \neq 0$ in Eq.~(\ref{eq:parametrization}). Eq.~(\ref{eq:foersterrate}) would then read
\begin{align}\label{eq:foersterratemitwinkel}
    &\gamma_{F}^{\mathcal{E}_{M}^{12}}(\vartheta)\nonumber\\
    &=\sum_{\mu\ell}
    \frac{(d\pll\varphi_{r=0}^{\mu\ell})^2(d^{12})^2M}{8\epsilon_{0}^2(\epsilon_1+\epsilon_2)^2\hbar^2}
    % \big(
    % 3\cos^2\vartheta
    % -
    % 2
    % \big)
    \Theta(Q_0^\parallel)
    \Big(\cos^2{\vartheta}+2\sin^2{\vartheta}\Big)
    \nonumber\\
    &\quad
    \times
    % \nonumber\\
    % &\qquad\qquad\qquad
    % \times
    \frac{(Q_0^\parallel)^2(1+\delta_{23}
    e^{-Q_0^\parallel(R + \frac{3}{2}R_{\Delta T})})^2
    e^{-2Q_0^\parallel (\ell-1)R_{\Delta T}}}
    {(1-\delta_{21}\delta_{23}e^{4Q_0^\parallel R_{\Delta T}})^2}
    e^{-2Q_0^\parallel R}
\end{align}
Fig.~\ref{fig:angles} gives a plot of this equation, showing that the rate overall increases with increasing angles, which is due to the fact that the acceptor is not a point dipole but a whole two-dimensional plane. We do not see a change in the momentum selection rules, since all momenta are summed over anyway due to the symmetry breaking. However, the molecules tend to align with the TMDC in experiments, as already mentioned, which means that $\vartheta \approx 0$ is a good approximation in realistic scenarios.
\begin{figure}[t!]
    \centering
    \includegraphics[width=\linewidth]{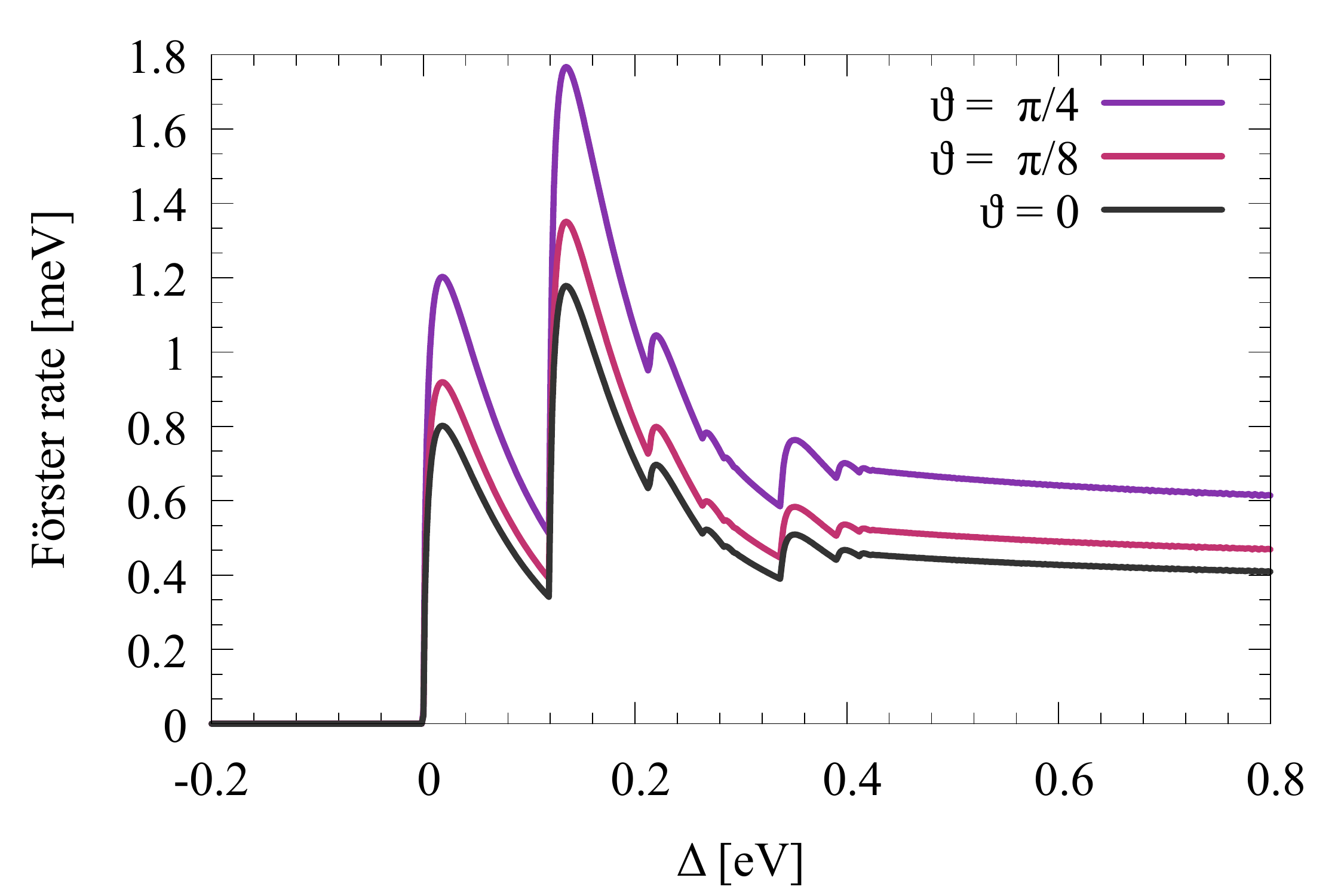}
    \caption{\changes{Finite angles between the molecular dipole and the TMDC plane for the example of a monolayer MoS$_2$. The angle causes an overall increase of the Förster rate, but not a change in the momentum selection rules.}
    \label{fig:angles}}
\end{figure}
}
\section{Analytical Förster rate vs full numerical solution}\label{app:comparenumericalanalytical}
\begin{figure}[t!]
    \centering
    \includegraphics[width=\linewidth]{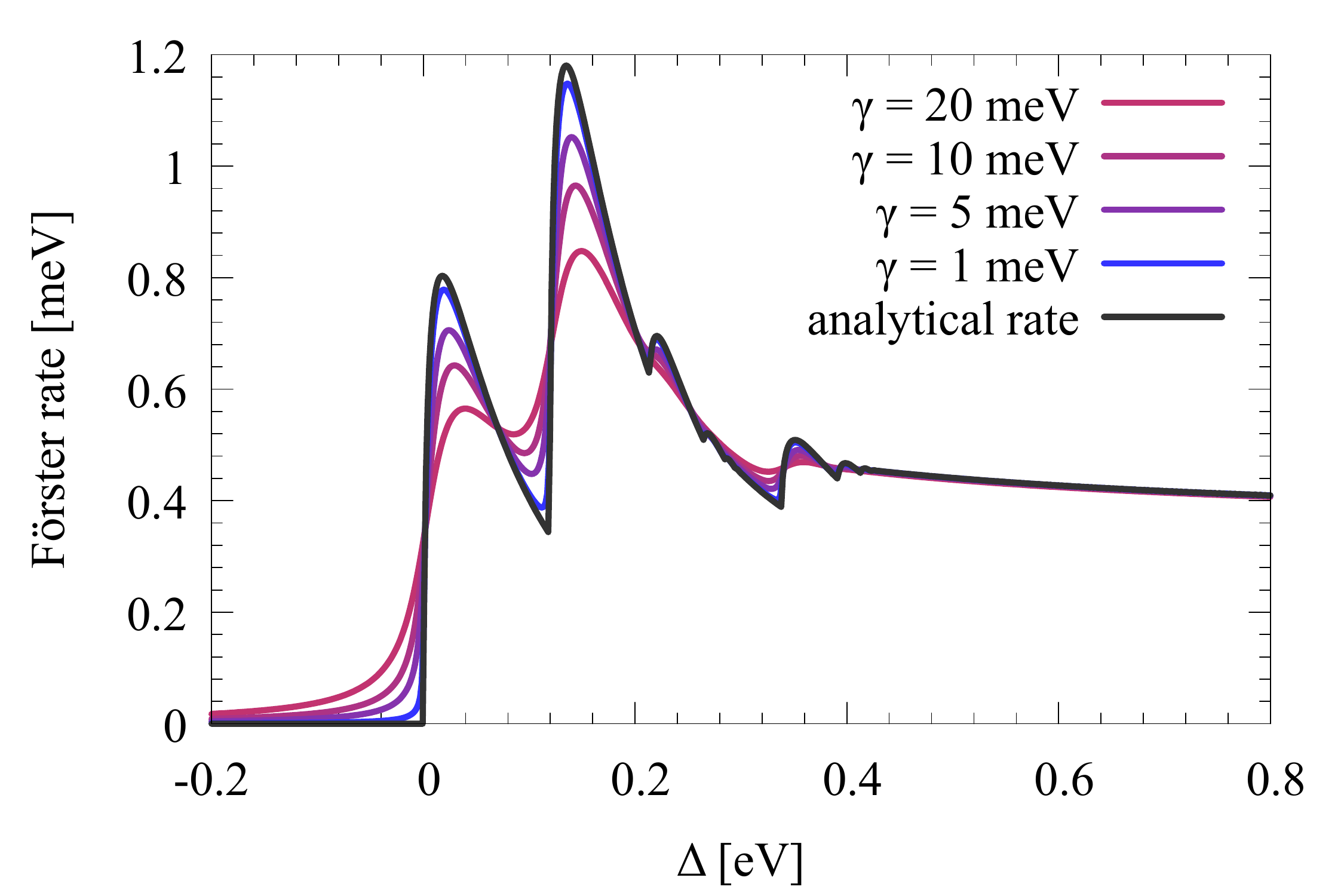}
    \caption{Analytical Förster rate as computed in the main part, Eq.~(\ref{eq:foersterrate})) vs imaginary part of the numerical solution of the full equation (Eq.~(\ref{eq:foerstereigenenergy})) for different non-radiative dephasing rates $\gamma^{\mu\ell}$, related to temperatures up to room temperature~\cite{selig2016excitonic}.
    \label{fig:vergleich-ana}}
\end{figure}

If not stated differently, the plots in this paper all show results for the numerical evaluation of Eq.~(\ref{eq:foerstereigenenergy}) at room temperature, i.e. non-radiative dephasing in the TMDC of $\gamma^{\mu\ell}\approx\unit[20]{meV}$~\cite{selig2016excitonic,christiansen2017phononen}. Fig.~\ref{fig:vergleich-ana} shows that for smaller $\gamma^{\mu\ell}$, i.e. lower temperatures, the rate from the numerical integration is converging against the analytical approximation, Eq.~(\ref{eq:foersterrate}).

\section{Förster Hamiltonian}\label{app:Hamiltonianappendix}
We can identify $V_{\QQ\pll \mu\ell 12}^{T\rightarrow M}(z_T,z_M)=\frac{1}{A}\big(V_{12\QQ\pll \mu\ell }^{M\rightarrow T}(z_M,z_T)\big)^*$ and hence write down a Hamiltonian that directly gives the coupled equations (\ref{eq:foerster_bloch_eq_sigma}),(\ref{eq:foerster_bloch_eq_p}).
\begin{align}
    H
    &=
    \mathcal{E}_{M}^{12} 
    \hat\sigma_{21}
    \hat\sigma_{12}
    +
    \sum_{\mu\ell\QQ\pll}
    \Big(\mathcal{E}^{\mu\ell} + \frac{\hbar^2\QQ\pll^2}{2M}
    \Big)
    \hat p_{\QQ\pll}^{\dagger\mu\ell}\hat p_{\QQ\pll}^{\mu\ell}\nonumber\\
    &\qquad
    +
    \sum_{\mu\ell\QQ\pll}
    \big(
    V_{12\QQ\pll \mu\ell }^{M\rightarrow T}\big)^*
    \hat p_{\QQ\pll}^{\mu\ell} \hat\sigma_{21}
    +h.c.\nonumber\\
    &\qquad
    -
    \EE_0\cdot
    \Big(
    \bv{d}^{21}\hat\sigma_{21}
    +
    \sum_{\mu\ell}
    (\bv{d}\varphi_{r=0}^{\mu\ell})^*\hat p_0^{\dagger\mu\ell}
    +
    h.c.
    \Big)
\end{align}
\section{$R^{-4}$-dependence for long distances}\label{app:distanceappendix}
\begin{figure}[t!]
    \centering
    \includegraphics[width=\linewidth]{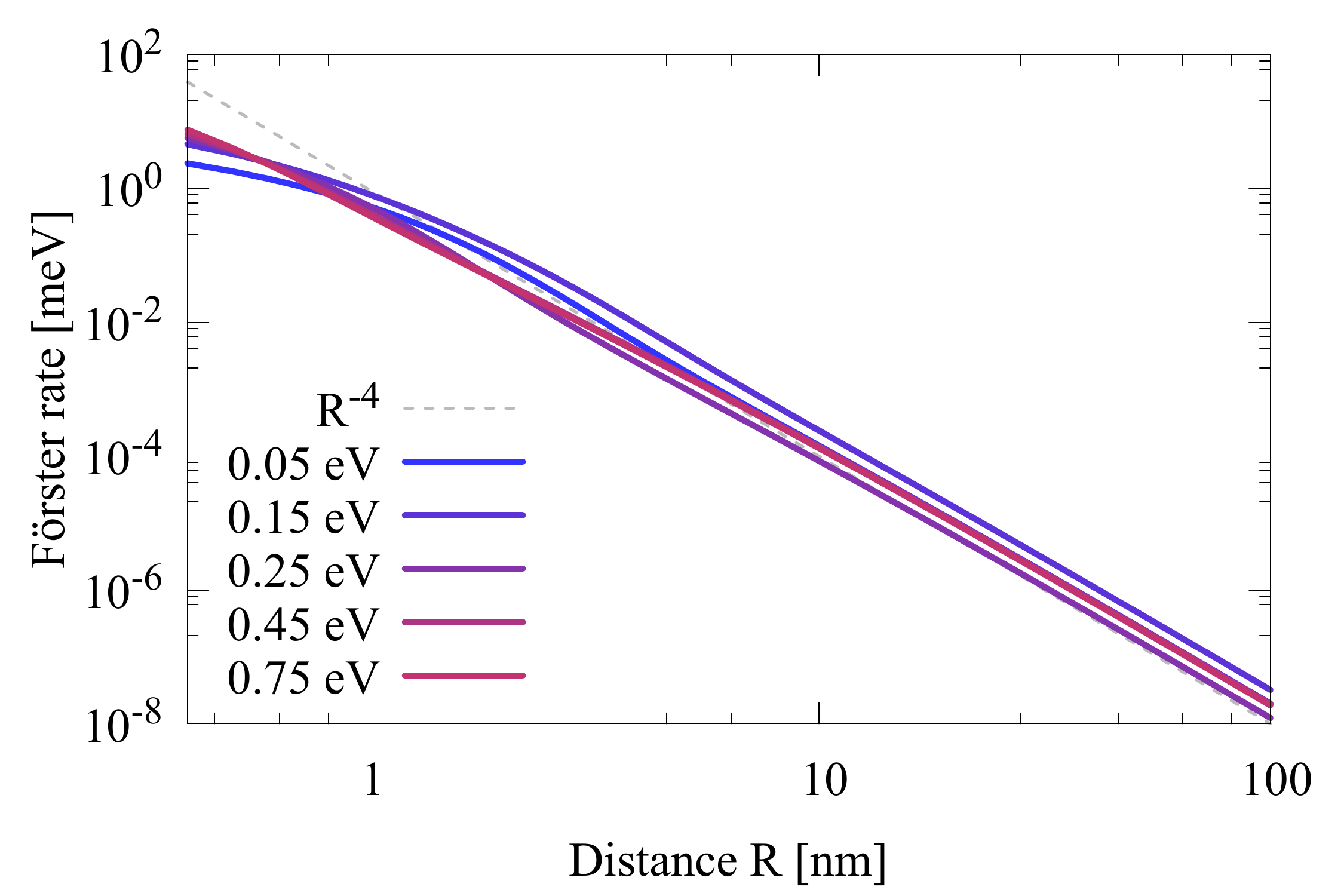}
    \caption{Loglogplot of the Förster rate. In the limit of long distances, the rate in Eq.~(\ref{eq:foerstereigenenergy}) gives a power-law dependence as suggested also by previous works. This accounts for detunings that only allow transfer into bound states as well as for transfer into the unbound exciton continuum. 
    \label{fig:loglogplot}}
\end{figure}
Previous works on comparable systems \cite{Swathi2009,Malic2014,berciaudgruppe2015distance,PhysRevB.99.035420} find a $R^4$-dependence for the Förster rate in the long distance limit. This can be traced back to Eqs.~(29) and (44) in~\cite{Swathi2009}, where the rate of transfer, referred to as k, is found to be proportional to the integral (in their case: $\gamma\rightarrow k$, $Q \rightarrow q$):
\begin{align}
    \gamma \propto
    \int_0^{\hbar \Omega/\nu_f}
    dQ
    \:
    Q^3
    e^{-2QR}
\end{align}
For large values of R, it is then approximated, that $Q\simeq\frac{1}{2R}$, and thus the upper limit of the integration can be extended to $\infty$ without any relevant effect. Thus the integral can be solved by integration by parts to
\begin{align}\label{distanceintegral_sebastian}
\int_0^{\infty}
    dQ
    \:
    Q^3
    e^{-2QR}
    =
    \frac{3}{8}
    \frac{1}{R^4}
\end{align}
The mentioned previous work does not take bound excitonic states into account. We can nevertheless show that the power-law dependence exists in the limit of long enough distances, also with bound states, as can be seen in Fig.~\ref{fig:loglogplot}. The sum in Eq.~(\ref{eq:foerstereigenenergy}) of the main text has a similar Q-R-dependence as Eq.~(\ref{distanceintegral_sebastian}), for the plots it was written in integral form and numerically solved.
\section{Full absorption}\label{app:full_spectrum}
\begin{figure}[t!]
    \centering
    \includegraphics[width=\linewidth]{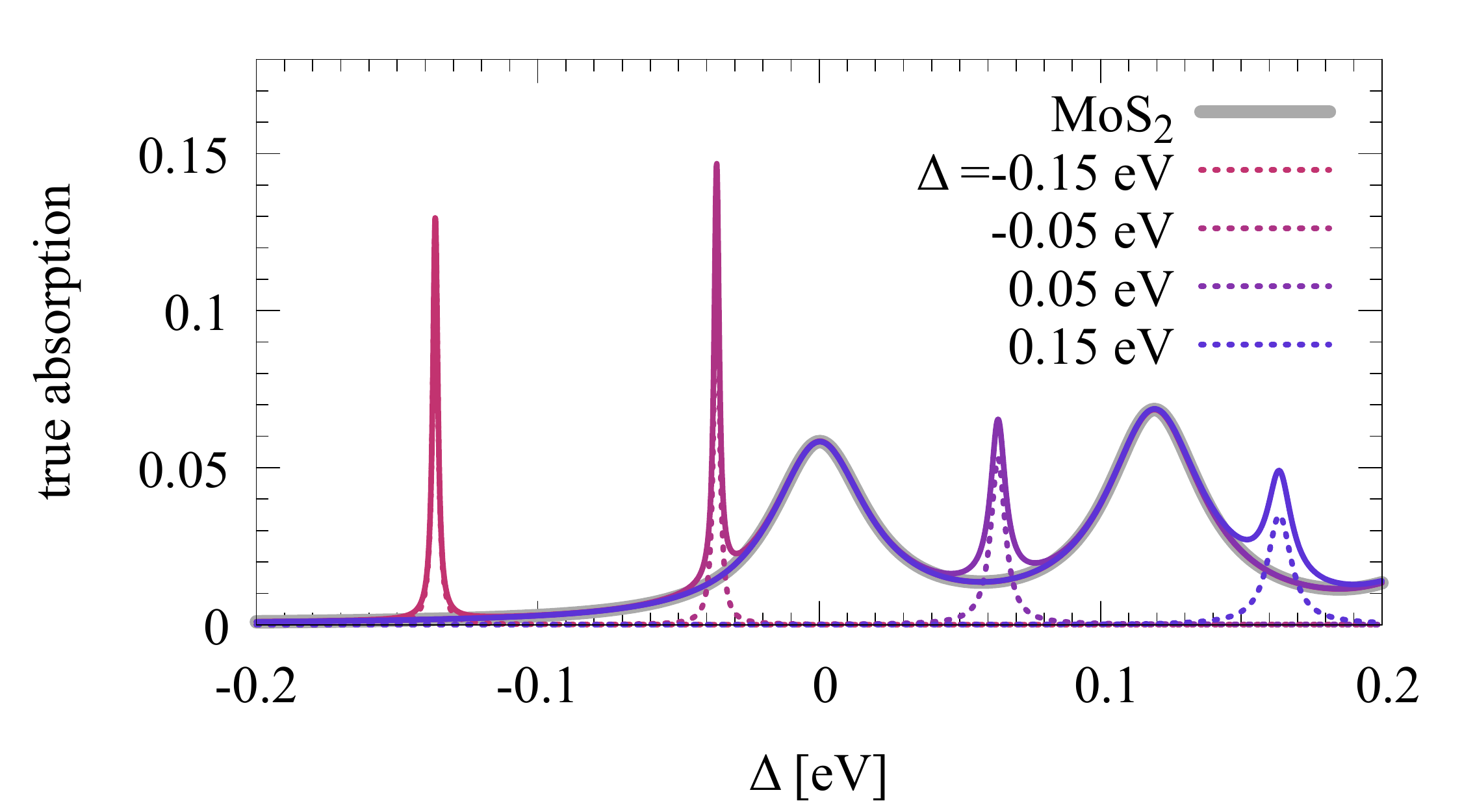}
    \caption{The full absorption can in good approximation be divided into the part of the TMDC and the Molecular part, as it is done in Fig.~\ref{fig:abs_spec}. Dashed lines show the absorption of the molecule for different molecular transition energies, full lines indicate the respective full spectra, while the pristine TMDC absorption is depicted in gray.
    \label{fig:full_abs}}
\end{figure}

As the denominator in Eq.~(\ref{eq:absorption}) is nonlinear, it is in principle not clear whether the absorption of molecules and TMDC layer can be substrated from each other, as it was done for clarity in Fig.~\ref{fig:abs_spec}. We therefore show here that $\chi$ is small enough to justify this, and show the full spectrum in Fig.~\ref{fig:full_abs} for completeness.

\section{Parameters}\label{app:parameterappendix}
\noindent
Parameters for MoS$_2$: 
\begin{align*}
&d\varphi_{r=0}^{\mu\ell}  &0.2 \text{ e nm/nm}\\ 
&M = M_{MoS_2} &0.97\cdot 5.685680 \text{ eV fs}^2\text{nm}^{-2}\\
&\epsilon_2 = \epsilon_{MoS_2}
&13.36\\
& \text{Thickness } R_{\Delta T}
& 0.6 \text{ nm}\\
& \gamma^{\mu\ell} 
& 20 \text{ meV}
\end{align*}
Parameters for the Molecular sheet:
\begin{align*}
    &\text{Molec. Density }n
    &0.05 \text{ nm$^{-2}$}\\
    &d^{12} = d_{Mol}
    &0.187 \text{ e nm}\\
    &\epsilon_{Mol}
    &1 \\ 
    &\gamma_{Mol}
    & 1 \text{ meV}
\end{align*}
Parameters for the substrate:
\begin{align}
    &\epsilon_{SiO_2}
    &3.9
\end{align}
\bibliography{Bib}

\begin{thebibliography}{82}%
\makeatletter
\providecommand \@ifxundefined [1]{%
 \@ifx{#1\undefined}
}%
\providecommand \@ifnum [1]{%
 \ifnum #1\expandafter \@firstoftwo
 \else \expandafter \@secondoftwo
 \fi
}%
\providecommand \@ifx [1]{%
 \ifx #1\expandafter \@firstoftwo
 \else \expandafter \@secondoftwo
 \fi
}%
\providecommand \natexlab [1]{#1}%
\providecommand \enquote  [1]{``#1''}%
\providecommand \bibnamefont  [1]{#1}%
\providecommand \bibfnamefont [1]{#1}%
\providecommand \citenamefont [1]{#1}%
\providecommand \href@noop [0]{\@secondoftwo}%
\providecommand \href [0]{\begingroup \@sanitize@url \@href}%
\providecommand \@href[1]{\@@startlink{#1}\@@href}%
\providecommand \@@href[1]{\endgroup#1\@@endlink}%
\providecommand \@sanitize@url [0]{\catcode `\\12\catcode `\$12\catcode
  `\&12\catcode `\#12\catcode `\^12\catcode `\_12\catcode `\%12\relax}%
\providecommand \@@startlink[1]{}%
\providecommand \@@endlink[0]{}%
\providecommand \url  [0]{\begingroup\@sanitize@url \@url }%
\providecommand \@url [1]{\endgroup\@href {#1}{\urlprefix }}%
\providecommand \urlprefix  [0]{URL }%
\providecommand \Eprint [0]{\href }%
\providecommand \doibase [0]{http://dx.doi.org/}%
\providecommand \selectlanguage [0]{\@gobble}%
\providecommand \bibinfo  [0]{\@secondoftwo}%
\providecommand \bibfield  [0]{\@secondoftwo}%
\providecommand \translation [1]{[#1]}%
\providecommand \BibitemOpen [0]{}%
\providecommand \bibitemStop [0]{}%
\providecommand \bibitemNoStop [0]{.\EOS\space}%
\providecommand \EOS [0]{\spacefactor3000\relax}%
\providecommand \BibitemShut  [1]{\csname bibitem#1\endcsname}%
\let\auto@bib@innerbib\@empty
%</preamble>
\bibitem [{\citenamefont {Blumstengel}\ \emph {et~al.}(2006)\citenamefont
  {Blumstengel}, \citenamefont {Sadofev}, \citenamefont {Xu}, \citenamefont
  {Puls},\ and\ \citenamefont {Henneberger}}]{blumstengel2006OM}%
  \BibitemOpen
  \bibfield  {author} {\bibinfo {author} {\bibfnamefont {S.}~\bibnamefont
  {Blumstengel}}, \bibinfo {author} {\bibfnamefont {S.}~\bibnamefont
  {Sadofev}}, \bibinfo {author} {\bibfnamefont {C.}~\bibnamefont {Xu}},
  \bibinfo {author} {\bibfnamefont {J.}~\bibnamefont {Puls}}, \ and\ \bibinfo
  {author} {\bibfnamefont {F.}~\bibnamefont {Henneberger}},\ }\href@noop {}
  {\bibfield  {journal} {\bibinfo  {journal} {Phys. Rev. Lett.}\ }\textbf
  {\bibinfo {volume} {97}},\ \bibinfo {pages} {237401} (\bibinfo {year}
  {2006})}\BibitemShut {NoStop}%
\bibitem [{\citenamefont {Zhang}\ \emph {et~al.}(2007)\citenamefont {Zhang},
  \citenamefont {Atay}, \citenamefont {Tischler}, \citenamefont {Bradley},
  \citenamefont {Bulovi{\'{c}}},\ and\ \citenamefont
  {Nurmikko}}]{nurmikkogroup2007OM}%
  \BibitemOpen
  \bibfield  {author} {\bibinfo {author} {\bibfnamefont {Q.}~\bibnamefont
  {Zhang}}, \bibinfo {author} {\bibfnamefont {T.}~\bibnamefont {Atay}},
  \bibinfo {author} {\bibfnamefont {J.~R.}\ \bibnamefont {Tischler}}, \bibinfo
  {author} {\bibfnamefont {M.~S.}\ \bibnamefont {Bradley}}, \bibinfo {author}
  {\bibfnamefont {V.}~\bibnamefont {Bulovi{\'{c}}}}, \ and\ \bibinfo {author}
  {\bibfnamefont {A.~V.}\ \bibnamefont {Nurmikko}},\ }\href@noop {} {\bibfield
  {journal} {\bibinfo  {journal} {Nature Nanotechnology}\ }\textbf {\bibinfo
  {volume} {2}},\ \bibinfo {pages} {555} (\bibinfo {year} {2007})}\BibitemShut
  {NoStop}%
\bibitem [{\citenamefont {Vaynzof}\ \emph {et~al.}(2012)\citenamefont
  {Vaynzof}, \citenamefont {Bakulin}, \citenamefont {G\'elinas},\ and\
  \citenamefont {Friend}}]{friendgruppe2012OM}%
  \BibitemOpen
  \bibfield  {author} {\bibinfo {author} {\bibfnamefont {Y.}~\bibnamefont
  {Vaynzof}}, \bibinfo {author} {\bibfnamefont {A.~A.}\ \bibnamefont
  {Bakulin}}, \bibinfo {author} {\bibfnamefont {S.}~\bibnamefont {G\'elinas}},
  \ and\ \bibinfo {author} {\bibfnamefont {R.~H.}\ \bibnamefont {Friend}},\
  }\href@noop {} {\bibfield  {journal} {\bibinfo  {journal} {Phys. Rev. Lett.}\
  }\textbf {\bibinfo {volume} {108}},\ \bibinfo {pages} {246605} (\bibinfo
  {year} {2012})}\BibitemShut {NoStop}%
\bibitem [{\citenamefont {Gaudreau}\ \emph {et~al.}(2013)\citenamefont
  {Gaudreau}, \citenamefont {Tielrooij}, \citenamefont {Prawiroatmodjo},
  \citenamefont {Osmond}, \citenamefont {de~Abajo},\ and\ \citenamefont
  {Koppens}}]{koppensgruppe2013OM}%
  \BibitemOpen
  \bibfield  {author} {\bibinfo {author} {\bibfnamefont {L.}~\bibnamefont
  {Gaudreau}}, \bibinfo {author} {\bibfnamefont {K.~J.}\ \bibnamefont
  {Tielrooij}}, \bibinfo {author} {\bibfnamefont {G.~E. D.~K.}\ \bibnamefont
  {Prawiroatmodjo}}, \bibinfo {author} {\bibfnamefont {J.}~\bibnamefont
  {Osmond}}, \bibinfo {author} {\bibfnamefont {F.~J.~G.}\ \bibnamefont
  {de~Abajo}}, \ and\ \bibinfo {author} {\bibfnamefont {F.~H.~L.}\ \bibnamefont
  {Koppens}},\ }\href@noop {} {\bibfield  {journal} {\bibinfo  {journal} {Nano
  Letters}\ }\textbf {\bibinfo {volume} {13}},\ \bibinfo {pages} {2030}
  (\bibinfo {year} {2013})}\BibitemShut {NoStop}%
\bibitem [{\citenamefont {Schlesinger}\ \emph {et~al.}(2015)\citenamefont
  {Schlesinger}, \citenamefont {Bianchi}, \citenamefont {Blumstengel},
  \citenamefont {Christodoulou}, \citenamefont {Ovsyannikov}, \citenamefont
  {Kobin}, \citenamefont {Moudgil}, \citenamefont {Barlow}, \citenamefont
  {Hecht}, \citenamefont {Marder}, \citenamefont {Henneberger},\ and\
  \citenamefont {Koch}}]{kochgruppe2015OM}%
  \BibitemOpen
  \bibfield  {author} {\bibinfo {author} {\bibfnamefont {R.}~\bibnamefont
  {Schlesinger}}, \bibinfo {author} {\bibfnamefont {F.}~\bibnamefont
  {Bianchi}}, \bibinfo {author} {\bibfnamefont {S.}~\bibnamefont
  {Blumstengel}}, \bibinfo {author} {\bibfnamefont {C.}~\bibnamefont
  {Christodoulou}}, \bibinfo {author} {\bibfnamefont {R.}~\bibnamefont
  {Ovsyannikov}}, \bibinfo {author} {\bibfnamefont {B.}~\bibnamefont {Kobin}},
  \bibinfo {author} {\bibfnamefont {K.}~\bibnamefont {Moudgil}}, \bibinfo
  {author} {\bibfnamefont {S.}~\bibnamefont {Barlow}}, \bibinfo {author}
  {\bibfnamefont {S.}~\bibnamefont {Hecht}}, \bibinfo {author} {\bibfnamefont
  {S.~R.}\ \bibnamefont {Marder}}, \bibinfo {author} {\bibfnamefont
  {F.}~\bibnamefont {Henneberger}}, \ and\ \bibinfo {author} {\bibfnamefont
  {N.}~\bibnamefont {Koch}},\ }\href@noop {} {\bibfield  {journal} {\bibinfo
  {journal} {Nature Communications}\ }\textbf {\bibinfo {volume} {6}},\
  \bibinfo {pages} {6754} (\bibinfo {year} {2015})}\BibitemShut {NoStop}%
\bibitem [{\citenamefont {Tielrooij}\ \emph {et~al.}(2015)\citenamefont
  {Tielrooij}, \citenamefont {Orona}, \citenamefont {Ferrier}, \citenamefont
  {Badioli}, \citenamefont {Navickaite}, \citenamefont {Coop}, \citenamefont
  {Nanot}, \citenamefont {Kalinic}, \citenamefont {Cesca}, \citenamefont
  {Gaudreau}, \citenamefont {Ma}, \citenamefont {Centeno}, \citenamefont
  {Pesquera}, \citenamefont {Zurutuza}, \citenamefont {de~Riedmatten},
  \citenamefont {Goldner}, \citenamefont {Garc{\'i}a~de Abajo}, \citenamefont
  {Jarillo-Herrero},\ and\ \citenamefont {Koppens}}]{koppensgruppe2015OM}%
  \BibitemOpen
  \bibfield  {author} {\bibinfo {author} {\bibfnamefont {K.~J.}\ \bibnamefont
  {Tielrooij}}, \bibinfo {author} {\bibfnamefont {L.}~\bibnamefont {Orona}},
  \bibinfo {author} {\bibfnamefont {A.}~\bibnamefont {Ferrier}}, \bibinfo
  {author} {\bibfnamefont {M.}~\bibnamefont {Badioli}}, \bibinfo {author}
  {\bibfnamefont {G.}~\bibnamefont {Navickaite}}, \bibinfo {author}
  {\bibfnamefont {S.}~\bibnamefont {Coop}}, \bibinfo {author} {\bibfnamefont
  {S.}~\bibnamefont {Nanot}}, \bibinfo {author} {\bibfnamefont
  {B.}~\bibnamefont {Kalinic}}, \bibinfo {author} {\bibfnamefont
  {T.}~\bibnamefont {Cesca}}, \bibinfo {author} {\bibfnamefont
  {L.}~\bibnamefont {Gaudreau}}, \bibinfo {author} {\bibfnamefont
  {Q.}~\bibnamefont {Ma}}, \bibinfo {author} {\bibfnamefont {A.}~\bibnamefont
  {Centeno}}, \bibinfo {author} {\bibfnamefont {A.}~\bibnamefont {Pesquera}},
  \bibinfo {author} {\bibfnamefont {A.}~\bibnamefont {Zurutuza}}, \bibinfo
  {author} {\bibfnamefont {H.}~\bibnamefont {de~Riedmatten}}, \bibinfo {author}
  {\bibfnamefont {P.}~\bibnamefont {Goldner}}, \bibinfo {author} {\bibfnamefont
  {F.~J.}\ \bibnamefont {Garc{\'i}a~de Abajo}}, \bibinfo {author}
  {\bibfnamefont {P.}~\bibnamefont {Jarillo-Herrero}}, \ and\ \bibinfo {author}
  {\bibfnamefont {F.~H.~L.}\ \bibnamefont {Koppens}},\ }\href@noop {}
  {\bibfield  {journal} {\bibinfo  {journal} {Nature Physics}\ }\textbf
  {\bibinfo {volume} {11}},\ \bibinfo {pages} {281} (\bibinfo {year}
  {2015})}\BibitemShut {NoStop}%
\bibitem [{\citenamefont {Flatten}\ \emph {et~al.}(2017)\citenamefont
  {Flatten}, \citenamefont {Coles}, \citenamefont {He}, \citenamefont {Lidzey},
  \citenamefont {Taylor}, \citenamefont {Warner},\ and\ \citenamefont
  {Smith}}]{smithgroup2017OM}%
  \BibitemOpen
  \bibfield  {author} {\bibinfo {author} {\bibfnamefont {L.~C.}\ \bibnamefont
  {Flatten}}, \bibinfo {author} {\bibfnamefont {D.~M.}\ \bibnamefont {Coles}},
  \bibinfo {author} {\bibfnamefont {Z.}~\bibnamefont {He}}, \bibinfo {author}
  {\bibfnamefont {D.~G.}\ \bibnamefont {Lidzey}}, \bibinfo {author}
  {\bibfnamefont {R.~A.}\ \bibnamefont {Taylor}}, \bibinfo {author}
  {\bibfnamefont {J.~H.}\ \bibnamefont {Warner}}, \ and\ \bibinfo {author}
  {\bibfnamefont {J.~M.}\ \bibnamefont {Smith}},\ }\href@noop {} {\bibfield
  {journal} {\bibinfo  {journal} {Nature Communications}\ }\textbf {\bibinfo
  {volume} {8}},\ \bibinfo {pages} {14097} (\bibinfo {year}
  {2017})}\BibitemShut {NoStop}%
\bibitem [{\citenamefont {Specht}\ \emph {et~al.}(2018)\citenamefont {Specht},
  \citenamefont {Verdenhalven}, \citenamefont {Bieniek}, \citenamefont {Rinke},
  \citenamefont {Knorr},\ and\ \citenamefont {Richter}}]{judith2018}%
  \BibitemOpen
  \bibfield  {author} {\bibinfo {author} {\bibfnamefont {J.~F.}\ \bibnamefont
  {Specht}}, \bibinfo {author} {\bibfnamefont {E.}~\bibnamefont
  {Verdenhalven}}, \bibinfo {author} {\bibfnamefont {B.}~\bibnamefont
  {Bieniek}}, \bibinfo {author} {\bibfnamefont {P.}~\bibnamefont {Rinke}},
  \bibinfo {author} {\bibfnamefont {A.}~\bibnamefont {Knorr}}, \ and\ \bibinfo
  {author} {\bibfnamefont {M.}~\bibnamefont {Richter}},\ }\href {\doibase
  10.1103/PhysRevApplied.9.044025} {\bibfield  {journal} {\bibinfo  {journal}
  {Phys. Rev. Applied}\ }\textbf {\bibinfo {volume} {9}},\ \bibinfo {pages}
  {044025} (\bibinfo {year} {2018})}\BibitemShut {NoStop}%
\bibitem [{\citenamefont {Zhao}\ \emph
  {et~al.}(2019{\natexlab{a}})\citenamefont {Zhao}, \citenamefont {Zhao},
  \citenamefont {Song}, \citenamefont {Zhou}, \citenamefont {Lv}, \citenamefont
  {Tao}, \citenamefont {Feng}, \citenamefont {Song}, \citenamefont {Ma},
  \citenamefont {Zhang}, \citenamefont {Xiao}, \citenamefont {Wang},
  \citenamefont {Lien}, \citenamefont {Amani}, \citenamefont {Kim},
  \citenamefont {Chen}, \citenamefont {Wu}, \citenamefont {Ni}, \citenamefont
  {Wang}, \citenamefont {Shi}, \citenamefont {Ma}, \citenamefont {Zhang},
  \citenamefont {Xu}, \citenamefont {Troisi}, \citenamefont {Javey},\ and\
  \citenamefont {Wang}}]{Zhao2019}%
  \BibitemOpen
  \bibfield  {author} {\bibinfo {author} {\bibfnamefont {H.}~\bibnamefont
  {Zhao}}, \bibinfo {author} {\bibfnamefont {Y.}~\bibnamefont {Zhao}}, \bibinfo
  {author} {\bibfnamefont {Y.}~\bibnamefont {Song}}, \bibinfo {author}
  {\bibfnamefont {M.}~\bibnamefont {Zhou}}, \bibinfo {author} {\bibfnamefont
  {W.}~\bibnamefont {Lv}}, \bibinfo {author} {\bibfnamefont {L.}~\bibnamefont
  {Tao}}, \bibinfo {author} {\bibfnamefont {Y.}~\bibnamefont {Feng}}, \bibinfo
  {author} {\bibfnamefont {B.}~\bibnamefont {Song}}, \bibinfo {author}
  {\bibfnamefont {Y.}~\bibnamefont {Ma}}, \bibinfo {author} {\bibfnamefont
  {J.}~\bibnamefont {Zhang}}, \bibinfo {author} {\bibfnamefont
  {J.}~\bibnamefont {Xiao}}, \bibinfo {author} {\bibfnamefont {Y.}~\bibnamefont
  {Wang}}, \bibinfo {author} {\bibfnamefont {D.-H.}\ \bibnamefont {Lien}},
  \bibinfo {author} {\bibfnamefont {M.}~\bibnamefont {Amani}}, \bibinfo
  {author} {\bibfnamefont {H.}~\bibnamefont {Kim}}, \bibinfo {author}
  {\bibfnamefont {X.}~\bibnamefont {Chen}}, \bibinfo {author} {\bibfnamefont
  {Z.}~\bibnamefont {Wu}}, \bibinfo {author} {\bibfnamefont {Z.}~\bibnamefont
  {Ni}}, \bibinfo {author} {\bibfnamefont {P.}~\bibnamefont {Wang}}, \bibinfo
  {author} {\bibfnamefont {Y.}~\bibnamefont {Shi}}, \bibinfo {author}
  {\bibfnamefont {H.}~\bibnamefont {Ma}}, \bibinfo {author} {\bibfnamefont
  {X.}~\bibnamefont {Zhang}}, \bibinfo {author} {\bibfnamefont {J.-B.}\
  \bibnamefont {Xu}}, \bibinfo {author} {\bibfnamefont {A.}~\bibnamefont
  {Troisi}}, \bibinfo {author} {\bibfnamefont {A.}~\bibnamefont {Javey}}, \
  and\ \bibinfo {author} {\bibfnamefont {X.}~\bibnamefont {Wang}},\ }\href@noop
  {} {\bibfield  {journal} {\bibinfo  {journal} {Nature Communications}\
  }\textbf {\bibinfo {volume} {10}},\ \bibinfo {pages} {5589} (\bibinfo {year}
  {2019}{\natexlab{a}})}\BibitemShut {NoStop}%
\bibitem [{\citenamefont {Karpińska}\ \emph {et~al.}(2021)\citenamefont
  {Karpińska}, \citenamefont {Liang}, \citenamefont {Kempt}, \citenamefont
  {Finzel}, \citenamefont {Kamminga}, \citenamefont {Dyksik}, \citenamefont
  {Zhang}, \citenamefont {Knodlseder}, \citenamefont {Maude}, \citenamefont
  {Baranowski}, \citenamefont {Klopotowski}, \citenamefont {Ye}, \citenamefont
  {Kuc},\ and\ \citenamefont {Plochocka}}]{plochocka2021HIOS}%
  \BibitemOpen
  \bibfield  {author} {\bibinfo {author} {\bibfnamefont {M.}~\bibnamefont
  {Karpińska}}, \bibinfo {author} {\bibfnamefont {M.}~\bibnamefont {Liang}},
  \bibinfo {author} {\bibfnamefont {R.}~\bibnamefont {Kempt}}, \bibinfo
  {author} {\bibfnamefont {K.}~\bibnamefont {Finzel}}, \bibinfo {author}
  {\bibfnamefont {M.}~\bibnamefont {Kamminga}}, \bibinfo {author}
  {\bibfnamefont {M.}~\bibnamefont {Dyksik}}, \bibinfo {author} {\bibfnamefont
  {N.}~\bibnamefont {Zhang}}, \bibinfo {author} {\bibfnamefont
  {C.}~\bibnamefont {Knodlseder}}, \bibinfo {author} {\bibfnamefont {D.~K.}\
  \bibnamefont {Maude}}, \bibinfo {author} {\bibfnamefont {M.}~\bibnamefont
  {Baranowski}}, \bibinfo {author} {\bibfnamefont {L.}~\bibnamefont
  {Klopotowski}}, \bibinfo {author} {\bibfnamefont {J.}~\bibnamefont {Ye}},
  \bibinfo {author} {\bibfnamefont {A.}~\bibnamefont {Kuc}}, \ and\ \bibinfo
  {author} {\bibfnamefont {P.}~\bibnamefont {Plochocka}},\ }\href@noop {}
  {\bibfield  {journal} {\bibinfo  {journal} {ACS Applied Materials \&
  Interfaces}\ } (\bibinfo {year} {2021})}\BibitemShut {NoStop}%
\bibitem [{\citenamefont {Park}\ \emph {et~al.}(2021)\citenamefont {Park},
  \citenamefont {Mutz}, \citenamefont {Kovalenko}, \citenamefont {Schultz},
  \citenamefont {Shin}, \citenamefont {Aljarb}, \citenamefont {Li},
  \citenamefont {Tung}, \citenamefont {Amsalem}, \citenamefont
  {List-Kratochvil}, \citenamefont {Stähler}, \citenamefont {Xu},
  \citenamefont {Blumstengel},\ and\ \citenamefont {Koch}}]{kochgruppe2021}%
  \BibitemOpen
  \bibfield  {author} {\bibinfo {author} {\bibfnamefont {S.}~\bibnamefont
  {Park}}, \bibinfo {author} {\bibfnamefont {N.}~\bibnamefont {Mutz}}, \bibinfo
  {author} {\bibfnamefont {S.~A.}\ \bibnamefont {Kovalenko}}, \bibinfo {author}
  {\bibfnamefont {T.}~\bibnamefont {Schultz}}, \bibinfo {author} {\bibfnamefont
  {D.}~\bibnamefont {Shin}}, \bibinfo {author} {\bibfnamefont {A.}~\bibnamefont
  {Aljarb}}, \bibinfo {author} {\bibfnamefont {L.-J.}\ \bibnamefont {Li}},
  \bibinfo {author} {\bibfnamefont {V.}~\bibnamefont {Tung}}, \bibinfo {author}
  {\bibfnamefont {P.}~\bibnamefont {Amsalem}}, \bibinfo {author} {\bibfnamefont
  {E.~J.~W.}\ \bibnamefont {List-Kratochvil}}, \bibinfo {author} {\bibfnamefont
  {J.}~\bibnamefont {Stähler}}, \bibinfo {author} {\bibfnamefont
  {X.}~\bibnamefont {Xu}}, \bibinfo {author} {\bibfnamefont {S.}~\bibnamefont
  {Blumstengel}}, \ and\ \bibinfo {author} {\bibfnamefont {N.}~\bibnamefont
  {Koch}},\ }\href {\doibase https://doi.org/10.1002/advs.202100215} {\bibfield
   {journal} {\bibinfo  {journal} {Advanced Science}\ }\textbf {\bibinfo
  {volume} {8}},\ \bibinfo {pages} {2100215} (\bibinfo {year}
  {2021})}\BibitemShut {NoStop}%
\bibitem [{\citenamefont {Feierabend}\ \emph {et~al.}(2021)\citenamefont
  {Feierabend}, \citenamefont {Lumsargis}, \citenamefont {Thompson},
  \citenamefont {Wang}, \citenamefont {Dou}, \citenamefont {Huang},\ and\
  \citenamefont {Malic}}]{feierabend2021dark}%
  \BibitemOpen
  \bibfield  {author} {\bibinfo {author} {\bibfnamefont {M.}~\bibnamefont
  {Feierabend}}, \bibinfo {author} {\bibfnamefont {V.}~\bibnamefont
  {Lumsargis}}, \bibinfo {author} {\bibfnamefont {J.~J.}\ \bibnamefont
  {Thompson}}, \bibinfo {author} {\bibfnamefont {K.}~\bibnamefont {Wang}},
  \bibinfo {author} {\bibfnamefont {L.}~\bibnamefont {Dou}}, \bibinfo {author}
  {\bibfnamefont {L.}~\bibnamefont {Huang}}, \ and\ \bibinfo {author}
  {\bibfnamefont {E.}~\bibnamefont {Malic}},\ }\href@noop {} {\bibfield
  {journal} {\bibinfo  {journal} {arXiv preprint arXiv:2111.12400}\ } (\bibinfo
  {year} {2021})}\BibitemShut {NoStop}%
\bibitem [{\citenamefont {Chernikov}\ \emph {et~al.}(2014)\citenamefont
  {Chernikov}, \citenamefont {Berkelbach}, \citenamefont {Hill}, \citenamefont
  {Rigosi}, \citenamefont {Li}, \citenamefont {Aslan}, \citenamefont
  {Reichman}, \citenamefont {Hybertsen},\ and\ \citenamefont
  {Heinz}}]{chernikov2014rydbergseries}%
  \BibitemOpen
  \bibfield  {author} {\bibinfo {author} {\bibfnamefont {A.}~\bibnamefont
  {Chernikov}}, \bibinfo {author} {\bibfnamefont {T.~C.}\ \bibnamefont
  {Berkelbach}}, \bibinfo {author} {\bibfnamefont {H.~M.}\ \bibnamefont
  {Hill}}, \bibinfo {author} {\bibfnamefont {A.}~\bibnamefont {Rigosi}},
  \bibinfo {author} {\bibfnamefont {Y.}~\bibnamefont {Li}}, \bibinfo {author}
  {\bibfnamefont {O.~B.}\ \bibnamefont {Aslan}}, \bibinfo {author}
  {\bibfnamefont {D.~R.}\ \bibnamefont {Reichman}}, \bibinfo {author}
  {\bibfnamefont {M.~S.}\ \bibnamefont {Hybertsen}}, \ and\ \bibinfo {author}
  {\bibfnamefont {T.~F.}\ \bibnamefont {Heinz}},\ }\href@noop {} {\bibfield
  {journal} {\bibinfo  {journal} {Phys. Rev. Lett.}\ }\textbf {\bibinfo
  {volume} {113}},\ \bibinfo {pages} {076802} (\bibinfo {year}
  {2014})}\BibitemShut {NoStop}%
\bibitem [{\citenamefont {Bergh\"auser}\ and\ \citenamefont
  {Malic}(2014)}]{gunnar2014wannier}%
  \BibitemOpen
  \bibfield  {author} {\bibinfo {author} {\bibfnamefont {G.}~\bibnamefont
  {Bergh\"auser}}\ and\ \bibinfo {author} {\bibfnamefont {E.}~\bibnamefont
  {Malic}},\ }\href {\doibase 10.1103/PhysRevB.89.125309} {\bibfield  {journal}
  {\bibinfo  {journal} {Phys. Rev. B}\ }\textbf {\bibinfo {volume} {89}},\
  \bibinfo {pages} {125309} (\bibinfo {year} {2014})}\BibitemShut {NoStop}%
\bibitem [{\citenamefont {Steinhoff}\ \emph {et~al.}(2017)\citenamefont
  {Steinhoff}, \citenamefont {Florian}, \citenamefont {R{\"o}sner},
  \citenamefont {Sch{\"o}nhoff}, \citenamefont {Wehling},\ and\ \citenamefont
  {Jahnke}}]{Steinhoff2017fission}%
  \BibitemOpen
  \bibfield  {author} {\bibinfo {author} {\bibfnamefont {A.}~\bibnamefont
  {Steinhoff}}, \bibinfo {author} {\bibfnamefont {M.}~\bibnamefont {Florian}},
  \bibinfo {author} {\bibfnamefont {M.}~\bibnamefont {R{\"o}sner}}, \bibinfo
  {author} {\bibfnamefont {G.}~\bibnamefont {Sch{\"o}nhoff}}, \bibinfo {author}
  {\bibfnamefont {T.~O.}\ \bibnamefont {Wehling}}, \ and\ \bibinfo {author}
  {\bibfnamefont {F.}~\bibnamefont {Jahnke}},\ }\href@noop {} {\bibfield
  {journal} {\bibinfo  {journal} {Nature Communications}\ }\textbf {\bibinfo
  {volume} {8}},\ \bibinfo {pages} {1166} (\bibinfo {year} {2017})}\BibitemShut
  {NoStop}%
\bibitem [{\citenamefont {Wang}\ \emph {et~al.}(2018)\citenamefont {Wang},
  \citenamefont {Chernikov}, \citenamefont {Glazov}, \citenamefont {Heinz},
  \citenamefont {Marie}, \citenamefont {Amand},\ and\ \citenamefont
  {Urbaszek}}]{chernikov2018RevModPhys}%
  \BibitemOpen
  \bibfield  {author} {\bibinfo {author} {\bibfnamefont {G.}~\bibnamefont
  {Wang}}, \bibinfo {author} {\bibfnamefont {A.}~\bibnamefont {Chernikov}},
  \bibinfo {author} {\bibfnamefont {M.~M.}\ \bibnamefont {Glazov}}, \bibinfo
  {author} {\bibfnamefont {T.~F.}\ \bibnamefont {Heinz}}, \bibinfo {author}
  {\bibfnamefont {X.}~\bibnamefont {Marie}}, \bibinfo {author} {\bibfnamefont
  {T.}~\bibnamefont {Amand}}, \ and\ \bibinfo {author} {\bibfnamefont
  {B.}~\bibnamefont {Urbaszek}},\ }\href@noop {} {\bibfield  {journal}
  {\bibinfo  {journal} {Rev. Mod. Phys.}\ }\textbf {\bibinfo {volume} {90}},\
  \bibinfo {pages} {021001} (\bibinfo {year} {2018})}\BibitemShut {NoStop}%
\bibitem [{\citenamefont {Selig}\ \emph {et~al.}(2020)\citenamefont {Selig},
  \citenamefont {Katsch}, \citenamefont {Brem}, \citenamefont {Mkrtchian},
  \citenamefont {Malic},\ and\ \citenamefont {Knorr}}]{selig2020suppression}%
  \BibitemOpen
  \bibfield  {author} {\bibinfo {author} {\bibfnamefont {M.}~\bibnamefont
  {Selig}}, \bibinfo {author} {\bibfnamefont {F.}~\bibnamefont {Katsch}},
  \bibinfo {author} {\bibfnamefont {S.}~\bibnamefont {Brem}}, \bibinfo {author}
  {\bibfnamefont {G.~F.}\ \bibnamefont {Mkrtchian}}, \bibinfo {author}
  {\bibfnamefont {E.}~\bibnamefont {Malic}}, \ and\ \bibinfo {author}
  {\bibfnamefont {A.}~\bibnamefont {Knorr}},\ }\href@noop {} {\bibfield
  {journal} {\bibinfo  {journal} {Physical Review Research}\ }\textbf {\bibinfo
  {volume} {2}},\ \bibinfo {pages} {023322} (\bibinfo {year}
  {2020})}\BibitemShut {NoStop}%
\bibitem [{\citenamefont {Lengers}\ \emph {et~al.}(2020)\citenamefont
  {Lengers}, \citenamefont {Kuhn},\ and\ \citenamefont
  {Reiter}}]{reiter2020phonon}%
  \BibitemOpen
  \bibfield  {author} {\bibinfo {author} {\bibfnamefont {F.}~\bibnamefont
  {Lengers}}, \bibinfo {author} {\bibfnamefont {T.}~\bibnamefont {Kuhn}}, \
  and\ \bibinfo {author} {\bibfnamefont {D.~E.}\ \bibnamefont {Reiter}},\
  }\href {\doibase 10.1103/PhysRevB.101.155304} {\bibfield  {journal} {\bibinfo
   {journal} {Phys. Rev. B}\ }\textbf {\bibinfo {volume} {101}},\ \bibinfo
  {pages} {155304} (\bibinfo {year} {2020})}\BibitemShut {NoStop}%
\bibitem [{\citenamefont {Trovatello}\ \emph {et~al.}(2020)\citenamefont
  {Trovatello}, \citenamefont {Katsch}, \citenamefont {Borys}, \citenamefont
  {Selig}, \citenamefont {Yao}, \citenamefont {Borrego-Varillas}, \citenamefont
  {Scotognella}, \citenamefont {Kriegel}, \citenamefont {Yan}, \citenamefont
  {Zettl}, \citenamefont {Schuck}, \citenamefont {Knorr}, \citenamefont
  {Cerullo},\ and\ \citenamefont {Conte}}]{Trovatello2020excitonproperties}%
  \BibitemOpen
  \bibfield  {author} {\bibinfo {author} {\bibfnamefont {C.}~\bibnamefont
  {Trovatello}}, \bibinfo {author} {\bibfnamefont {F.}~\bibnamefont {Katsch}},
  \bibinfo {author} {\bibfnamefont {N.~J.}\ \bibnamefont {Borys}}, \bibinfo
  {author} {\bibfnamefont {M.}~\bibnamefont {Selig}}, \bibinfo {author}
  {\bibfnamefont {K.}~\bibnamefont {Yao}}, \bibinfo {author} {\bibfnamefont
  {R.}~\bibnamefont {Borrego-Varillas}}, \bibinfo {author} {\bibfnamefont
  {F.}~\bibnamefont {Scotognella}}, \bibinfo {author} {\bibfnamefont
  {I.}~\bibnamefont {Kriegel}}, \bibinfo {author} {\bibfnamefont
  {A.}~\bibnamefont {Yan}}, \bibinfo {author} {\bibfnamefont {A.}~\bibnamefont
  {Zettl}}, \bibinfo {author} {\bibfnamefont {P.~J.}\ \bibnamefont {Schuck}},
  \bibinfo {author} {\bibfnamefont {A.}~\bibnamefont {Knorr}}, \bibinfo
  {author} {\bibfnamefont {G.}~\bibnamefont {Cerullo}}, \ and\ \bibinfo
  {author} {\bibfnamefont {S.~D.}\ \bibnamefont {Conte}},\ }\href@noop {}
  {\bibfield  {journal} {\bibinfo  {journal} {Nature Communications}\ }\textbf
  {\bibinfo {volume} {11}},\ \bibinfo {pages} {5277} (\bibinfo {year}
  {2020})}\BibitemShut {NoStop}%
\bibitem [{\citenamefont {Katsch}\ \emph {et~al.}(2020)\citenamefont {Katsch},
  \citenamefont {Selig},\ and\ \citenamefont {Knorr}}]{katsch2020exciton}%
  \BibitemOpen
  \bibfield  {author} {\bibinfo {author} {\bibfnamefont {F.}~\bibnamefont
  {Katsch}}, \bibinfo {author} {\bibfnamefont {M.}~\bibnamefont {Selig}}, \
  and\ \bibinfo {author} {\bibfnamefont {A.}~\bibnamefont {Knorr}},\
  }\href@noop {} {\bibfield  {journal} {\bibinfo  {journal} {Physical Review
  Letters}\ }\textbf {\bibinfo {volume} {124}},\ \bibinfo {pages} {257402}
  (\bibinfo {year} {2020})}\BibitemShut {NoStop}%
\bibitem [{\citenamefont {Dong}\ \emph {et~al.}(2021)\citenamefont {Dong},
  \citenamefont {Puppin}, \citenamefont {Pincelli}, \citenamefont {Beaulieu},
  \citenamefont {Christiansen}, \citenamefont {Hübener}, \citenamefont
  {Nicholson}, \citenamefont {Xian}, \citenamefont {Dendzik}, \citenamefont
  {Deng}, \citenamefont {Windsor}, \citenamefont {Selig}, \citenamefont
  {Malic}, \citenamefont {Rubio}, \citenamefont {Knorr}, \citenamefont {Wolf},
  \citenamefont {Rettig},\ and\ \citenamefont {Ernstorfer}}]{shoudong2021}%
  \BibitemOpen
  \bibfield  {author} {\bibinfo {author} {\bibfnamefont {S.}~\bibnamefont
  {Dong}}, \bibinfo {author} {\bibfnamefont {M.}~\bibnamefont {Puppin}},
  \bibinfo {author} {\bibfnamefont {T.}~\bibnamefont {Pincelli}}, \bibinfo
  {author} {\bibfnamefont {S.}~\bibnamefont {Beaulieu}}, \bibinfo {author}
  {\bibfnamefont {D.}~\bibnamefont {Christiansen}}, \bibinfo {author}
  {\bibfnamefont {H.}~\bibnamefont {Hübener}}, \bibinfo {author}
  {\bibfnamefont {C.~W.}\ \bibnamefont {Nicholson}}, \bibinfo {author}
  {\bibfnamefont {R.~P.}\ \bibnamefont {Xian}}, \bibinfo {author}
  {\bibfnamefont {M.}~\bibnamefont {Dendzik}}, \bibinfo {author} {\bibfnamefont
  {Y.}~\bibnamefont {Deng}}, \bibinfo {author} {\bibfnamefont {Y.~W.}\
  \bibnamefont {Windsor}}, \bibinfo {author} {\bibfnamefont {M.}~\bibnamefont
  {Selig}}, \bibinfo {author} {\bibfnamefont {E.}~\bibnamefont {Malic}},
  \bibinfo {author} {\bibfnamefont {A.}~\bibnamefont {Rubio}}, \bibinfo
  {author} {\bibfnamefont {A.}~\bibnamefont {Knorr}}, \bibinfo {author}
  {\bibfnamefont {M.}~\bibnamefont {Wolf}}, \bibinfo {author} {\bibfnamefont
  {L.}~\bibnamefont {Rettig}}, \ and\ \bibinfo {author} {\bibfnamefont
  {R.}~\bibnamefont {Ernstorfer}},\ }\href@noop {} {\bibfield  {journal}
  {\bibinfo  {journal} {Natural Sciences}\ }\textbf {\bibinfo {volume} {1}},\
  \bibinfo {pages} {e10010} (\bibinfo {year} {2021})}\BibitemShut {NoStop}%
\bibitem [{\citenamefont {Selig}\ \emph {et~al.}(2022)\citenamefont {Selig},
  \citenamefont {Christiansen}, \citenamefont {Katzer}, \citenamefont
  {Ballottin}, \citenamefont {Christianen},\ and\ \citenamefont
  {Knorr}}]{selig2022impact}%
  \BibitemOpen
  \bibfield  {author} {\bibinfo {author} {\bibfnamefont {M.}~\bibnamefont
  {Selig}}, \bibinfo {author} {\bibfnamefont {D.}~\bibnamefont {Christiansen}},
  \bibinfo {author} {\bibfnamefont {M.}~\bibnamefont {Katzer}}, \bibinfo
  {author} {\bibfnamefont {M.~V.}\ \bibnamefont {Ballottin}}, \bibinfo {author}
  {\bibfnamefont {P.}~\bibnamefont {Christianen}}, \ and\ \bibinfo {author}
  {\bibfnamefont {A.}~\bibnamefont {Knorr}},\ }\href@noop {} {\bibfield
  {journal} {\bibinfo  {journal} {arXiv preprint arXiv:2201.03362}\ } (\bibinfo
  {year} {2022})}\BibitemShut {NoStop}%
\bibitem [{\citenamefont {Holler}\ \emph {et~al.}(2022)\citenamefont {Holler},
  \citenamefont {Selig}, \citenamefont {Kempf}, \citenamefont {Zipfel},
  \citenamefont {Nagler}, \citenamefont {Katzer}, \citenamefont {Katsch},
  \citenamefont {Ballottin}, \citenamefont {Mitioglu}, \citenamefont
  {Chernikov} \emph {et~al.}}]{holler2022interlayer}%
  \BibitemOpen
  \bibfield  {author} {\bibinfo {author} {\bibfnamefont {J.}~\bibnamefont
  {Holler}}, \bibinfo {author} {\bibfnamefont {M.}~\bibnamefont {Selig}},
  \bibinfo {author} {\bibfnamefont {M.}~\bibnamefont {Kempf}}, \bibinfo
  {author} {\bibfnamefont {J.}~\bibnamefont {Zipfel}}, \bibinfo {author}
  {\bibfnamefont {P.}~\bibnamefont {Nagler}}, \bibinfo {author} {\bibfnamefont
  {M.}~\bibnamefont {Katzer}}, \bibinfo {author} {\bibfnamefont
  {F.}~\bibnamefont {Katsch}}, \bibinfo {author} {\bibfnamefont {M.~V.}\
  \bibnamefont {Ballottin}}, \bibinfo {author} {\bibfnamefont {A.~A.}\
  \bibnamefont {Mitioglu}}, \bibinfo {author} {\bibfnamefont {A.}~\bibnamefont
  {Chernikov}},  \emph {et~al.},\ }\href@noop {} {\bibfield  {journal}
  {\bibinfo  {journal} {Physical Review B}\ }\textbf {\bibinfo {volume}
  {105}},\ \bibinfo {pages} {085303} (\bibinfo {year} {2022})}\BibitemShut
  {NoStop}%
\bibitem [{\citenamefont {Prasai}\ \emph {et~al.}(2015)\citenamefont {Prasai},
  \citenamefont {Klots}, \citenamefont {Newaz}, \citenamefont {Niezgoda},
  \citenamefont {Orfield}, \citenamefont {Escobar}, \citenamefont {Wynn},
  \citenamefont {Efimov}, \citenamefont {Jennings}, \citenamefont {Rosenthal},\
  and\ \citenamefont {Bolotin}}]{Bolotingruppe2015QD}%
  \BibitemOpen
  \bibfield  {author} {\bibinfo {author} {\bibfnamefont {D.}~\bibnamefont
  {Prasai}}, \bibinfo {author} {\bibfnamefont {A.~R.}\ \bibnamefont {Klots}},
  \bibinfo {author} {\bibfnamefont {A.}~\bibnamefont {Newaz}}, \bibinfo
  {author} {\bibfnamefont {J.~S.}\ \bibnamefont {Niezgoda}}, \bibinfo {author}
  {\bibfnamefont {N.~J.}\ \bibnamefont {Orfield}}, \bibinfo {author}
  {\bibfnamefont {C.~A.}\ \bibnamefont {Escobar}}, \bibinfo {author}
  {\bibfnamefont {A.}~\bibnamefont {Wynn}}, \bibinfo {author} {\bibfnamefont
  {A.}~\bibnamefont {Efimov}}, \bibinfo {author} {\bibfnamefont {G.~K.}\
  \bibnamefont {Jennings}}, \bibinfo {author} {\bibfnamefont {S.~J.}\
  \bibnamefont {Rosenthal}}, \ and\ \bibinfo {author} {\bibfnamefont {K.~I.}\
  \bibnamefont {Bolotin}},\ }\href@noop {} {\bibfield  {journal} {\bibinfo
  {journal} {Nano Letters}\ }\textbf {\bibinfo {volume} {15}},\ \bibinfo
  {pages} {4374} (\bibinfo {year} {2015})}\BibitemShut {NoStop}%
\bibitem [{\citenamefont {Raja}\ \emph {et~al.}(2016)\citenamefont {Raja},
  \citenamefont {Montoya-Castillo}, \citenamefont {Zultak}, \citenamefont
  {Zhang}, \citenamefont {Ye}, \citenamefont {Roquelet}, \citenamefont
  {Chenet}, \citenamefont {Van Der~Zande}, \citenamefont {Huang}, \citenamefont
  {Jockusch} \emph {et~al.}}]{raja2016energy}%
  \BibitemOpen
  \bibfield  {author} {\bibinfo {author} {\bibfnamefont {A.}~\bibnamefont
  {Raja}}, \bibinfo {author} {\bibfnamefont {A.}~\bibnamefont
  {Montoya-Castillo}}, \bibinfo {author} {\bibfnamefont {J.}~\bibnamefont
  {Zultak}}, \bibinfo {author} {\bibfnamefont {X.-X.}\ \bibnamefont {Zhang}},
  \bibinfo {author} {\bibfnamefont {Z.}~\bibnamefont {Ye}}, \bibinfo {author}
  {\bibfnamefont {C.}~\bibnamefont {Roquelet}}, \bibinfo {author}
  {\bibfnamefont {D.~A.}\ \bibnamefont {Chenet}}, \bibinfo {author}
  {\bibfnamefont {A.~M.}\ \bibnamefont {Van Der~Zande}}, \bibinfo {author}
  {\bibfnamefont {P.}~\bibnamefont {Huang}}, \bibinfo {author} {\bibfnamefont
  {S.}~\bibnamefont {Jockusch}},  \emph {et~al.},\ }\href@noop {} {\bibfield
  {journal} {\bibinfo  {journal} {Nano letters}\ }\textbf {\bibinfo {volume}
  {16}},\ \bibinfo {pages} {2328} (\bibinfo {year} {2016})}\BibitemShut
  {NoStop}%
\bibitem [{\citenamefont {Zhang}\ \emph {et~al.}(2019)\citenamefont {Zhang},
  \citenamefont {Lian}, \citenamefont {Yang}, \citenamefont {Zhang},\ and\
  \citenamefont {Zhu}}]{zhu2019QD}%
  \BibitemOpen
  \bibfield  {author} {\bibinfo {author} {\bibfnamefont {C.}~\bibnamefont
  {Zhang}}, \bibinfo {author} {\bibfnamefont {L.}~\bibnamefont {Lian}},
  \bibinfo {author} {\bibfnamefont {Z.}~\bibnamefont {Yang}}, \bibinfo {author}
  {\bibfnamefont {J.}~\bibnamefont {Zhang}}, \ and\ \bibinfo {author}
  {\bibfnamefont {H.}~\bibnamefont {Zhu}},\ }\href@noop {} {\bibfield
  {journal} {\bibinfo  {journal} {The Journal of Physical Chemistry Letters}\
  }\textbf {\bibinfo {volume} {10}},\ \bibinfo {pages} {7665} (\bibinfo {year}
  {2019})}\BibitemShut {NoStop}%
\bibitem [{\citenamefont {Tanoh}\ \emph {et~al.}(2020)\citenamefont {Tanoh},
  \citenamefont {Gauriot}, \citenamefont {Delport}, \citenamefont {Xiao},
  \citenamefont {Pandya}, \citenamefont {Sung}, \citenamefont {Allardice},
  \citenamefont {Li}, \citenamefont {Williams}, \citenamefont {Baldwin},
  \citenamefont {Stranks},\ and\ \citenamefont {Rao}}]{akshaygruppe2020QD}%
  \BibitemOpen
  \bibfield  {author} {\bibinfo {author} {\bibfnamefont {A.~O.~A.}\
  \bibnamefont {Tanoh}}, \bibinfo {author} {\bibfnamefont {N.}~\bibnamefont
  {Gauriot}}, \bibinfo {author} {\bibfnamefont {G.}~\bibnamefont {Delport}},
  \bibinfo {author} {\bibfnamefont {J.}~\bibnamefont {Xiao}}, \bibinfo {author}
  {\bibfnamefont {R.}~\bibnamefont {Pandya}}, \bibinfo {author} {\bibfnamefont
  {J.}~\bibnamefont {Sung}}, \bibinfo {author} {\bibfnamefont {J.}~\bibnamefont
  {Allardice}}, \bibinfo {author} {\bibfnamefont {Z.}~\bibnamefont {Li}},
  \bibinfo {author} {\bibfnamefont {C.~A.}\ \bibnamefont {Williams}}, \bibinfo
  {author} {\bibfnamefont {A.}~\bibnamefont {Baldwin}}, \bibinfo {author}
  {\bibfnamefont {S.~D.}\ \bibnamefont {Stranks}}, \ and\ \bibinfo {author}
  {\bibfnamefont {A.}~\bibnamefont {Rao}},\ }\href@noop {} {\bibfield
  {journal} {\bibinfo  {journal} {ACS Nano}\ }\textbf {\bibinfo {volume}
  {14}},\ \bibinfo {pages} {15374} (\bibinfo {year} {2020})}\BibitemShut
  {NoStop}%
\bibitem [{\citenamefont {Tisler}\ \emph {et~al.}(2011)\citenamefont {Tisler},
  \citenamefont {Reuter}, \citenamefont {L{\"a}mmle}, \citenamefont {Jelezko},
  \citenamefont {Balasubramanian}, \citenamefont {Hemmer}, \citenamefont
  {Reinhard},\ and\ \citenamefont {Wrachtrup}}]{Tisler2011NVcenter}%
  \BibitemOpen
  \bibfield  {author} {\bibinfo {author} {\bibfnamefont {J.}~\bibnamefont
  {Tisler}}, \bibinfo {author} {\bibfnamefont {R.}~\bibnamefont {Reuter}},
  \bibinfo {author} {\bibfnamefont {A.}~\bibnamefont {L{\"a}mmle}}, \bibinfo
  {author} {\bibfnamefont {F.}~\bibnamefont {Jelezko}}, \bibinfo {author}
  {\bibfnamefont {G.}~\bibnamefont {Balasubramanian}}, \bibinfo {author}
  {\bibfnamefont {P.~R.}\ \bibnamefont {Hemmer}}, \bibinfo {author}
  {\bibfnamefont {F.}~\bibnamefont {Reinhard}}, \ and\ \bibinfo {author}
  {\bibfnamefont {J.}~\bibnamefont {Wrachtrup}},\ }\href@noop {} {\bibfield
  {journal} {\bibinfo  {journal} {ACS Nano}\ }\textbf {\bibinfo {volume} {5}},\
  \bibinfo {pages} {7893} (\bibinfo {year} {2011})}\BibitemShut {NoStop}%
\bibitem [{\citenamefont {Kleemann}\ \emph {et~al.}(2017)\citenamefont
  {Kleemann}, \citenamefont {Chikkaraddy}, \citenamefont {Alexeev},
  \citenamefont {Kos}, \citenamefont {Carnegie}, \citenamefont {Deacon},
  \citenamefont {de~Pury}, \citenamefont {Gro{\ss}e}, \citenamefont {de~Nijs},
  \citenamefont {Mertens}, \citenamefont {Tartakovskii},\ and\ \citenamefont
  {Baumberg}}]{baumberggruppe2017plasmo}%
  \BibitemOpen
  \bibfield  {author} {\bibinfo {author} {\bibfnamefont {M.-E.}\ \bibnamefont
  {Kleemann}}, \bibinfo {author} {\bibfnamefont {R.}~\bibnamefont
  {Chikkaraddy}}, \bibinfo {author} {\bibfnamefont {E.~M.}\ \bibnamefont
  {Alexeev}}, \bibinfo {author} {\bibfnamefont {D.}~\bibnamefont {Kos}},
  \bibinfo {author} {\bibfnamefont {C.}~\bibnamefont {Carnegie}}, \bibinfo
  {author} {\bibfnamefont {W.}~\bibnamefont {Deacon}}, \bibinfo {author}
  {\bibfnamefont {A.~C.}\ \bibnamefont {de~Pury}}, \bibinfo {author}
  {\bibfnamefont {C.}~\bibnamefont {Gro{\ss}e}}, \bibinfo {author}
  {\bibfnamefont {B.}~\bibnamefont {de~Nijs}}, \bibinfo {author} {\bibfnamefont
  {J.}~\bibnamefont {Mertens}}, \bibinfo {author} {\bibfnamefont {A.~I.}\
  \bibnamefont {Tartakovskii}}, \ and\ \bibinfo {author} {\bibfnamefont
  {J.~J.}\ \bibnamefont {Baumberg}},\ }\href@noop {} {\bibfield  {journal}
  {\bibinfo  {journal} {Nature Communications}\ }\textbf {\bibinfo {volume}
  {8}},\ \bibinfo {pages} {1296} (\bibinfo {year} {2017})}\BibitemShut
  {NoStop}%
\bibitem [{\citenamefont {Geisler}\ \emph {et~al.}(2019)\citenamefont
  {Geisler}, \citenamefont {Cui}, \citenamefont {Wang}, \citenamefont
  {Rindzevicius}, \citenamefont {Gammelgaard}, \citenamefont {Jessen},
  \citenamefont {Gonçalves}, \citenamefont {Todisco}, \citenamefont
  {Bøggild}, \citenamefont {Boisen}, \citenamefont {Wubs}, \citenamefont
  {Mortensen}, \citenamefont {Xiao},\ and\ \citenamefont
  {Stenger}}]{stengergruppegoldaufws2_2019}%
  \BibitemOpen
  \bibfield  {author} {\bibinfo {author} {\bibfnamefont {M.}~\bibnamefont
  {Geisler}}, \bibinfo {author} {\bibfnamefont {X.}~\bibnamefont {Cui}},
  \bibinfo {author} {\bibfnamefont {J.}~\bibnamefont {Wang}}, \bibinfo {author}
  {\bibfnamefont {T.}~\bibnamefont {Rindzevicius}}, \bibinfo {author}
  {\bibfnamefont {L.}~\bibnamefont {Gammelgaard}}, \bibinfo {author}
  {\bibfnamefont {B.~S.}\ \bibnamefont {Jessen}}, \bibinfo {author}
  {\bibfnamefont {P.~A.~D.}\ \bibnamefont {Gonçalves}}, \bibinfo {author}
  {\bibfnamefont {F.}~\bibnamefont {Todisco}}, \bibinfo {author} {\bibfnamefont
  {P.}~\bibnamefont {Bøggild}}, \bibinfo {author} {\bibfnamefont
  {A.}~\bibnamefont {Boisen}}, \bibinfo {author} {\bibfnamefont
  {M.}~\bibnamefont {Wubs}}, \bibinfo {author} {\bibfnamefont {N.~A.}\
  \bibnamefont {Mortensen}}, \bibinfo {author} {\bibfnamefont {S.}~\bibnamefont
  {Xiao}}, \ and\ \bibinfo {author} {\bibfnamefont {N.}~\bibnamefont
  {Stenger}},\ }\href@noop {} {\bibfield  {journal} {\bibinfo  {journal} {ACS
  Photonics}\ }\textbf {\bibinfo {volume} {6}},\ \bibinfo {pages} {994}
  (\bibinfo {year} {2019})}\BibitemShut {NoStop}%
\bibitem [{\citenamefont {Xu}\ \emph {et~al.}(2021)\citenamefont {Xu},
  \citenamefont {Yong}, \citenamefont {He}, \citenamefont {Long}, \citenamefont
  {Cadore}, \citenamefont {Paradisanos}, \citenamefont {Ott}, \citenamefont
  {Soavi}, \citenamefont {Tongay}, \citenamefont {Cerullo}, \citenamefont
  {Ferrari}, \citenamefont {Prezhdo},\ and\ \citenamefont
  {Loh}}]{zhi-heng2021plasmo}%
  \BibitemOpen
  \bibfield  {author} {\bibinfo {author} {\bibfnamefont {C.}~\bibnamefont
  {Xu}}, \bibinfo {author} {\bibfnamefont {H.~W.}\ \bibnamefont {Yong}},
  \bibinfo {author} {\bibfnamefont {J.}~\bibnamefont {He}}, \bibinfo {author}
  {\bibfnamefont {R.}~\bibnamefont {Long}}, \bibinfo {author} {\bibfnamefont
  {A.~R.}\ \bibnamefont {Cadore}}, \bibinfo {author} {\bibfnamefont
  {I.}~\bibnamefont {Paradisanos}}, \bibinfo {author} {\bibfnamefont {A.~K.}\
  \bibnamefont {Ott}}, \bibinfo {author} {\bibfnamefont {G.}~\bibnamefont
  {Soavi}}, \bibinfo {author} {\bibfnamefont {S.}~\bibnamefont {Tongay}},
  \bibinfo {author} {\bibfnamefont {G.}~\bibnamefont {Cerullo}}, \bibinfo
  {author} {\bibfnamefont {A.~C.}\ \bibnamefont {Ferrari}}, \bibinfo {author}
  {\bibfnamefont {O.~V.}\ \bibnamefont {Prezhdo}}, \ and\ \bibinfo {author}
  {\bibfnamefont {Z.-H.}\ \bibnamefont {Loh}},\ }\href@noop {} {\bibfield
  {journal} {\bibinfo  {journal} {ACS Nano}\ }\textbf {\bibinfo {volume}
  {15}},\ \bibinfo {pages} {819} (\bibinfo {year} {2021})}\BibitemShut
  {NoStop}%
\bibitem [{\citenamefont {Carlson}\ \emph {et~al.}(2021)\citenamefont
  {Carlson}, \citenamefont {Salzwedel}, \citenamefont {Selig}, \citenamefont
  {Knorr},\ and\ \citenamefont {Hughes}}]{carlson2021strong}%
  \BibitemOpen
  \bibfield  {author} {\bibinfo {author} {\bibfnamefont {C.}~\bibnamefont
  {Carlson}}, \bibinfo {author} {\bibfnamefont {R.}~\bibnamefont {Salzwedel}},
  \bibinfo {author} {\bibfnamefont {M.}~\bibnamefont {Selig}}, \bibinfo
  {author} {\bibfnamefont {A.}~\bibnamefont {Knorr}}, \ and\ \bibinfo {author}
  {\bibfnamefont {S.}~\bibnamefont {Hughes}},\ }\href@noop {} {\bibfield
  {journal} {\bibinfo  {journal} {Phys. Rev. B}\ }\textbf {\bibinfo {volume}
  {104}},\ \bibinfo {pages} {125424} (\bibinfo {year} {2021})}\BibitemShut
  {NoStop}%
\bibitem [{\citenamefont {Zhang}\ \emph {et~al.}(2015)\citenamefont {Zhang},
  \citenamefont {You}, \citenamefont {Zhao},\ and\ \citenamefont
  {Heinz}}]{zhang2015experimental}%
  \BibitemOpen
  \bibfield  {author} {\bibinfo {author} {\bibfnamefont {X.~X.}\ \bibnamefont
  {Zhang}}, \bibinfo {author} {\bibfnamefont {Y.}~\bibnamefont {You}}, \bibinfo
  {author} {\bibfnamefont {S.~Y.~F.}\ \bibnamefont {Zhao}}, \ and\ \bibinfo
  {author} {\bibfnamefont {T.~F.}\ \bibnamefont {Heinz}},\ }\href@noop {}
  {\bibfield  {journal} {\bibinfo  {journal} {Physical review letters}\
  }\textbf {\bibinfo {volume} {115}},\ \bibinfo {pages} {257403} (\bibinfo
  {year} {2015})}\BibitemShut {NoStop}%
\bibitem [{\citenamefont {Arora}\ \emph {et~al.}(2015)\citenamefont {Arora},
  \citenamefont {Koperski}, \citenamefont {Nogajewski}, \citenamefont {Marcus},
  \citenamefont {Faugeras},\ and\ \citenamefont
  {Potemski}}]{arora2015excitonic}%
  \BibitemOpen
  \bibfield  {author} {\bibinfo {author} {\bibfnamefont {A.}~\bibnamefont
  {Arora}}, \bibinfo {author} {\bibfnamefont {M.}~\bibnamefont {Koperski}},
  \bibinfo {author} {\bibfnamefont {K.}~\bibnamefont {Nogajewski}}, \bibinfo
  {author} {\bibfnamefont {J.}~\bibnamefont {Marcus}}, \bibinfo {author}
  {\bibfnamefont {C.}~\bibnamefont {Faugeras}}, \ and\ \bibinfo {author}
  {\bibfnamefont {M.}~\bibnamefont {Potemski}},\ }\href@noop {} {\bibfield
  {journal} {\bibinfo  {journal} {Nanoscale}\ }\textbf {\bibinfo {volume}
  {7}},\ \bibinfo {pages} {10421} (\bibinfo {year} {2015})}\BibitemShut
  {NoStop}%
\bibitem [{\citenamefont {Selig}\ \emph {et~al.}(2016)\citenamefont {Selig},
  \citenamefont {Bergh{\"a}user}, \citenamefont {Raja}, \citenamefont {Nagler},
  \citenamefont {Sch{\"u}ller}, \citenamefont {Heinz}, \citenamefont {Korn},
  \citenamefont {Chernikov}, \citenamefont {Malic},\ and\ \citenamefont
  {Knorr}}]{selig2016excitonic}%
  \BibitemOpen
  \bibfield  {author} {\bibinfo {author} {\bibfnamefont {M.}~\bibnamefont
  {Selig}}, \bibinfo {author} {\bibfnamefont {G.}~\bibnamefont
  {Bergh{\"a}user}}, \bibinfo {author} {\bibfnamefont {A.}~\bibnamefont
  {Raja}}, \bibinfo {author} {\bibfnamefont {P.}~\bibnamefont {Nagler}},
  \bibinfo {author} {\bibfnamefont {C.}~\bibnamefont {Sch{\"u}ller}}, \bibinfo
  {author} {\bibfnamefont {T.~F.}\ \bibnamefont {Heinz}}, \bibinfo {author}
  {\bibfnamefont {T.}~\bibnamefont {Korn}}, \bibinfo {author} {\bibfnamefont
  {A.}~\bibnamefont {Chernikov}}, \bibinfo {author} {\bibfnamefont
  {E.}~\bibnamefont {Malic}}, \ and\ \bibinfo {author} {\bibfnamefont
  {A.}~\bibnamefont {Knorr}},\ }\href@noop {} {\bibfield  {journal} {\bibinfo
  {journal} {Nature communications}\ }\textbf {\bibinfo {volume} {7}},\
  \bibinfo {pages} {1} (\bibinfo {year} {2016})}\BibitemShut {NoStop}%
\bibitem [{\citenamefont {Christiansen}\ \emph {et~al.}(2017)\citenamefont
  {Christiansen}, \citenamefont {Selig}, \citenamefont {Bergh\"auser},
  \citenamefont {Schmidt}, \citenamefont {Niehues}, \citenamefont {Schneider},
  \citenamefont {Arora}, \citenamefont {de~Vasconcellos}, \citenamefont
  {Bratschitsch}, \citenamefont {Malic},\ and\ \citenamefont
  {Knorr}}]{christiansen2017phononen}%
  \BibitemOpen
  \bibfield  {author} {\bibinfo {author} {\bibfnamefont {D.}~\bibnamefont
  {Christiansen}}, \bibinfo {author} {\bibfnamefont {M.}~\bibnamefont {Selig}},
  \bibinfo {author} {\bibfnamefont {G.}~\bibnamefont {Bergh\"auser}}, \bibinfo
  {author} {\bibfnamefont {R.}~\bibnamefont {Schmidt}}, \bibinfo {author}
  {\bibfnamefont {I.}~\bibnamefont {Niehues}}, \bibinfo {author} {\bibfnamefont
  {R.}~\bibnamefont {Schneider}}, \bibinfo {author} {\bibfnamefont
  {A.}~\bibnamefont {Arora}}, \bibinfo {author} {\bibfnamefont {S.~M.}\
  \bibnamefont {de~Vasconcellos}}, \bibinfo {author} {\bibfnamefont
  {R.}~\bibnamefont {Bratschitsch}}, \bibinfo {author} {\bibfnamefont
  {E.}~\bibnamefont {Malic}}, \ and\ \bibinfo {author} {\bibfnamefont
  {A.}~\bibnamefont {Knorr}},\ }\href@noop {} {\bibfield  {journal} {\bibinfo
  {journal} {Physical review letters}\ }\textbf {\bibinfo {volume} {119}},\
  \bibinfo {pages} {187402} (\bibinfo {year} {2017})}\BibitemShut {NoStop}%
\bibitem [{\citenamefont {Selig}\ \emph {et~al.}(2018)\citenamefont {Selig},
  \citenamefont {Bergh{\"a}user}, \citenamefont {Richter}, \citenamefont
  {Bratschitsch}, \citenamefont {Knorr},\ and\ \citenamefont
  {Malic}}]{selig2018dark}%
  \BibitemOpen
  \bibfield  {author} {\bibinfo {author} {\bibfnamefont {M.}~\bibnamefont
  {Selig}}, \bibinfo {author} {\bibfnamefont {G.}~\bibnamefont
  {Bergh{\"a}user}}, \bibinfo {author} {\bibfnamefont {M.}~\bibnamefont
  {Richter}}, \bibinfo {author} {\bibfnamefont {R.}~\bibnamefont
  {Bratschitsch}}, \bibinfo {author} {\bibfnamefont {A.}~\bibnamefont {Knorr}},
  \ and\ \bibinfo {author} {\bibfnamefont {E.}~\bibnamefont {Malic}},\
  }\href@noop {} {\bibfield  {journal} {\bibinfo  {journal} {2D Materials}\
  }\textbf {\bibinfo {volume} {5}},\ \bibinfo {pages} {035017} (\bibinfo {year}
  {2018})}\BibitemShut {NoStop}%
\bibitem [{\citenamefont {Selig}\ \emph
  {et~al.}(2019{\natexlab{a}})\citenamefont {Selig}, \citenamefont {Katsch},
  \citenamefont {Schmidt}, \citenamefont {Michaelis~de Vasconcellos},
  \citenamefont {Bratschitsch}, \citenamefont {Malic},\ and\ \citenamefont
  {Knorr}}]{selig2019PRR}%
  \BibitemOpen
  \bibfield  {author} {\bibinfo {author} {\bibfnamefont {M.}~\bibnamefont
  {Selig}}, \bibinfo {author} {\bibfnamefont {F.}~\bibnamefont {Katsch}},
  \bibinfo {author} {\bibfnamefont {R.}~\bibnamefont {Schmidt}}, \bibinfo
  {author} {\bibfnamefont {S.}~\bibnamefont {Michaelis~de Vasconcellos}},
  \bibinfo {author} {\bibfnamefont {R.}~\bibnamefont {Bratschitsch}}, \bibinfo
  {author} {\bibfnamefont {E.}~\bibnamefont {Malic}}, \ and\ \bibinfo {author}
  {\bibfnamefont {A.}~\bibnamefont {Knorr}},\ }\href@noop {} {\bibfield
  {journal} {\bibinfo  {journal} {Phys. Rev. Research}\ }\textbf {\bibinfo
  {volume} {1}},\ \bibinfo {pages} {022007} (\bibinfo {year}
  {2019}{\natexlab{a}})}\BibitemShut {NoStop}%
\bibitem [{\citenamefont {Christiansen}\ \emph {et~al.}(2021)\citenamefont
  {Christiansen}, \citenamefont {Selig}, \citenamefont {Rossi},\ and\
  \citenamefont {Knorr}}]{christiansen2021strong}%
  \BibitemOpen
  \bibfield  {author} {\bibinfo {author} {\bibfnamefont {D.}~\bibnamefont
  {Christiansen}}, \bibinfo {author} {\bibfnamefont {M.}~\bibnamefont {Selig}},
  \bibinfo {author} {\bibfnamefont {M.}~\bibnamefont {Rossi}}, \ and\ \bibinfo
  {author} {\bibfnamefont {A.}~\bibnamefont {Knorr}},\ }\href {\doibase
  10.48550/arxiv.2112.03135} {\  (\bibinfo {year} {2021}),\
  10.48550/arxiv.2112.03135}\BibitemShut {NoStop}%
\bibitem [{\citenamefont {Jariwala}\ \emph {et~al.}(2017)\citenamefont
  {Jariwala}, \citenamefont {Marks},\ and\ \citenamefont
  {Hersam}}]{hersamgroup2017natmatreview}%
  \BibitemOpen
  \bibfield  {author} {\bibinfo {author} {\bibfnamefont {D.}~\bibnamefont
  {Jariwala}}, \bibinfo {author} {\bibfnamefont {T.~J.}\ \bibnamefont {Marks}},
  \ and\ \bibinfo {author} {\bibfnamefont {M.~C.}\ \bibnamefont {Hersam}},\
  }\href@noop {} {\bibfield  {journal} {\bibinfo  {journal} {Nature Materials}\
  }\textbf {\bibinfo {volume} {16}},\ \bibinfo {pages} {170} (\bibinfo {year}
  {2017})}\BibitemShut {NoStop}%
\bibitem [{\citenamefont {Padgaonkar}\ \emph {et~al.}(2020)\citenamefont
  {Padgaonkar}, \citenamefont {Olding}, \citenamefont {Lauhon}, \citenamefont
  {Hersam},\ and\ \citenamefont {Weiss}}]{weissgruppe2020review0d2d}%
  \BibitemOpen
  \bibfield  {author} {\bibinfo {author} {\bibfnamefont {S.}~\bibnamefont
  {Padgaonkar}}, \bibinfo {author} {\bibfnamefont {J.~N.}\ \bibnamefont
  {Olding}}, \bibinfo {author} {\bibfnamefont {L.~J.}\ \bibnamefont {Lauhon}},
  \bibinfo {author} {\bibfnamefont {M.~C.}\ \bibnamefont {Hersam}}, \ and\
  \bibinfo {author} {\bibfnamefont {E.~A.}\ \bibnamefont {Weiss}},\ }\href@noop
  {} {\bibfield  {journal} {\bibinfo  {journal} {Accounts of Chemical
  Research}\ }\textbf {\bibinfo {volume} {53}},\ \bibinfo {pages} {763}
  (\bibinfo {year} {2020})}\BibitemShut {NoStop}%
\bibitem [{\citenamefont {Förster}(1948)}]{foersteroriginal1948}%
  \BibitemOpen
  \bibfield  {author} {\bibinfo {author} {\bibfnamefont {T.}~\bibnamefont
  {Förster}},\ }\href@noop {} {\bibfield  {journal} {\bibinfo  {journal}
  {Annalen der Physik}\ }\textbf {\bibinfo {volume} {437}},\ \bibinfo {pages}
  {55} (\bibinfo {year} {1948})}\BibitemShut {NoStop}%
\bibitem [{\citenamefont {Froehlicher}\ \emph {et~al.}(2018)\citenamefont
  {Froehlicher}, \citenamefont {Lorchat},\ and\ \citenamefont
  {Berciaud}}]{froehlicher2018}%
  \BibitemOpen
  \bibfield  {author} {\bibinfo {author} {\bibfnamefont {G.}~\bibnamefont
  {Froehlicher}}, \bibinfo {author} {\bibfnamefont {E.}~\bibnamefont
  {Lorchat}}, \ and\ \bibinfo {author} {\bibfnamefont {S.}~\bibnamefont
  {Berciaud}},\ }\href {\doibase 10.1103/PhysRevX.8.011007} {\bibfield
  {journal} {\bibinfo  {journal} {Phys. Rev. X}\ }\textbf {\bibinfo {volume}
  {8}},\ \bibinfo {pages} {011007} (\bibinfo {year} {2018})}\BibitemShut
  {NoStop}%
\bibitem [{\citenamefont {Demir}\ \emph {et~al.}(2011)\citenamefont {Demir},
  \citenamefont {Nizamoglu}, \citenamefont {Erdem}, \citenamefont {Mutlugun},
  \citenamefont {Gaponik},\ and\ \citenamefont
  {Eychmüller}}]{DEMIR2011FRETQDLED}%
  \BibitemOpen
  \bibfield  {author} {\bibinfo {author} {\bibfnamefont {H.~V.}\ \bibnamefont
  {Demir}}, \bibinfo {author} {\bibfnamefont {S.}~\bibnamefont {Nizamoglu}},
  \bibinfo {author} {\bibfnamefont {T.}~\bibnamefont {Erdem}}, \bibinfo
  {author} {\bibfnamefont {E.}~\bibnamefont {Mutlugun}}, \bibinfo {author}
  {\bibfnamefont {N.}~\bibnamefont {Gaponik}}, \ and\ \bibinfo {author}
  {\bibfnamefont {A.}~\bibnamefont {Eychmüller}},\ }\href@noop {} {\bibfield
  {journal} {\bibinfo  {journal} {Nano Today}\ }\textbf {\bibinfo {volume}
  {6}},\ \bibinfo {pages} {632} (\bibinfo {year} {2011})}\BibitemShut {NoStop}%
\bibitem [{\citenamefont {Dagher}\ \emph {et~al.}(2018)\citenamefont {Dagher},
  \citenamefont {Kleinman}, \citenamefont {Ng},\ and\ \citenamefont
  {Juncker}}]{junkergruppe2018fret}%
  \BibitemOpen
  \bibfield  {author} {\bibinfo {author} {\bibfnamefont {M.}~\bibnamefont
  {Dagher}}, \bibinfo {author} {\bibfnamefont {M.}~\bibnamefont {Kleinman}},
  \bibinfo {author} {\bibfnamefont {A.}~\bibnamefont {Ng}}, \ and\ \bibinfo
  {author} {\bibfnamefont {D.}~\bibnamefont {Juncker}},\ }\href@noop {}
  {\bibfield  {journal} {\bibinfo  {journal} {Nature Nanotechnology}\ }\textbf
  {\bibinfo {volume} {13}},\ \bibinfo {pages} {925} (\bibinfo {year}
  {2018})}\BibitemShut {NoStop}%
\bibitem [{\citenamefont {Swathi}\ and\ \citenamefont
  {Sebastian}(2009)}]{Swathi2009}%
  \BibitemOpen
  \bibfield  {author} {\bibinfo {author} {\bibfnamefont {R.~S.}\ \bibnamefont
  {Swathi}}\ and\ \bibinfo {author} {\bibfnamefont {K.~L.}\ \bibnamefont
  {Sebastian}},\ }\href@noop {} {\bibfield  {journal} {\bibinfo  {journal}
  {Journal of Chemical Sciences}\ }\textbf {\bibinfo {volume} {121}},\ \bibinfo
  {pages} {777} (\bibinfo {year} {2009})}\BibitemShut {NoStop}%
\bibitem [{\citenamefont {Malic}\ \emph {et~al.}(2014)\citenamefont {Malic},
  \citenamefont {Appel}, \citenamefont {Hofmann},\ and\ \citenamefont
  {Rubio}}]{Malic2014}%
  \BibitemOpen
  \bibfield  {author} {\bibinfo {author} {\bibfnamefont {E.}~\bibnamefont
  {Malic}}, \bibinfo {author} {\bibfnamefont {H.}~\bibnamefont {Appel}},
  \bibinfo {author} {\bibfnamefont {O.~T.}\ \bibnamefont {Hofmann}}, \ and\
  \bibinfo {author} {\bibfnamefont {A.}~\bibnamefont {Rubio}},\ }\href@noop {}
  {\bibfield  {journal} {\bibinfo  {journal} {The Journal of Physical Chemistry
  C}\ }\textbf {\bibinfo {volume} {118}},\ \bibinfo {pages} {9283} (\bibinfo
  {year} {2014})}\BibitemShut {NoStop}%
\bibitem [{\citenamefont {Federspiel}\ \emph {et~al.}(2015)\citenamefont
  {Federspiel}, \citenamefont {Froehlicher}, \citenamefont {Nasilowski},
  \citenamefont {Pedetti}, \citenamefont {Mahmood}, \citenamefont {Doudin},
  \citenamefont {Park}, \citenamefont {Lee}, \citenamefont {Halley},
  \citenamefont {Dubertret}, \citenamefont {Gilliot},\ and\ \citenamefont
  {Berciaud}}]{berciaudgruppe2015distance}%
  \BibitemOpen
  \bibfield  {author} {\bibinfo {author} {\bibfnamefont {F.}~\bibnamefont
  {Federspiel}}, \bibinfo {author} {\bibfnamefont {G.}~\bibnamefont
  {Froehlicher}}, \bibinfo {author} {\bibfnamefont {M.}~\bibnamefont
  {Nasilowski}}, \bibinfo {author} {\bibfnamefont {S.}~\bibnamefont {Pedetti}},
  \bibinfo {author} {\bibfnamefont {A.}~\bibnamefont {Mahmood}}, \bibinfo
  {author} {\bibfnamefont {B.}~\bibnamefont {Doudin}}, \bibinfo {author}
  {\bibfnamefont {S.}~\bibnamefont {Park}}, \bibinfo {author} {\bibfnamefont
  {J.-O.}\ \bibnamefont {Lee}}, \bibinfo {author} {\bibfnamefont
  {D.}~\bibnamefont {Halley}}, \bibinfo {author} {\bibfnamefont
  {B.}~\bibnamefont {Dubertret}}, \bibinfo {author} {\bibfnamefont
  {P.}~\bibnamefont {Gilliot}}, \ and\ \bibinfo {author} {\bibfnamefont
  {S.}~\bibnamefont {Berciaud}},\ }\href@noop {} {\bibfield  {journal}
  {\bibinfo  {journal} {Nano Letters}\ }\textbf {\bibinfo {volume} {15}},\
  \bibinfo {pages} {1252} (\bibinfo {year} {2015})}\BibitemShut {NoStop}%
\bibitem [{\citenamefont {Selig}\ \emph
  {et~al.}(2019{\natexlab{b}})\citenamefont {Selig}, \citenamefont {Malic},
  \citenamefont {Ahn}, \citenamefont {Koch},\ and\ \citenamefont
  {Knorr}}]{PhysRevB.99.035420}%
  \BibitemOpen
  \bibfield  {author} {\bibinfo {author} {\bibfnamefont {M.}~\bibnamefont
  {Selig}}, \bibinfo {author} {\bibfnamefont {E.}~\bibnamefont {Malic}},
  \bibinfo {author} {\bibfnamefont {K.~J.}\ \bibnamefont {Ahn}}, \bibinfo
  {author} {\bibfnamefont {N.}~\bibnamefont {Koch}}, \ and\ \bibinfo {author}
  {\bibfnamefont {A.}~\bibnamefont {Knorr}},\ }\href {\doibase
  10.1103/PhysRevB.99.035420} {\bibfield  {journal} {\bibinfo  {journal} {Phys.
  Rev. B}\ }\textbf {\bibinfo {volume} {99}},\ \bibinfo {pages} {035420}
  (\bibinfo {year} {2019}{\natexlab{b}})}\BibitemShut {NoStop}%
\bibitem [{\citenamefont {Mazzamuto}\ \emph {et~al.}(2014)\citenamefont
  {Mazzamuto}, \citenamefont {Tabani}, \citenamefont {Pazzagli}, \citenamefont
  {Rizvi}, \citenamefont {Reserbat-Plantey}, \citenamefont {Sch{\"a}dler},
  \citenamefont {Navickaite}, \citenamefont {Gaudreau}, \citenamefont
  {Cataliotti}, \citenamefont {Koppens} \emph {et~al.}}]{mazzamuto2014single}%
  \BibitemOpen
  \bibfield  {author} {\bibinfo {author} {\bibfnamefont {G.}~\bibnamefont
  {Mazzamuto}}, \bibinfo {author} {\bibfnamefont {A.}~\bibnamefont {Tabani}},
  \bibinfo {author} {\bibfnamefont {S.}~\bibnamefont {Pazzagli}}, \bibinfo
  {author} {\bibfnamefont {S.}~\bibnamefont {Rizvi}}, \bibinfo {author}
  {\bibfnamefont {A.}~\bibnamefont {Reserbat-Plantey}}, \bibinfo {author}
  {\bibfnamefont {K.}~\bibnamefont {Sch{\"a}dler}}, \bibinfo {author}
  {\bibfnamefont {G.}~\bibnamefont {Navickaite}}, \bibinfo {author}
  {\bibfnamefont {L.}~\bibnamefont {Gaudreau}}, \bibinfo {author}
  {\bibfnamefont {F.}~\bibnamefont {Cataliotti}}, \bibinfo {author}
  {\bibfnamefont {F.}~\bibnamefont {Koppens}},  \emph {et~al.},\ }\href@noop {}
  {\bibfield  {journal} {\bibinfo  {journal} {New Journal of Physics}\ }\textbf
  {\bibinfo {volume} {16}},\ \bibinfo {pages} {113007} (\bibinfo {year}
  {2014})}\BibitemShut {NoStop}%
\bibitem [{\citenamefont {Malic}\ \emph {et~al.}(2018)\citenamefont {Malic},
  \citenamefont {Selig}, \citenamefont {Feierabend}, \citenamefont {Brem},
  \citenamefont {Christiansen}, \citenamefont {Wendler}, \citenamefont
  {Knorr},\ and\ \citenamefont {Bergh\"auser}}]{Malic2018Dark}%
  \BibitemOpen
  \bibfield  {author} {\bibinfo {author} {\bibfnamefont {E.}~\bibnamefont
  {Malic}}, \bibinfo {author} {\bibfnamefont {M.}~\bibnamefont {Selig}},
  \bibinfo {author} {\bibfnamefont {M.}~\bibnamefont {Feierabend}}, \bibinfo
  {author} {\bibfnamefont {S.}~\bibnamefont {Brem}}, \bibinfo {author}
  {\bibfnamefont {D.}~\bibnamefont {Christiansen}}, \bibinfo {author}
  {\bibfnamefont {F.}~\bibnamefont {Wendler}}, \bibinfo {author} {\bibfnamefont
  {A.}~\bibnamefont {Knorr}}, \ and\ \bibinfo {author} {\bibfnamefont
  {G.}~\bibnamefont {Bergh\"auser}},\ }\href@noop {} {\bibfield  {journal}
  {\bibinfo  {journal} {Phys. Rev. Materials}\ }\textbf {\bibinfo {volume}
  {2}},\ \bibinfo {pages} {014002} (\bibinfo {year} {2018})}\BibitemShut
  {NoStop}%
\bibitem [{\citenamefont {Lindlau}\ \emph {et~al.}(2018)\citenamefont
  {Lindlau}, \citenamefont {Selig}, \citenamefont {Neumann}, \citenamefont
  {Colombier}, \citenamefont {F{\"o}rste}, \citenamefont {Funk}, \citenamefont
  {F{\"o}rg}, \citenamefont {Kim}, \citenamefont {Bergh{\"a}user},
  \citenamefont {Taniguchi} \emph {et~al.}}]{lindlau2018role}%
  \BibitemOpen
  \bibfield  {author} {\bibinfo {author} {\bibfnamefont {J.}~\bibnamefont
  {Lindlau}}, \bibinfo {author} {\bibfnamefont {M.}~\bibnamefont {Selig}},
  \bibinfo {author} {\bibfnamefont {A.}~\bibnamefont {Neumann}}, \bibinfo
  {author} {\bibfnamefont {L.}~\bibnamefont {Colombier}}, \bibinfo {author}
  {\bibfnamefont {J.}~\bibnamefont {F{\"o}rste}}, \bibinfo {author}
  {\bibfnamefont {V.}~\bibnamefont {Funk}}, \bibinfo {author} {\bibfnamefont
  {M.}~\bibnamefont {F{\"o}rg}}, \bibinfo {author} {\bibfnamefont
  {J.}~\bibnamefont {Kim}}, \bibinfo {author} {\bibfnamefont {G.}~\bibnamefont
  {Bergh{\"a}user}}, \bibinfo {author} {\bibfnamefont {T.}~\bibnamefont
  {Taniguchi}},  \emph {et~al.},\ }\href@noop {} {\bibfield  {journal}
  {\bibinfo  {journal} {Nature communications}\ }\textbf {\bibinfo {volume}
  {9}},\ \bibinfo {pages} {1} (\bibinfo {year} {2018})}\BibitemShut {NoStop}%
\bibitem [{\citenamefont {Brem}\ \emph {et~al.}(2020)\citenamefont {Brem},
  \citenamefont {Ekman}, \citenamefont {Christiansen}, \citenamefont {Katsch},
  \citenamefont {Selig}, \citenamefont {Robert}, \citenamefont {Marie},
  \citenamefont {Urbaszek}, \citenamefont {Knorr},\ and\ \citenamefont
  {Malic}}]{brem2020phonon}%
  \BibitemOpen
  \bibfield  {author} {\bibinfo {author} {\bibfnamefont {S.}~\bibnamefont
  {Brem}}, \bibinfo {author} {\bibfnamefont {A.}~\bibnamefont {Ekman}},
  \bibinfo {author} {\bibfnamefont {D.}~\bibnamefont {Christiansen}}, \bibinfo
  {author} {\bibfnamefont {F.}~\bibnamefont {Katsch}}, \bibinfo {author}
  {\bibfnamefont {M.}~\bibnamefont {Selig}}, \bibinfo {author} {\bibfnamefont
  {C.}~\bibnamefont {Robert}}, \bibinfo {author} {\bibfnamefont
  {X.}~\bibnamefont {Marie}}, \bibinfo {author} {\bibfnamefont
  {B.}~\bibnamefont {Urbaszek}}, \bibinfo {author} {\bibfnamefont
  {A.}~\bibnamefont {Knorr}}, \ and\ \bibinfo {author} {\bibfnamefont
  {E.}~\bibnamefont {Malic}},\ }\href@noop {} {\bibfield  {journal} {\bibinfo
  {journal} {Nano letters}\ }\textbf {\bibinfo {volume} {20}},\ \bibinfo
  {pages} {2849} (\bibinfo {year} {2020})}\BibitemShut {NoStop}%
\bibitem [{\citenamefont {Jarvinen}\ \emph {et~al.}(2013)\citenamefont
  {Jarvinen}, \citenamefont {H{\"a}m{\"a}l{\"a}inen}, \citenamefont {Banerjee},
  \citenamefont {Hakkinen}, \citenamefont {Ijas}, \citenamefont {Harju},\ and\
  \citenamefont {Liljeroth}}]{jarvinen2013molecular}%
  \BibitemOpen
  \bibfield  {author} {\bibinfo {author} {\bibfnamefont {P.}~\bibnamefont
  {Jarvinen}}, \bibinfo {author} {\bibfnamefont {S.~K.}\ \bibnamefont
  {H{\"a}m{\"a}l{\"a}inen}}, \bibinfo {author} {\bibfnamefont {K.}~\bibnamefont
  {Banerjee}}, \bibinfo {author} {\bibfnamefont {P.}~\bibnamefont {Hakkinen}},
  \bibinfo {author} {\bibfnamefont {M.}~\bibnamefont {Ijas}}, \bibinfo {author}
  {\bibfnamefont {A.}~\bibnamefont {Harju}}, \ and\ \bibinfo {author}
  {\bibfnamefont {P.}~\bibnamefont {Liljeroth}},\ }\href@noop {} {\bibfield
  {journal} {\bibinfo  {journal} {Nano letters}\ }\textbf {\bibinfo {volume}
  {13}},\ \bibinfo {pages} {3199} (\bibinfo {year} {2013})}\BibitemShut
  {NoStop}%
\bibitem [{\citenamefont {Tsai}\ \emph {et~al.}(2015)\citenamefont {Tsai},
  \citenamefont {Omrani}, \citenamefont {Coh}, \citenamefont {Oh},
  \citenamefont {Wickenburg}, \citenamefont {Son}, \citenamefont {Wong},
  \citenamefont {Riss}, \citenamefont {Jung}, \citenamefont {Nguyen},
  \citenamefont {Rodgers}, \citenamefont {Aikawa}, \citenamefont {Taniguchi},
  \citenamefont {Watanabe}, \citenamefont {Zettl}, \citenamefont {Louie},
  \citenamefont {Lu}, \citenamefont {Cohen},\ and\ \citenamefont
  {Crommie}}]{Tsai2015dipolealignment}%
  \BibitemOpen
  \bibfield  {author} {\bibinfo {author} {\bibfnamefont {H.-Z.}\ \bibnamefont
  {Tsai}}, \bibinfo {author} {\bibfnamefont {A.~A.}\ \bibnamefont {Omrani}},
  \bibinfo {author} {\bibfnamefont {S.}~\bibnamefont {Coh}}, \bibinfo {author}
  {\bibfnamefont {H.}~\bibnamefont {Oh}}, \bibinfo {author} {\bibfnamefont
  {S.}~\bibnamefont {Wickenburg}}, \bibinfo {author} {\bibfnamefont {Y.-W.}\
  \bibnamefont {Son}}, \bibinfo {author} {\bibfnamefont {D.}~\bibnamefont
  {Wong}}, \bibinfo {author} {\bibfnamefont {A.}~\bibnamefont {Riss}}, \bibinfo
  {author} {\bibfnamefont {H.~S.}\ \bibnamefont {Jung}}, \bibinfo {author}
  {\bibfnamefont {G.~D.}\ \bibnamefont {Nguyen}}, \bibinfo {author}
  {\bibfnamefont {G.~F.}\ \bibnamefont {Rodgers}}, \bibinfo {author}
  {\bibfnamefont {A.~S.}\ \bibnamefont {Aikawa}}, \bibinfo {author}
  {\bibfnamefont {T.}~\bibnamefont {Taniguchi}}, \bibinfo {author}
  {\bibfnamefont {K.}~\bibnamefont {Watanabe}}, \bibinfo {author}
  {\bibfnamefont {A.}~\bibnamefont {Zettl}}, \bibinfo {author} {\bibfnamefont
  {S.~G.}\ \bibnamefont {Louie}}, \bibinfo {author} {\bibfnamefont
  {J.}~\bibnamefont {Lu}}, \bibinfo {author} {\bibfnamefont {M.~L.}\
  \bibnamefont {Cohen}}, \ and\ \bibinfo {author} {\bibfnamefont {M.~F.}\
  \bibnamefont {Crommie}},\ }\href@noop {} {\bibfield  {journal} {\bibinfo
  {journal} {ACS Nano}\ }\textbf {\bibinfo {volume} {9}},\ \bibinfo {pages}
  {12168} (\bibinfo {year} {2015})}\BibitemShut {NoStop}%
\bibitem [{\citenamefont {Zhao}\ \emph
  {et~al.}(2019{\natexlab{b}})\citenamefont {Zhao}, \citenamefont {Zhao},
  \citenamefont {Song}, \citenamefont {Zhou}, \citenamefont {Lv}, \citenamefont
  {Tao}, \citenamefont {Feng}, \citenamefont {Song}, \citenamefont {Ma},
  \citenamefont {Zhang} \emph {et~al.}}]{zhao2019strong}%
  \BibitemOpen
  \bibfield  {author} {\bibinfo {author} {\bibfnamefont {H.}~\bibnamefont
  {Zhao}}, \bibinfo {author} {\bibfnamefont {Y.}~\bibnamefont {Zhao}}, \bibinfo
  {author} {\bibfnamefont {Y.}~\bibnamefont {Song}}, \bibinfo {author}
  {\bibfnamefont {M.}~\bibnamefont {Zhou}}, \bibinfo {author} {\bibfnamefont
  {W.}~\bibnamefont {Lv}}, \bibinfo {author} {\bibfnamefont {L.}~\bibnamefont
  {Tao}}, \bibinfo {author} {\bibfnamefont {Y.}~\bibnamefont {Feng}}, \bibinfo
  {author} {\bibfnamefont {B.}~\bibnamefont {Song}}, \bibinfo {author}
  {\bibfnamefont {Y.}~\bibnamefont {Ma}}, \bibinfo {author} {\bibfnamefont
  {J.}~\bibnamefont {Zhang}},  \emph {et~al.},\ }\href@noop {} {\bibfield
  {journal} {\bibinfo  {journal} {Nature communications}\ }\textbf {\bibinfo
  {volume} {10}},\ \bibinfo {pages} {1} (\bibinfo {year}
  {2019}{\natexlab{b}})}\BibitemShut {NoStop}%
\bibitem [{\citenamefont {Bondarev}\ \emph {et~al.}(2004)\citenamefont
  {Bondarev}, \citenamefont {Knyukshto}, \citenamefont {Stepuro}, \citenamefont
  {Stupak},\ and\ \citenamefont {Turban}}]{bondarev2004fluorescence}%
  \BibitemOpen
  \bibfield  {author} {\bibinfo {author} {\bibfnamefont {S.}~\bibnamefont
  {Bondarev}}, \bibinfo {author} {\bibfnamefont {V.}~\bibnamefont {Knyukshto}},
  \bibinfo {author} {\bibfnamefont {V.}~\bibnamefont {Stepuro}}, \bibinfo
  {author} {\bibfnamefont {A.}~\bibnamefont {Stupak}}, \ and\ \bibinfo {author}
  {\bibfnamefont {A.}~\bibnamefont {Turban}},\ }\href@noop {} {\bibfield
  {journal} {\bibinfo  {journal} {Journal of Applied Spectroscopy}\ }\textbf
  {\bibinfo {volume} {71}},\ \bibinfo {pages} {194} (\bibinfo {year}
  {2004})}\BibitemShut {NoStop}%
\bibitem [{\citenamefont {Katsch}\ \emph {et~al.}(2018)\citenamefont {Katsch},
  \citenamefont {Selig}, \citenamefont {Carmele},\ and\ \citenamefont
  {Knorr}}]{katsch2018theory}%
  \BibitemOpen
  \bibfield  {author} {\bibinfo {author} {\bibfnamefont {F.}~\bibnamefont
  {Katsch}}, \bibinfo {author} {\bibfnamefont {M.}~\bibnamefont {Selig}},
  \bibinfo {author} {\bibfnamefont {A.}~\bibnamefont {Carmele}}, \ and\
  \bibinfo {author} {\bibfnamefont {A.}~\bibnamefont {Knorr}},\ }\href@noop {}
  {\bibfield  {journal} {\bibinfo  {journal} {physica status solidi (b)}\
  }\textbf {\bibinfo {volume} {255}},\ \bibinfo {pages} {1800185} (\bibinfo
  {year} {2018})}\BibitemShut {NoStop}%
\bibitem [{\citenamefont {Mak}\ \emph {et~al.}(2010)\citenamefont {Mak},
  \citenamefont {Lee}, \citenamefont {Hone}, \citenamefont {Shan},\ and\
  \citenamefont {Heinz}}]{mak2010atomically}%
  \BibitemOpen
  \bibfield  {author} {\bibinfo {author} {\bibfnamefont {K.~F.}\ \bibnamefont
  {Mak}}, \bibinfo {author} {\bibfnamefont {C.}~\bibnamefont {Lee}}, \bibinfo
  {author} {\bibfnamefont {J.}~\bibnamefont {Hone}}, \bibinfo {author}
  {\bibfnamefont {J.}~\bibnamefont {Shan}}, \ and\ \bibinfo {author}
  {\bibfnamefont {T.~F.}\ \bibnamefont {Heinz}},\ }\href@noop {} {\bibfield
  {journal} {\bibinfo  {journal} {Physical review letters}\ }\textbf {\bibinfo
  {volume} {105}},\ \bibinfo {pages} {136805} (\bibinfo {year}
  {2010})}\BibitemShut {NoStop}%
\bibitem [{\citenamefont {Splendiani}\ \emph {et~al.}(2010)\citenamefont
  {Splendiani}, \citenamefont {Sun}, \citenamefont {Zhang}, \citenamefont {Li},
  \citenamefont {Kim}, \citenamefont {Chim}, \citenamefont {Galli},\ and\
  \citenamefont {Wang}}]{splendiani2010emerging}%
  \BibitemOpen
  \bibfield  {author} {\bibinfo {author} {\bibfnamefont {A.}~\bibnamefont
  {Splendiani}}, \bibinfo {author} {\bibfnamefont {L.}~\bibnamefont {Sun}},
  \bibinfo {author} {\bibfnamefont {Y.}~\bibnamefont {Zhang}}, \bibinfo
  {author} {\bibfnamefont {T.}~\bibnamefont {Li}}, \bibinfo {author}
  {\bibfnamefont {J.}~\bibnamefont {Kim}}, \bibinfo {author} {\bibfnamefont
  {C.-Y.}\ \bibnamefont {Chim}}, \bibinfo {author} {\bibfnamefont
  {G.}~\bibnamefont {Galli}}, \ and\ \bibinfo {author} {\bibfnamefont
  {F.}~\bibnamefont {Wang}},\ }\href@noop {} {\bibfield  {journal} {\bibinfo
  {journal} {Nano letters}\ }\textbf {\bibinfo {volume} {10}},\ \bibinfo
  {pages} {1271} (\bibinfo {year} {2010})}\BibitemShut {NoStop}%
\bibitem [{\citenamefont {Wurstbauer}\ \emph {et~al.}(2017)\citenamefont
  {Wurstbauer}, \citenamefont {Miller}, \citenamefont {Parzinger},\ and\
  \citenamefont {Holleitner}}]{wurstbauer2017light}%
  \BibitemOpen
  \bibfield  {author} {\bibinfo {author} {\bibfnamefont {U.}~\bibnamefont
  {Wurstbauer}}, \bibinfo {author} {\bibfnamefont {B.}~\bibnamefont {Miller}},
  \bibinfo {author} {\bibfnamefont {E.}~\bibnamefont {Parzinger}}, \ and\
  \bibinfo {author} {\bibfnamefont {A.~W.}\ \bibnamefont {Holleitner}},\
  }\href@noop {} {\bibfield  {journal} {\bibinfo  {journal} {Journal of Physics
  D: Applied Physics}\ }\textbf {\bibinfo {volume} {50}},\ \bibinfo {pages}
  {173001} (\bibinfo {year} {2017})}\BibitemShut {NoStop}%
\bibitem [{\citenamefont {Gerber}\ \emph {et~al.}(2019)\citenamefont {Gerber},
  \citenamefont {Courtade}, \citenamefont {Shree}, \citenamefont {Robert},
  \citenamefont {Taniguchi}, \citenamefont {Watanabe}, \citenamefont
  {Balocchi}, \citenamefont {Renucci}, \citenamefont {Lagarde}, \citenamefont
  {Marie} \emph {et~al.}}]{gerber2019interlayer}%
  \BibitemOpen
  \bibfield  {author} {\bibinfo {author} {\bibfnamefont {I.~C.}\ \bibnamefont
  {Gerber}}, \bibinfo {author} {\bibfnamefont {E.}~\bibnamefont {Courtade}},
  \bibinfo {author} {\bibfnamefont {S.}~\bibnamefont {Shree}}, \bibinfo
  {author} {\bibfnamefont {C.}~\bibnamefont {Robert}}, \bibinfo {author}
  {\bibfnamefont {T.}~\bibnamefont {Taniguchi}}, \bibinfo {author}
  {\bibfnamefont {K.}~\bibnamefont {Watanabe}}, \bibinfo {author}
  {\bibfnamefont {A.}~\bibnamefont {Balocchi}}, \bibinfo {author}
  {\bibfnamefont {P.}~\bibnamefont {Renucci}}, \bibinfo {author} {\bibfnamefont
  {D.}~\bibnamefont {Lagarde}}, \bibinfo {author} {\bibfnamefont
  {X.}~\bibnamefont {Marie}},  \emph {et~al.},\ }\href@noop {} {\bibfield
  {journal} {\bibinfo  {journal} {Physical Review B}\ }\textbf {\bibinfo
  {volume} {99}},\ \bibinfo {pages} {035443} (\bibinfo {year}
  {2019})}\BibitemShut {NoStop}%
\bibitem [{\citenamefont {Kira}\ and\ \citenamefont
  {Koch}(2006)}]{kira2006many}%
  \BibitemOpen
  \bibfield  {author} {\bibinfo {author} {\bibfnamefont {M.}~\bibnamefont
  {Kira}}\ and\ \bibinfo {author} {\bibfnamefont {S.}~\bibnamefont {Koch}},\
  }\href {\doibase http://dx.doi.org/10.1016/j.pquantelec.2006.12.002}
  {\bibfield  {journal} {\bibinfo  {journal} {Progress in Quantum Electronics}\
  }\textbf {\bibinfo {volume} {30}},\ \bibinfo {pages} {155 } (\bibinfo {year}
  {2006})}\BibitemShut {NoStop}%
\bibitem [{\citenamefont {Korm{\'a}nyos}\ \emph {et~al.}(2015)\citenamefont
  {Korm{\'a}nyos}, \citenamefont {Burkard}, \citenamefont {Gmitra},
  \citenamefont {Fabian}, \citenamefont {Z{\'o}lyomi}, \citenamefont
  {Drummond},\ and\ \citenamefont {Fal’ko}}]{kormanyos2015k}%
  \BibitemOpen
  \bibfield  {author} {\bibinfo {author} {\bibfnamefont {A.}~\bibnamefont
  {Korm{\'a}nyos}}, \bibinfo {author} {\bibfnamefont {G.}~\bibnamefont
  {Burkard}}, \bibinfo {author} {\bibfnamefont {M.}~\bibnamefont {Gmitra}},
  \bibinfo {author} {\bibfnamefont {J.}~\bibnamefont {Fabian}}, \bibinfo
  {author} {\bibfnamefont {V.}~\bibnamefont {Z{\'o}lyomi}}, \bibinfo {author}
  {\bibfnamefont {N.~D.}\ \bibnamefont {Drummond}}, \ and\ \bibinfo {author}
  {\bibfnamefont {V.}~\bibnamefont {Fal’ko}},\ }\href@noop {} {\bibfield
  {journal} {\bibinfo  {journal} {2D Materials}\ }\textbf {\bibinfo {volume}
  {2}},\ \bibinfo {pages} {022001} (\bibinfo {year} {2015})}\BibitemShut
  {NoStop}%
\bibitem [{\citenamefont {Deilmann}\ and\ \citenamefont
  {Thygesen}(2018)}]{deilmann2018interlayer}%
  \BibitemOpen
  \bibfield  {author} {\bibinfo {author} {\bibfnamefont {T.}~\bibnamefont
  {Deilmann}}\ and\ \bibinfo {author} {\bibfnamefont {K.~S.}\ \bibnamefont
  {Thygesen}},\ }\href@noop {} {\bibfield  {journal} {\bibinfo  {journal} {Nano
  letters}\ }\textbf {\bibinfo {volume} {18}},\ \bibinfo {pages} {2984}
  (\bibinfo {year} {2018})}\BibitemShut {NoStop}%
\bibitem [{\citenamefont {Xiao}\ \emph {et~al.}(2012)\citenamefont {Xiao},
  \citenamefont {Liu}, \citenamefont {Feng}, \citenamefont {Xu},\ and\
  \citenamefont {Yao}}]{xiao2012coupled}%
  \BibitemOpen
  \bibfield  {author} {\bibinfo {author} {\bibfnamefont {D.}~\bibnamefont
  {Xiao}}, \bibinfo {author} {\bibfnamefont {G.-B.}\ \bibnamefont {Liu}},
  \bibinfo {author} {\bibfnamefont {W.}~\bibnamefont {Feng}}, \bibinfo {author}
  {\bibfnamefont {X.}~\bibnamefont {Xu}}, \ and\ \bibinfo {author}
  {\bibfnamefont {W.}~\bibnamefont {Yao}},\ }\href@noop {} {\bibfield
  {journal} {\bibinfo  {journal} {Physical review letters}\ }\textbf {\bibinfo
  {volume} {108}},\ \bibinfo {pages} {196802} (\bibinfo {year}
  {2012})}\BibitemShut {NoStop}%
\bibitem [{\citenamefont {May}\ and\ \citenamefont
  {K{\"u}hn}(2008)}]{may2008charge}%
  \BibitemOpen
  \bibfield  {author} {\bibinfo {author} {\bibfnamefont {V.}~\bibnamefont
  {May}}\ and\ \bibinfo {author} {\bibfnamefont {O.}~\bibnamefont {K{\"u}hn}},\
  }\href@noop {} {\emph {\bibinfo {title} {Charge and energy transfer dynamics
  in molecular systems}}}\ (\bibinfo  {publisher} {John Wiley \& Sons},\
  \bibinfo {year} {2008})\BibitemShut {NoStop}%
\bibitem [{\citenamefont {Verdenhalven}\ \emph {et~al.}(2014)\citenamefont
  {Verdenhalven}, \citenamefont {Knorr}, \citenamefont {Richter}, \citenamefont
  {Bieniek},\ and\ \citenamefont {Rinke}}]{verdenhalven2014dipolemomentmolek}%
  \BibitemOpen
  \bibfield  {author} {\bibinfo {author} {\bibfnamefont {E.}~\bibnamefont
  {Verdenhalven}}, \bibinfo {author} {\bibfnamefont {A.}~\bibnamefont {Knorr}},
  \bibinfo {author} {\bibfnamefont {M.}~\bibnamefont {Richter}}, \bibinfo
  {author} {\bibfnamefont {B.}~\bibnamefont {Bieniek}}, \ and\ \bibinfo
  {author} {\bibfnamefont {P.}~\bibnamefont {Rinke}},\ }\href@noop {}
  {\bibfield  {journal} {\bibinfo  {journal} {Phys. Rev. B}\ }\textbf {\bibinfo
  {volume} {89}},\ \bibinfo {pages} {235314} (\bibinfo {year}
  {2014})}\BibitemShut {NoStop}%
\bibitem [{\citenamefont {Rytova}(1967)}]{rytova1967}%
  \BibitemOpen
  \bibfield  {author} {\bibinfo {author} {\bibfnamefont {N.~S.}\ \bibnamefont
  {Rytova}},\ }\href@noop {} {\bibfield  {journal} {\bibinfo  {journal} {Moscow
  University Physics Bulletin}\ }\textbf {\bibinfo {volume} {3}},\ \bibinfo
  {pages} {18} (\bibinfo {year} {1967})}\BibitemShut {NoStop}%
\bibitem [{\citenamefont {Rytova}(2020)}]{rytova2020}%
  \BibitemOpen
  \bibfield  {author} {\bibinfo {author} {\bibfnamefont {N.~S.}\ \bibnamefont
  {Rytova}},\ }\href@noop {} {\enquote {\bibinfo {title} {Screened potential of
  a point charge in a thin film},}\ } (\bibinfo {year} {2020}),\ \Eprint
  {http://arxiv.org/abs/1806.00976} {arXiv:1806.00976 [cond-mat.mes-hall]}
  \BibitemShut {NoStop}%
\bibitem [{\citenamefont {Christiansen}\ \emph {et~al.}(2022)\citenamefont
  {Christiansen}, \citenamefont {Selig}, \citenamefont {Biegert},\ and\
  \citenamefont {Knorr}}]{christiansen2022mesoscale}%
  \BibitemOpen
  \bibfield  {author} {\bibinfo {author} {\bibfnamefont {D.}~\bibnamefont
  {Christiansen}}, \bibinfo {author} {\bibfnamefont {M.}~\bibnamefont {Selig}},
  \bibinfo {author} {\bibfnamefont {J.}~\bibnamefont {Biegert}}, \ and\
  \bibinfo {author} {\bibfnamefont {A.}~\bibnamefont {Knorr}},\ }\href
  {\doibase 10.48550/arxiv.2212.04727} {\  (\bibinfo {year} {2022}),\
  10.48550/arxiv.2212.04727}\BibitemShut {NoStop}%
\bibitem [{\citenamefont {Berkelbach}\ \emph {et~al.}(2013)\citenamefont
  {Berkelbach}, \citenamefont {Hybertsen},\ and\ \citenamefont
  {Reichman}}]{berkelbach2013theory}%
  \BibitemOpen
  \bibfield  {author} {\bibinfo {author} {\bibfnamefont {T.~C.}\ \bibnamefont
  {Berkelbach}}, \bibinfo {author} {\bibfnamefont {M.~S.}\ \bibnamefont
  {Hybertsen}}, \ and\ \bibinfo {author} {\bibfnamefont {D.~R.}\ \bibnamefont
  {Reichman}},\ }\href@noop {} {\bibfield  {journal} {\bibinfo  {journal}
  {Physical Review B}\ }\textbf {\bibinfo {volume} {88}},\ \bibinfo {pages}
  {045318} (\bibinfo {year} {2013})}\BibitemShut {NoStop}%
\bibitem [{\citenamefont {Sigl}\ \emph {et~al.}(2022)\citenamefont {Sigl},
  \citenamefont {Troue}, \citenamefont {Katzer}, \citenamefont {Selig},
  \citenamefont {Sigger}, \citenamefont {Kiemle}, \citenamefont
  {Brotons-Gisbert}, \citenamefont {Watanabe}, \citenamefont {Taniguchi},
  \citenamefont {Gerardot} \emph {et~al.}}]{sigl2022optical}%
  \BibitemOpen
  \bibfield  {author} {\bibinfo {author} {\bibfnamefont {L.}~\bibnamefont
  {Sigl}}, \bibinfo {author} {\bibfnamefont {M.}~\bibnamefont {Troue}},
  \bibinfo {author} {\bibfnamefont {M.}~\bibnamefont {Katzer}}, \bibinfo
  {author} {\bibfnamefont {M.}~\bibnamefont {Selig}}, \bibinfo {author}
  {\bibfnamefont {F.}~\bibnamefont {Sigger}}, \bibinfo {author} {\bibfnamefont
  {J.}~\bibnamefont {Kiemle}}, \bibinfo {author} {\bibfnamefont
  {M.}~\bibnamefont {Brotons-Gisbert}}, \bibinfo {author} {\bibfnamefont
  {K.}~\bibnamefont {Watanabe}}, \bibinfo {author} {\bibfnamefont
  {T.}~\bibnamefont {Taniguchi}}, \bibinfo {author} {\bibfnamefont {B.~D.}\
  \bibnamefont {Gerardot}},  \emph {et~al.},\ }\href@noop {} {\bibfield
  {journal} {\bibinfo  {journal} {Physical Review B}\ }\textbf {\bibinfo
  {volume} {105}},\ \bibinfo {pages} {035417} (\bibinfo {year}
  {2022})}\BibitemShut {NoStop}%
\bibitem [{\citenamefont {Rasmussen}\ and\ \citenamefont
  {Thygesen}(2015)}]{rasmussen_thygesen_2015}%
  \BibitemOpen
  \bibfield  {author} {\bibinfo {author} {\bibfnamefont {F.~A.}\ \bibnamefont
  {Rasmussen}}\ and\ \bibinfo {author} {\bibfnamefont {K.~S.}\ \bibnamefont
  {Thygesen}},\ }\href@noop {} {\bibfield  {journal} {\bibinfo  {journal} {The
  Journal of Physical Chemistry C}\ }\textbf {\bibinfo {volume} {119}},\
  \bibinfo {pages} {13169} (\bibinfo {year} {2015})}\BibitemShut {NoStop}%
\bibitem [{\citenamefont {Knorr}\ \emph {et~al.}(1996)\citenamefont {Knorr},
  \citenamefont {Hughes}, \citenamefont {Stroucken},\ and\ \citenamefont
  {Koch}}]{knorr1996theory}%
  \BibitemOpen
  \bibfield  {author} {\bibinfo {author} {\bibfnamefont {A.}~\bibnamefont
  {Knorr}}, \bibinfo {author} {\bibfnamefont {S.}~\bibnamefont {Hughes}},
  \bibinfo {author} {\bibfnamefont {T.}~\bibnamefont {Stroucken}}, \ and\
  \bibinfo {author} {\bibfnamefont {S.}~\bibnamefont {Koch}},\ }\href@noop {}
  {\bibfield  {journal} {\bibinfo  {journal} {Chemical physics}\ }\textbf
  {\bibinfo {volume} {210}},\ \bibinfo {pages} {27} (\bibinfo {year}
  {1996})}\BibitemShut {NoStop}%
\bibitem [{\citenamefont {Stroucken}\ \emph {et~al.}(1996)\citenamefont
  {Stroucken}, \citenamefont {Knorr}, \citenamefont {Thomas},\ and\
  \citenamefont {Koch}}]{stroucken1996coherent}%
  \BibitemOpen
  \bibfield  {author} {\bibinfo {author} {\bibfnamefont {T.}~\bibnamefont
  {Stroucken}}, \bibinfo {author} {\bibfnamefont {A.}~\bibnamefont {Knorr}},
  \bibinfo {author} {\bibfnamefont {P.}~\bibnamefont {Thomas}}, \ and\ \bibinfo
  {author} {\bibfnamefont {S.}~\bibnamefont {Koch}},\ }\href@noop {} {\bibfield
   {journal} {\bibinfo  {journal} {Physical Review B}\ }\textbf {\bibinfo
  {volume} {53}},\ \bibinfo {pages} {2026} (\bibinfo {year}
  {1996})}\BibitemShut {NoStop}%
\bibitem [{\citenamefont {Pazzagli}\ \emph {et~al.}(2018)\citenamefont
  {Pazzagli}, \citenamefont {Lombardi}, \citenamefont {Martella}, \citenamefont
  {Colautti}, \citenamefont {Tiribilli}, \citenamefont {Cataliotti},\ and\
  \citenamefont {Toninelli}}]{pazzagli2018self}%
  \BibitemOpen
  \bibfield  {author} {\bibinfo {author} {\bibfnamefont {S.}~\bibnamefont
  {Pazzagli}}, \bibinfo {author} {\bibfnamefont {P.}~\bibnamefont {Lombardi}},
  \bibinfo {author} {\bibfnamefont {D.}~\bibnamefont {Martella}}, \bibinfo
  {author} {\bibfnamefont {M.}~\bibnamefont {Colautti}}, \bibinfo {author}
  {\bibfnamefont {B.}~\bibnamefont {Tiribilli}}, \bibinfo {author}
  {\bibfnamefont {F.~S.}\ \bibnamefont {Cataliotti}}, \ and\ \bibinfo {author}
  {\bibfnamefont {C.}~\bibnamefont {Toninelli}},\ }\href@noop {} {\bibfield
  {journal} {\bibinfo  {journal} {ACS nano}\ }\textbf {\bibinfo {volume}
  {12}},\ \bibinfo {pages} {4295} (\bibinfo {year} {2018})}\BibitemShut
  {NoStop}%
\bibitem [{\citenamefont {Kira}\ \emph {et~al.}(1999)\citenamefont {Kira},
  \citenamefont {Jahnke}, \citenamefont {Hoyer},\ and\ \citenamefont
  {Koch}}]{Kira1999}%
  \BibitemOpen
  \bibfield  {author} {\bibinfo {author} {\bibfnamefont {M.}~\bibnamefont
  {Kira}}, \bibinfo {author} {\bibfnamefont {F.}~\bibnamefont {Jahnke}},
  \bibinfo {author} {\bibfnamefont {W.}~\bibnamefont {Hoyer}}, \ and\ \bibinfo
  {author} {\bibfnamefont {S.}~\bibnamefont {Koch}},\ }\href@noop {} {\bibfield
   {journal} {\bibinfo  {journal} {Progress in Quantum Electronics}\ }\textbf
  {\bibinfo {volume} {23}},\ \bibinfo {pages} {189} (\bibinfo {year}
  {1999})}\BibitemShut {NoStop}%
\bibitem [{\citenamefont {Lakowicz}(2006)}]{lakowicz2006bookluminescence}%
  \BibitemOpen
  \bibfield  {author} {\bibinfo {author} {\bibfnamefont {J.~R.}\ \bibnamefont
  {Lakowicz}},\ }\href@noop {} {\emph {\bibinfo {title} {Principles of
  fluorescence spectroscopy}}}\ (\bibinfo  {publisher} {Springer},\ \bibinfo
  {year} {2006})\BibitemShut {NoStop}%
\bibitem [{\citenamefont {Preus}\ and\ \citenamefont
  {Wilhelmsson}(2012)}]{wilhelmsson2012FRETmeasure}%
  \BibitemOpen
  \bibfield  {author} {\bibinfo {author} {\bibfnamefont {S.}~\bibnamefont
  {Preus}}\ and\ \bibinfo {author} {\bibfnamefont {L.~M.}\ \bibnamefont
  {Wilhelmsson}},\ }\href@noop {} {\bibfield  {journal} {\bibinfo  {journal}
  {ChemBioChem}\ }\textbf {\bibinfo {volume} {13}},\ \bibinfo {pages} {1990}
  (\bibinfo {year} {2012})}\BibitemShut {NoStop}%
\bibitem [{\citenamefont {Valeur}\ and\ \citenamefont
  {Berberan-Santos}(2013)}]{valeur2013measurefret}%
  \BibitemOpen
  \bibfield  {author} {\bibinfo {author} {\bibfnamefont {B.}~\bibnamefont
  {Valeur}}\ and\ \bibinfo {author} {\bibfnamefont {M.~N.}\ \bibnamefont
  {Berberan-Santos}},\ }\href@noop {} {\emph {\bibinfo {title} {Molecular
  fluorescence principles and applications}}}\ (\bibinfo  {publisher}
  {Wiley-VCH},\ \bibinfo {year} {2013})\BibitemShut {NoStop}%
\bibitem [{\citenamefont {Cohen-Tannoudji}\ \emph {et~al.}(1997)\citenamefont
  {Cohen-Tannoudji}, \citenamefont {Dupont-Roc},\ and\ \citenamefont
  {Grynberg}}]{cohen1997photons}%
  \BibitemOpen
  \bibfield  {author} {\bibinfo {author} {\bibfnamefont {C.}~\bibnamefont
  {Cohen-Tannoudji}}, \bibinfo {author} {\bibfnamefont {J.}~\bibnamefont
  {Dupont-Roc}}, \ and\ \bibinfo {author} {\bibfnamefont {G.}~\bibnamefont
  {Grynberg}},\ }\href@noop {} {\emph {\bibinfo {title} {Photons and
  Atoms-Introduction to Quantum Electrodynamics}}}\ (\bibinfo  {publisher}
  {WILEY‐VCH Verlag},\ \bibinfo {year} {1997})\BibitemShut {NoStop}%
\end{thebibliography}%
\end{document}